\RequirePackage{fix-cm}
\documentclass[final]{svjour3}     
\smartqed  

\usepackage{amsmath, amssymb, stmaryrd}
\usepackage{color}
\usepackage{xspace}
\usepackage{graphicx}
\usepackage[vlined, ruled]{algorithm2e}
\usepackage{enumerate}
\usepackage{multirow}
\usepackage{wrapfig}

\usepackage[misc,geometry]{ifsym}

\newcommand{\hide}[1]{}
\newcommand{\eat}[1]{}
\newcommand\pnote[1]{[\textcolor{red}{pbour:} \textcolor{cyan}{#1}]}

\newcommand{\stitle}[1]{\vspace{0.1cm}\noindent\textbf{#1}}

\newcommand\dasfaa{\ensuremath{\mathtt{PRETTI}}\xspace}
\newcommand\new{\ensuremath{\mathtt{OPJ}}\xspace}
\newcommand\limit{\ensuremath{\mathtt{LIMIT}}\xspace}
\newcommand\limita{\ensuremath{\mathtt{LIMIT{\small+}}}\xspace}
\newcommand\loracle{\ensuremath{\mathtt{L\!-\!ORACLE}}\xspace}
\newcommand\toracle{\ensuremath{\mathtt{T\!-\!ORACLE}}\xspace}

\newcommand{\<}{\ensuremath{\langle}}
\renewcommand{\>}{\ensuremath{\rangle}}

\begin{document}

\title{Set Containment Join Revisited\thanks{\eat{This work was conducted while P. Bouros was with the University of Hong Kong\\}To appear at the Knowledge and Information Systems Journal (KAIS).}
}


\author{Panagiotis Bouros         \and
            Nikos Mamoulis \and 
            Shen Ge \and
            Manolis Terrovitis
}

\authorrunning{P. Bouros et al.} 

\institute{Panagiotis Bouros\eat{ (\Letter)} \at
              Department of Computer Science, Aarhus University, \eat{IT-Parken, Aabogade 34, 8200 Aarhus~N, }Denmark\\
              \email{pbour@cs.au.dk}           
           \and
           N. Mamoulis
			\and
			S. Ge \at
           Department of Computer Science, The University of Hong Kong\eat{, Pokfulam Road}, Hong Kong SAR, China\\
			\email{\{nikos,sge\}@cs.hku.hk}
			\and
			M. Terrovitis \at
			Institute for the Management of Information Systems, Research Center ``Athena''\eat{, Artemidos 6 \& Epidavrou, 15125 Marousi}, Greece\\
			\email{mter@imis.athena-innovation.gr}
}


\maketitle

\begin{abstract}
Given two collections of set objects $R$ and $S$, 
the $R\Join_{\subseteq}S$ set containment join 
returns all object pairs $(r,s) \in R\times S$ such that $r\subseteq s$.
Besides being a basic operator in all modern data management systems with a wide range of applications,
the join can be used to evaluate
complex SQL queries based on relational division and as a module of
data mining algorithms.
The state-of-the-art algorithm for set containment joins (\dasfaa)
builds an inverted index on the right-hand collection $S$ and a prefix
tree on the left-hand collection $R$ that groups set objects with common
prefixes and thus, avoids redundant processing.
In this paper, we present a framework which improves
\dasfaa in two directions.
First, we limit the prefix
tree construction by proposing an
adaptive methodology based on a cost model;
this way, we can greatly reduce the space and time cost of the join.
Second, we \eat{propose to}
partition the objects of each collection based on their first contained
item, assuming that the set objects are internally sorted.
We show that we can process the partitions and evaluate the join while
building the prefix tree and the inverted index progressively.
This allows us to significantly reduce not only the join cost, but also the maximum memory requirements
during the join.
An experimental evaluation using both real and synthetic datasets shows that our
framework outperforms \dasfaa by a wide margin.
\keywords{Set-valued data \and containment join \and query processing \and inverted index \and prefix tree}
\end{abstract}

\section{Introduction}
%
%

%
%
%
%
Sets are ubiquitous in computer science and most importantly in the field of data management; they model among others transactions and scientific data, click streams and Web search data,  text. 
Contemporary data management systems allow the definition of set-valued (or multi-valued) data attributes and support operations
such as containment queries \cite{AgrawalAK10,IbrahimF13,\eat{JiaXLLM12,Savnik13,}TerrovitisBVSM11,TerrovitisPVS06,ZhangCSCGT12}. Joins
are also extended to include predicates on sets (containment,
similarity, equality, etc.)
\cite{HelmerM97}. 
In this paper, we focus on the
efficient evaluation of an important join operator: the set containment
join. Formally, let $R$, $S$ be two collections of set objects, the $R\Join_{\subseteq}S$ set containment
join returns
all pairs of \eat{set }objects $(r,s) \in R\times S$ such that $r\subseteq s$.

\stitle{Application examples/scenarios}.
Set containment joins find application in a wide range of domains for knowledge and data management. In decision support scenarios, the\eat{such a join operator} join is employed to identify resources that match a \emph{set} of preferences or qualifications, e.g., on\eat{in the context of} real estate or job agencies.
C\eat{Particularly, c}onsider a recruitment agency which besides publishing job-offers also performs a first level filtering of the candidates. \eat{Under this scenario, t}The agency retains a collection of job-offers $R$ where an object $r$ contains the set of required skills for each job, and a collection of job-seekers $S$ with $s$ capturing the
skills of each candidate. The $R\! \Join_{\subseteq}\! S$ \eat{set containment }join returns all pairs of jobs and qualifying candidates for them which the agency then forwards to \eat{every }job-offerers for making the final decision.
Containment joins can also support critical operations in data warehousing. For instance, the join can be used to compare different versions of set-valued records for entities that evolve over time (e.g., sets of products in the inventories of all departments in a company). By identifying records that subsume each other (i.e., a set containment join between two versions), the evolution of the data is monitored and possibly hidden correlations and anomalies are discovered.

In the core of traditional
database systems
and data engineering, set containment joins can be employed  to evaluate 
complex SQL queries based on division
\cite{CaoB05,RantzauSMW03}. Consider for example  Figure~\ref{fig:divex}
which shows two relational tables. The first table
shows students and the courses they have passed, while the second table
shows the required courses to be taken and passed in order for a
student to acquire a skill. 
For example, Maria has passed Operating
systems and Programming. As the courses required for a Systems
Programming skill are Operating systems and Programming, it can be
said that Maria has acquired this skill. 
Consider the query ``for each student find the
skills s/he has acquired'' expressed in SQL below:
%
\begin{normalsize}
\begin{tabbing}
~~~~~~\textbf{select} P1.Student, R1.Skill\\
~~~~~~\textbf{from} Passes \textbf{as} P1, Requires \textbf{as} R1\\
~~~~~~\textbf{where not exist}\=s (\textbf{select} R2.Course\\
~~~~~~\>~~~\textbf{from} Requires \textbf{as} R2\\
~~~~~~\>~~~\textbf{where} R1.Skill = R2.Skill\\
~~~~~~\>~~~\textbf{and not exist}\=s (\textbf{select} P2.Course\\
~~~~~~\>\>~~~\textbf{from} Passes \textbf{as} P2\\
~~~~~~\>\>~~~\textbf{where} P2.Student=P1.Student\\
~~~~~~\>\>~~~\textbf{and} P2.Course=R2.Course));
\end{tabbing}
\end{normalsize}
It is not hard to see that this query is in fact a set containment
join between tables Requires and Passes, considering each skill and student
as the set of courses they require or have passed, respectively. This
example demonstrates the usefulness of set containment joins even
in classic databases with relations in 1NF.

\eat{
\begin{figure}
\begin{center}
\begin{tabular}{cl}
\hline
\textbf{Student} & \textbf{Course}\\
\hline\hline
John & Algorithms\\
Peter & Databases\\
Maria & Op. Systems\\
Peter & Programming\\
John & Databases\\
Maria & Programming\\
Peter & Op. Systems\\
\hline
\end{tabular}\\
(a) table Passes
\\
\vspace{2ex}
\begin{tabular}{cl}
\hline
\textbf{Skill} & \textbf{Course}\\
\hline\hline
DBA & Databases\\
DBWeb & Databases \\
DBWeb & Programming \\
Sys. Prog. & Programming\\
Sys. Prog. & Op. Systems\\
\hline
\end{tabular}\\
(b) table Requires
\caption{Example of relational division based on set containment join: ``for each student find the skills s/he has acquired''}\label{fig:divex}
\end{center}
\end{figure}
}
\begin{figure}[ht]
\begin{center}
\begin{tabular}{cc}
\begin{tabular}{cl}
\hline
\textbf{Student} & \textbf{Course}\\
\hline\hline
John & Algorithms\\
Peter & Databases\\
Maria & Op. Systems\\
Peter & Programming\\
John & Databases\\
Maria & Programming\\
Peter & Op. Systems\\
\hline
\end{tabular}
&
\vspace{2ex}
\begin{tabular}{cl}
\hline
\textbf{Skill} & \textbf{Course}\\
\hline\hline
DBA & Databases\\
DBWeb & Databases \\
DBWeb & Programming \\
Sys. Prog. & Programming\\
Sys. Prog. & Op. Systems\\
\hline
\end{tabular}\\
(a) table Passes &(b) table Requires
\end{tabular}
\caption{Example of relational division based on set containment join: ``for each student find the skills s/he has acquired''}\label{fig:divex}
\end{center}
\end{figure}

In the context of data mining, containment join can act as a module during frequent itemset
mining \cite{Rantzau03}. Consider the classic Apriori algorithm
\cite{AgrawalS94} which is well-known for its generality and
adaptiveness to mining problems in most data domains; besides, studies like \cite{ZhengKM01} report that Apriori can be faster than FP-growth-like algorithms for certain support threshold ranges and datasets.
At each level, the Apriori algorithm (i) generates a set of candidate frequent
itemsets (having specific cardinality) and (ii) counts their support
in the database. Candidates verification (i.e., step (ii)), which  is typically
more expensive than candidates generation (i.e., step (i)), can be
enhanced by applying a set containment join between
the collection of candidates and the collection of database
transactions. The difference is that we do
not output the qualifying pairs, but instead count the number of pairs
where each candidate participates (i.e., a join followed by aggregation).

\stitle{Motivation}.
The above examples highlight not only the range of \eat{traditional and modern }applications for set containment join but also the importance of optimizing its evaluation.
Even though
this operation received significant attention in the past 
with
a number of algorithms proposed being either signature \cite{HelmerM97,MelnikG02,MelnikG03,RamasamyPNK00} or inverted index based \cite{JampaniP05,Mamoulis03}, to our knowledge, since then, there have
not been any new techniques that improve the state-of-the-art
algorithm \dasfaa \cite{JampaniP05}.
\dasfaa  evaluates the join by employing
an inverted index $I_S$ on the right-hand collection $S$
and a prefix tree $T_R$ on the left-hand collection $R$ that
groups set objects with common prefixes in order to avoid redundant
processing. The experiment analysis in \cite{JampaniP05} showed that \dasfaa outperforms previous inverted index-based \cite{Mamoulis03} and signature-based methods \cite{MelnikG03,RamasamyPNK00}, but as we discuss in this paper, there is still a lot of room for improvement primarily due to the following two shortcomings of \dasfaa. 
First,
the prefix tree can be too expensive to build and 
store, 
especially if $R$ contains sets of high
cardinality or very long. Second,
 \dasfaa completely traverses the prefix tree during join
evaluation, which may be unnecessary, especially if the set of
remaining candidates is small.

\stitle{Contributions}. 
Initially, we tackle the aforementioned shortcomings of \dasfaa by proposing an \emph{adaptive} evaluation methodology. In brief, we avoid building the entire prefix tree $T_R$ on left-hand collection $R$ which significantly reduces the requirements in both space and indexing time. Under this \emph{limited} prefix tree denoted by $\ell T_R$, the evaluation of set containment join becomes a two-phase procedure that involves (i) \emph{candidates generation} by traversing the prefix tree, and (ii) \emph{candidates verification}. Then, we propose a \emph{cost model} to switch \emph{on-the-fly} to candidates verification if the cost of verifying the remaining join candidates in \eat{the }current subtree is expected to be lower than prefix-tree based evaluation, i.e., candidates generation.

Next, we propose the {\em Order and Partition Join} (\new) paradigm
which considers the items of each set object in a particular order (e.g.,
in decreasing order of their frequency in the objects of $R\cup S$).
\eat{The two r}Collection $R$ and $S$ are divided into partitions such that
$R_i$ ($S_i$) contains all objects in $R$ ($S$) for which the first item is $i$. 
Then, for each item $i$ in order, \new processes partitions $R_i$ and $S_i$
by (i) updating \eat{the }inverted index $I_S$ to include all objects in
$S_i$ and (ii) creating \eat{a }prefix tree $T_{R_i}$ for partition $R_i$
and joining it with $I_S$. As the inverted index is \emph{incrementally} built, 
its lists are initially shorter and the join is faster. 
Further, the overall memory requirements 
are reduced since each $T_{R_i}$ is constructed and
processed separately, but most importantly, it can be discarded right
after joining it with $I_S$.

As an additional contribution of our study, we reveal that
ordering the set items in increasing order of their frequency (in contrast with
decreasing frequency proposed in \cite{JampaniP05}) in fact
improves query performance.
Although such an ordering may lead to a larger
prefix tree (compared to \dasfaa), it dramatically reduces the number
of candidates during query processing and enables our adaptive
technique to achieve high performance gains.

We focus on main-memory evaluation of set containment joins (i.e., we
optimize the main module of \dasfaa, which joins two in-memory
partitions); note that our solution is easily integrated in the
block-based\eat{ evaluation} approaches of \cite{JampaniP05,Mamoulis03}. The fact
that we limit the size of the prefix tree and that we use the \new paradigm, 
allows our method to operate with
larger partitions compared to \dasfaa
in an external-memory
problem, thus making our overall improvements even higher. 
Our thorough experimental evaluation using real datasets of
different characteristics 
shows that our framework always outperforms \dasfaa, being up to
more than one order of magnitude times faster and saving at least 50\% of memory.

\stitle{Outline}.
The rest of the paper is organized as follows. Section~\ref{sec:dasfaa} describes in detail the state-of-the-art set containment join algorithm \dasfaa. Our adaptive evaluation methodology and the \new novel join paradigm\eat{ that employs the ordering of the objects in the right-hand collection $S$} are presented in Sections~\ref{sec:limits} and \ref{sec:new}, respectively. Section~\ref{sec:exps} presents our experimental evaluation. Finally,  Section~\ref{sec:related} reviews related work and Section~\ref{sec:concl} concludes the paper.
\section{Background on Set Containment Join: The {\dasfaa} Algorithm}
\label{sec:dasfaa}

In this section, we describe in detail the state-of-the-art method \dasfaa \cite{JampaniP05} for computing the $R \bowtie_{\subseteq} S$ set containment join of two collections $R$ and $S$. The method has the following key features:
\begin{enumerate}[(i)]
\item The left-hand collection $R$ is indexed by a prefix tree $T_R$ and the
  right-hand collection $S$ by an inverted index $I_S$. Both index structures
  are built on-the-fly, which 
  enables the generality of the algorithm (for example, it can be
  applied for arbitrary data partitions instead of entire collections, and/or
  on data produced by underlying operators without interesting orders).
  %
%
\item \dasfaa traverses the prefix tree $T_R$ in a depth-first
  manner. While following a path on the tree, the algorithm intersects
  the corresponding lists of inverted index $I_S$\eat{ of right-hand
  collection $S$}. The join algorithm is identical to the one proposed in
  \cite{Mamoulis03} (see Section \ref{sec:related}); however, due to grouping the objects under $T_R$, \dasfaa performs the intersections for
  all sets in $R$ with a common prefix only once.
\end{enumerate}

\begin{algorithm}
\SetKwFunction{ContructPrefixTree}{ContructPrefixTree}
\SetKwFunction{ConstructInvertedIndex}{ConstructInvertedIndex}
\SetKwFunction{ProcessNode}{ProcessNode}
\SetKwFunction{Verify}{Verify}
\SetKwInOut{Input}{input}
\SetKwInOut{Output}{output}
\LinesNumbered
\Input{Collections $R$ and $S$; every object $r \in R$ is internally sorted such that the most frequent item appears first}
\Output{the set $J$ of all object pairs $(r,s)$ such that $r \in R$, $s \in S$ and $r \subseteq s$}
\BlankLine
\nlset{1}$T_R \leftarrow \ContructPrefixTree(R)$\;
\nlset{2}$I_S \leftarrow \ConstructInvertedIndex(S)$\;
\ForEach{child node $c$ of the root in $T_R$}
{
	$CL \leftarrow \{s | s \in S\}$\tcp*[r]{Candidates list}
	$\ProcessNode(c, CL, I_S,J)$\;
}
\Return $J$;
%
\\
\vspace{3ex}
\bf{Function} $\ProcessNode(n, CL, I_S, J)$\\
$CL' \leftarrow CL \cap I_S[n.item]$\tcp*[r]{List intersection}
\ForEach{object $r \in n.RL$}
{
  \ForEach{object $s \in CL'$}
  {

		$J \leftarrow J \cup (r,s)$\;
	}
}
\ForEach{child node $c$ of $n$}
{
	$\ProcessNode(c,CL',I_S,J)$\tcp*[r]{Recursion}
}
\caption{$\dasfaa(R,S)$}
\label{algo:dasfaa}
\end{algorithm}

\begin{figure}
\begin{center}
\begin{tabular}{@{~~}cc@{~~}}
\begin{tabular}{l}
$r_1\!\!: \{G,F,E,C,B\}$\\
$r_2\!\!: \{G,F,D,B\}$\\
$r_3\!\!: \{G,D,A\}$\\
$r_4\!\!: \{F,D,C,B\}$\\
$r_5\!\!: \{G,F,E\}$\\
$r_6\!\!: \{E,C\}$\\
$r_7\!\!: \{G,F,E\}$\\
\\
\\
\\
\\
\\
\end{tabular}
\eat{
&
\begin{tabular}{l}
$s_1\!\!: \{A,D,C\}$\\
$s_2\!\!: \{G,A,F,C,D,E\}$\\
$s_3\!\!: \{D,B\}$\\
$s_4\!\!: \{F,G,B,C\}$\\
$s_5\!\!: \{G,E,B,F\}$\\
$s_6\!\!: \{B,D,F,E,C\}$\\
$s_7\!\!: \{D,G,E,B,C\}$\\
$s_8\!\!: \{E,G,B,C,D\}$\\
$s_9\!\!: \{G,F,E,D\}$\\
$s_{10}\!\!: \{G,F,E,D\}$\\
$s_{11}\!\!: \{F,G\}$\\
$s_{12}\!\!: \{G,E,F\}$\\
\end{tabular}
}
&\begin{tabular}{l}
$s_1\!\!: \{D,C,A\}$\\
$s_2\!\!: \{G,F,E,D,C,A\}$\\
$s_3\!\!: \{D,B\}$\\
$s_4\!\!: \{G,F,C,B\}$\\
$s_5\!\!: \{G,F,E,B\}$\\
$s_6\!\!: \{F,E,D,C,B\}$\\
$s_7\!\!: \{G,E,D,C,B\}$\\
$s_8\!\!: \{G,E,D,C,B\}$\\
$s_9\!\!: \{G,F,E,D\}$\\
$s_{10}\!\!: \{G,F,E,D\}$\\
$s_{11}\!\!: \{G,F\}$\\
$s_{12}\!\!: \{G,F,E\}$\\
\end{tabular}\\\\
\normalsize{(a) left-hand collection $R$} &\normalsize{(b) right-hand collection $S$} \\
\end{tabular}
\end{center}
\caption{Example of two collections $R$ and $S$}
\label{fig:running_example_collections}
\end{figure}
\begin{figure}
\begin{center}
\begin{tabular}{cc}
\begin{tabular}{l}
\includegraphics[width=0.5\linewidth]{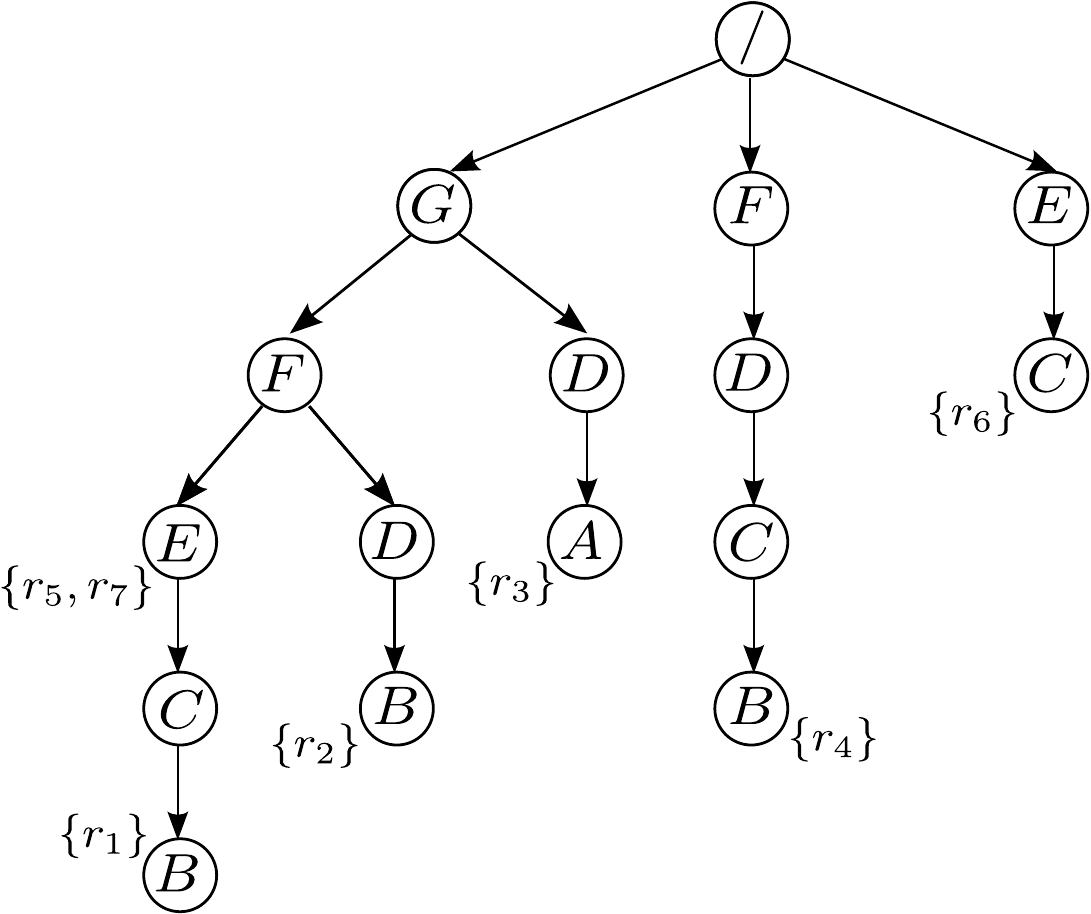}\\
\end{tabular}
&
\begin{tabular}{l}
$A\!\!: \{s_1,\!s_2\}$\\
$B\!\!: \{s_3,s_4,\!s_5,\!s_6,\!s_7,\!s_8\}$\\
$C\!\!: \{s_1,\!s_2,\!s_4,\!s_6,\!s_7,\!s_8\}$\\
$D\!\!: \{s_1,\!s_2,\!s_3,\!s_6,\!s_7,\!s_8,\!s_9,\!s_{10}\}$\\
$E\!\!: \{s_2,\!s_5,\!s_6,\!s_7,\!s_8,\!s_9,\!s_{10},\!s_{12}\}$\\
$F\!\!: \{s_2,\!s_4,\!s_5,\!s_6,\!s_9,\!s_{10},\!s_{11},\!s_{12}\}$\\
$G\!\!: \{s_2,\!s_4,\!s_5,\!s_7,\!s_8,\!s_9,\!s_{10},\!s_{11},\!s_{12}\}$\\
\end{tabular}
\\
(a) prefix tree $T_R$&(b) inverted index $I_S$
\end{tabular}
\end{center}
\caption{Indices of\eat{built by} \dasfaa for the collections in
  Figure~\ref{fig:running_example_collections}}
\label{fig:running_example_indices}
\end{figure}

Algorithm~\ref{algo:dasfaa} illustrates the pseudocode of \dasfaa.
%
During the initialization phase (Lines
1--2), \dasfaa builds  prefix tree $T_R$ and inverted
index $I_S$ for input collections $R$ and $S$, respectively. To
construct $T_R$, every object $r$ in $R$ is internally sorted, so that
its items appear in decreasing order of their frequency in $R$
(this ordering is expected to achieve the highest path
compression for $T_R$).\footnote{Our experiments show that an
  increasing frequency order is in practice more beneficial. Yet, for
  the sake of readability, we present both \dasfaa and our methodology considering a decreasing order.}
Each node $n$ of prefix tree $T_R$ is a triple $(item,path,RL)$
where $n.item$ is an item, 
$n.path$ is
the sequence of the items in the nodes from the root
of $T_R$ to $n$ (including $n.item$)\eat{
, i.e., $n.path = \{/,\ldots,n.item\}$
\note{excluding $n$, otherwise the description
  below is not correct}\pnote{Edw na valoume auto pou einai pio intuitive: dyo epiloges (i) auto pou exeis, (ii) $n.path$ includes $n.item$, i.e., $n.path = \{/,\ldots,n.item\}$ kai tote de xreiazetai na allaksoume tipota sto ypoloipo keimeno. Poio sou moiazei pio intuitive esena?}}, and finally, $n.RL$ is the set of objects in
$R$ whose content is equal to $n.path$.
For example, Figure~\ref{fig:running_example_indices}(a) depicts
prefix tree $T_R$ for collection $R$ in
Figure~\ref{fig:running_example_collections}(a). 
Set $n.RL$ is shown next to every node $n$ unless it is empty. 
The inverted index $I_S$ on collection $S$ associates each item $i$ in the domain of $S$ to a \emph{postings} list denoted by $I_S[i]$. The $I_S[i]$ postings list has an entry for every object $s \in S$ that contains item $i$. Figure~\ref{fig:running_example_indices}(b) pictures inverted index $I_S$ for collection $S$ in Figure~\ref{fig:running_example_collections}(b).

The second phase of the algorithm involves the computation of the join
result set $J$ (Lines 3--5). \dasfaa traverses 
the subtree rooted at every child node $c$ of $T_R$'s root
by recursively calling the $\ProcessNode$ function. For a node $n$,
$\ProcessNode$ receives as input from its parent node $p$ in $T_R$, a
candidates list $CL$. List $CL$ includes all objects $s\in S$
that contain every item in $p.path$, i.e., $p.path \subseteq s$.
Note that for every child of the root in $T_R$, $CL=S$. 
Next, $\ProcessNode$ intersects $CL$ with 
inverted list $I_S[n.item]$ to find the objects in $S$ that
contain 
$n.path$ and stores them in $CL'$ (Line
8). 
At this point, every pair of objects in $n.RL\times CL'$ is guaranteed
to be a join result (Lines 9--11). Finally, the algorithm calls $\ProcessNode$ for every child node of $n$ (Line 12--13).

\begin{example}
\label{ex:dasfaa}
{\em
We demonstrate \dasfaa for the set containment join of collections $R$
and $S$\eat{ shown} in Figure~\ref{fig:running_example_collections}. The algorithm
constructs prefix tree $T_R$ and inverted index $I_S$ shown in
Figures~\ref{fig:running_example_indices}(a) and \ref{fig:running_example_indices}(b), respectively. To
construct $T_R$ note that the items inside every object $r \in R$ are internally
sorted in decreasing order of global item frequency in $R$
(this is not necessary for the objects in $S$).
First, \dasfaa traverses the leftmost subtree of $T_R$ under the node
labeled by item $G$. Considering paths $\</,G\>$ and $\</,G,F\>$,
the algorithm intersects candidates list $CL$ (initially
containing every object in $S$, i.e., $\{s_1,\ldots,s_{12}\}$) first
with $I_S[G]$ and then with $I_S[F]$, and produces candidates list
$\{s_2,\!s_4,\!s_5,\!s_9,$ $\!s_{10},\!s_{11},\!s_{12}\}$, i.e., \eat{the
subset of }the objects in $S$ that contain both $G$ and $F$. The $RL$
lists of the nodes examined so far are empty and thus, no result pair
is reported. Next, path $\</,G,F,E\>$ is considered where 
$CL$ is intersected with $I_S[E]$ producing $CL' =
\{s_2,\!s_5,\!s_9,\!s_{10},\!s_{12}\}$. At current node, $RL = \{r_5,\!r_7\}$, and thus, \dasfaa reports result pairs $(r_5,s_2)$, $(r_5,s_5)$, $(r_5,s_9)$, $(r_5,s_{10})$, $(r_5,s_{12})$, $(r_7,s_2)$, $(r_7,s_5)$, $(r_7,s_9)$, $(r_7,s_{10})$, $(r_7,s_{12})$. 
The algorithm proceeds in this manner to examine the rest of the prefix tree nodes performing in total $15$ list intersections. The result of the join contains $16$ pairs of objects. \hfill $\blacksquare$
}
\end{example}

Finally, to deal with the case where the available main memory is not
sufficient for computing the entire set containment join of the input
collections, a partition-based join strategy was also proposed in
\cite{JampaniP05}. Particularly, the input collections $R$ and $S$ are
horizontally partitioned so that the prefix tree and the inverted
index for each pair of partitions $(R_i,S_j)$ from $R$ and $S$,
respectively, fit in memory. Then, in a nested-loop fashion, each
partition $R_i$ is joined in memory with every partition $S_j$ in $S$
invoking $\dasfaa(R_i,S_j)$.
\eat{
\note{why should $S$ be partitioned? the inverted index can be in
  secondary memory without any implications. I think also this makes
  our life easier because we do not have to argue how we deal with $S$
  if it does not fit in memory, and we can only focus on $R$ (which we
  handle with the reduced tree size).}
\\
\pnote{To partitioning tou $S$ den einai kati pou egrapsa egw; to lene
  sto \cite{JampaniP05}. Episis theloume i lysi na einai generic,
  efarmosimi akoma kai an den yaprxei index stin $S$ dioti p.x. den
  yparxei $S$ einai apotelesma kapoiou allou operator. An veloume ton
  $I_S$ sto disko allazei ligo to geniko plano}
}
\section{An Adaptive Methodology}
\label{sec:limits}
By employing a prefix tree on the left-hand collection $R$, \dasfaa avoids redundant intersections and thus outperforms previous methods that used only inverted indices, e.g., \cite{Mamoulis03}. However, we observe two important shortcomings of the \dasfaa algorithm. 
First, the cost of building and storing the prefix tree\eat{ $T_R$} on $R$ can be high
especially if $R$ contains sets of high cardinality. This raises a challenge when the available memory is limited which is only partially addressed by the partition-based \eat{nested loop }join strategy\eat{ proposed} in \cite{JampaniP05}. Second, after a candidates list $CL$ becomes short\eat{small}, continuing the traversal of the prefix tree to obtain the join results for $CL$  may incur many unnecessary in practice inverted list intersections.
\eat{
By employing a prefix tree on the left-hand collection $R$, \dasfaa avoids redundant intersections and therefore, it outperforms previous methods that used only inverted indices, e.g., \cite{Mamoulis03}. However, we observe two shortcomings of the \dasfaa algorithm. 
First, the cost of building and storing the prefix tree on $R$ can be high
especially if $R$ contains sets of high cardinality.
In this case, there could be too many nodes $n$ of the tree, for which $n.RL=\emptyset$, i.e., there are no objects in $R$ equal to $n.path$.
This raises a challenge when the available memory is limited. This problem is partially addressed by the partition-based \eat{nested loop }join strategy proposed in \cite{JampaniP05}; however, partitioning negatively affects the efficiency of the join method since common prefixes of objects across different partitions of the left-hand input collection are not considered together. For the second shortcoming of \dasfaa, consider 
object $r_1$ in Figure~\ref{fig:running_example_collections}(a) and the prefix tree $T_R$ in Figure~\ref{fig:running_example_indices}(a). After following path $\{/,G,F,E\}$ and intersecting the corresponding postings lists of the inverted index $I_S$ in Figure~\ref{fig:running_example_indices}(b), the candidates list is $CL = \{s_2,\!s_5,\!s_9,\!s_{10},\!s_{12}\}$. Next, \dasfaa will intersect $CL$ with $I_S[C]$ and $I_S[B]$ only to find out that no objects in $S$ can be joined with $r_1$; however, we could do much faster by directly verifying whether $r_1$ can join with each member of $CL$.
This toy example shows that in some cases performing list intersections could actually slow down the join process. \note{this example is not convincing. why would the cost of verifying 5 candidates be lower compared to the cost of intersecting 2 lists?}
}
This section presents an adaptive methodology which builds upon and improves \dasfaa. In Section~\ref{sec:limit} we primarily target the first shortcoming of \dasfaa proposing the \limit algorithm, while in Section~\ref{sec:limita} we propose an extension to \limit, termed \limita, that additionally deals with the second shortcoming.

\subsection{The \limit Algorithm}
\label{sec:limit}
To deal with\eat{ the first shortcoming of the \dasfaa algorithm, i.e.,} the high building and storage cost of the prefix tree $T_R$, \cite{JampaniP05} suggests to partition $R$, as discussed in the previous section.
Instead, we propose to build $T_R$ only up to a predefined maximum depth $\ell$, called {\em limit}. Hence\eat{Consequently}, computing set containment join becomes a two-phase process that  involves a \emph{candidate generation} and a \emph{verification} stage; for every candidate pair $(r,s)$ with $|r| > \ell$ we need to compare the suffixes of objects $r$ and $s$ beyond $\ell$ in order to determine whether $r \subseteq s$. This approach is adopted by the \limit algorithm.

\begin{figure}
\begin{center}
\begin{tabular}{cc}
\begin{tabular}{l}
\hspace*{-0.6cm}
\includegraphics[width=0.45\linewidth]{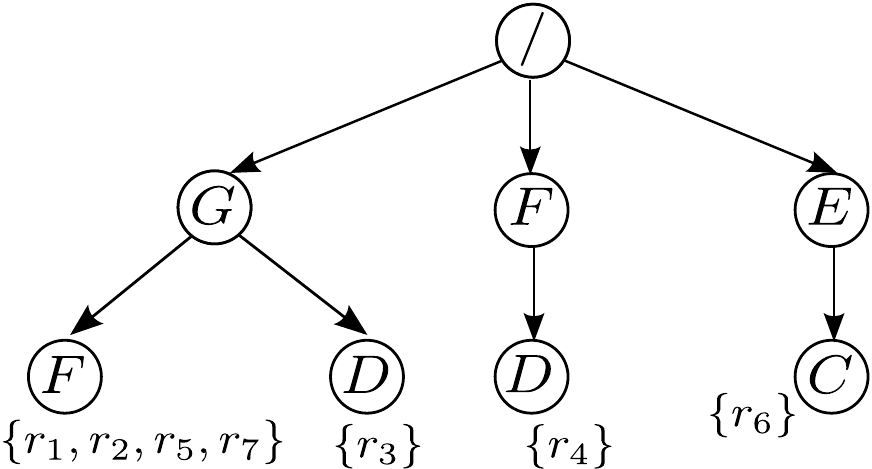}\\
\end{tabular}
&
\hspace*{-0.8cm}
\begin{tabular}{l}
\includegraphics[width=0.5\linewidth]{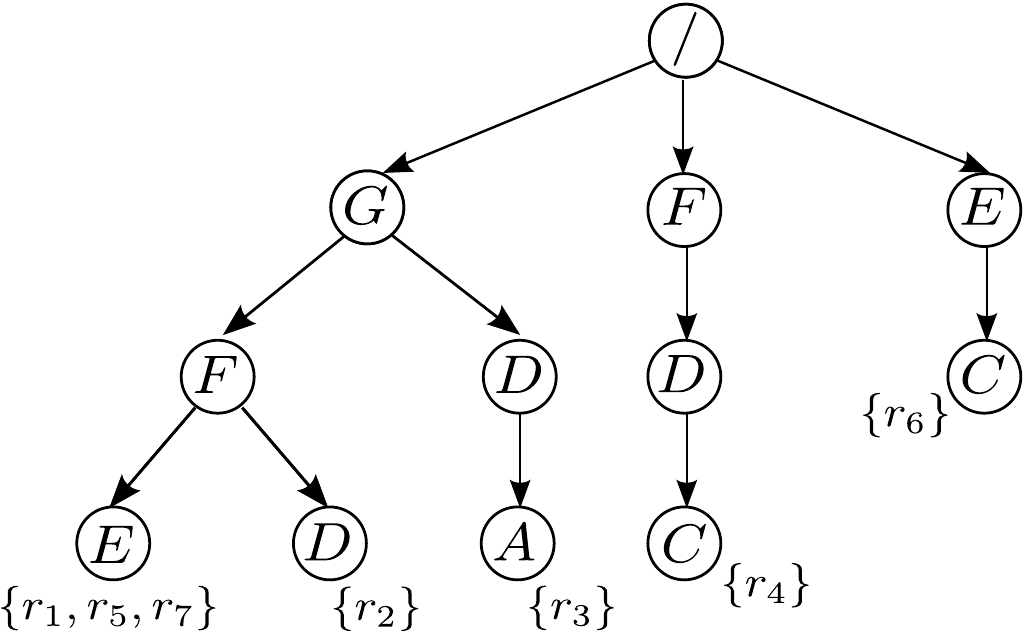}\\
\end{tabular}
\\
(a) $\ell = 2$ &(b) $\ell = 3$
\end{tabular}
\end{center}
\caption{\emph{Limited} prefix tree $\ell T_R$ for collection $R$ in Figure~\ref{fig:running_example_collections}}
\label{fig:running_example_limited_prefix_trees}
\end{figure}

Algorithm~\ref{algo:limit} illustrates the pseudocode of 
\limit\eat{ for set containment joins}. Compared to \dasfaa (Algorithm~\ref{algo:dasfaa}), 
\limit differs in two ways. First in Line~1, \limit constructs \emph{limited} prefix tree $\ell T_R$ on the left-hand collection $R$ w.r.t. limit $\ell$. The $\ell T_R$ prefix tree has almost identical structure to \eat{the prefix tree }\emph{unlimited} $T_R$ built by \dasfaa except that the $n.RL$ list of a {\em leaf} node $n$ contains every object $r \in R$ with $r \supseteq n.path$ instead of $r = n.path$.
Figures~\ref{fig:running_example_limited_prefix_trees}(a) and (b) illustrate the limited versions of the prefix tree in Figure~\ref{fig:running_example_indices}(b) for $\ell = 2$ and $\ell = 3$, respectively. 
Second, the $\ProcessNode$ function 
distinguishes between two cases of objects in $n.RL$ (Lines 11--14). 
If, for a object $r\in n.RL$, $|r| \leq \ell$ holds, then $r = n.path$ 
and, similar to \dasfaa, pair $(r,s)$ is guaranteed to be part of the join result $J$ (Line 12). Otherwise, $r \supset n.path$ 
holds and $\ProcessNode$ invokes the $\Verify$ function which compares the suffixes of objects $r$ and $s$ beyond $\ell$ (Line 14). 
Intuitively, the latter case arises only for \emph{leaf} nodes according to the definition of the \emph{limited} prefix tree.
To achieve a low verification cost, the objects of \emph{both} $R$ and $S$ collections are internally sorted, i.e., the items appear in decreasing order of their frequency in $R \cup S$,  which enables \Verify to operate in a merge-sort manner.

\begin{algorithm}[t]
\SetKwFunction{ContructPrefixTree}{ContructPrefixTree}
\SetKwFunction{ConstructInvertedIndex}{ConstructInvertedIndex}
\SetKwFunction{ProcessNode}{ProcessNode}
\SetKwFunction{Verify}{Verify}
\SetKwInOut{Input}{input}
\SetKwInOut{Output}{output}
\LinesNumbered
\Input{Collections $R$ and $S$, limit $\ell$; every object $r\! \in\! R$ and $s\! \in\! S$ is internally sorted such that the most frequent item in $R \cup S$ appears first}
\Output{the set $J$ of all object pairs $(r,s)$ such that $r \in R$, $s \in S$ and $r \subseteq s$}
\BlankLine
\nlset{1}$\ell T_R \leftarrow\ContructPrefixTree(R,\ell)$\;
\nlset{2}$I_S \leftarrow \ConstructInvertedIndex(S)$\;
\ForEach{child node $c$ of the root in $T_R$}
{
	$CL \leftarrow \{s | s \in S\}$\tcp*[r]{Candidates list}
	$\ProcessNode(c, \ell, CL, I_S, J)$\;
}
\Return $J$;
%
\\
\vspace{3ex}
\bf{Function} $\ProcessNode(n, \ell, CL, I_S, J)$\\
$CL' \leftarrow CL \cap I_S[n.item]$\tcp*[r]{List intersection}
\ForEach{object $s \in CL'$}
{
	\ForEach{object $r \in n.RL$}
	{
		\If{$|r| \leq \ell$}
		{
			$J \leftarrow J \cup (r,s)$\;
		}
		\Else
		{
			$\Verify(r,s,\ell,J)$\tcp*[r]{Compare object suffixes}
		}
	}
}
\ForEach{child node $c$ of $n$}
{
	$\ProcessNode(c, \ell, CL',I_S, J)$\tcp*[r]{Recursion}
}
\caption{$\limit(R,S,\ell)$}
\label{algo:limit}
\end{algorithm}

\begin{example}
\label{ex:limit}
{\em
We demonstrate \limit using collections $R$ and $S$ in Figure~\ref{fig:running_example_collections};
in contrast to \dasfaa and Example~\ref{ex:dasfaa}, the objects of both collections are internally sorted. Consider first the case of $\ell = 2$. \limit constructs \emph{limited} prefix tree $\ell T_R$ shown in Figure~\ref{fig:running_example_limited_prefix_trees}(a) for collection $R$ in Figure~\ref{fig:running_example_collections}(a), and inverted index $I_S$ in Figure~\ref{fig:running_example_indices}(b). Then, similar to \dasfaa, it traverses $\ell T_R$.
When considering path $\</,G,F\>$, candidates list $CL' = \{s_2,\!s_4,\!s_5,\!s_9,\!s_{10},$ $\!s_{11},\!s_{12}\}$ is produced. The $RL = \{r_1,\!r_2,\!r_5,\!r_7\}$ set of current node ($F$) is non-empty and thus, the algorithm examines every pair of objects from $RL \times CL'$ to report join results. As all objects in $RL$ are of length larger than limit $\ell = 2$, \limit compares the suffixes beyond length $\ell = 2$ of all candidates by calling \Verify, and finally, reports results $(r_5,s_2)$, $(r_5,s_5)$, $(r_5,s_9)$, $(r_5,s_{10})$, $(r_5,s_{12})$, $(r_7,s_2)$, $(r_7,s_5)$, $(r_7,s_9)$, $(r_7,s_{10})$, $(r_7,s_{12})$. At the next steps, the algorithm proceeds in a similar way to examine the rest of the prefix tree nodes performing $4$ list intersections and verifying $37$ candidate pairs by comparing their suffixes.
\hide{
Next, consider limit $\ell = 3$. In this case, \limit constructs the limited prefix tree $\ell T_R$ shown in Figure~\ref{fig:running_example_limited_prefix_trees}(b). Compared to the case of $\ell = 2$, when traversing the left-most subtree of $T_R$, the algorithm considers also path $\{/,G,F,E\}$ and thus, it additionally intersects $CL = \{s_2,\!s_4,\!s_5,\!s_9,\!s_{10},\!s_{11},\!s_{12}\}$ with $I_S[E]$ producing $CL' = \{s_2,\!s_5,\!s_9,\!s_{10},\!s_{12}\}$, similar to \dasfaa. However, at this point, \limit has to deal with two types of objects inside the $RL$ list of current node. Specifically, on one hand it treats $\{r_5,\!r_7\}$ exactly as \dasfaa does since $|r_5| = |r_7| = \ell$, and thus, without any verification it reports result pairs $(r_5,s_2)$, $(r_5,s_5)$, $(r_5,s_9)$, $(r_5,s_{10})$, $(r_5,s_{12})$, $(r_7,s_2)$, $(r_7,s_5)$, $(r_7,s_9)$, $(r_7,s_{10})$, $(r_7,s_{12})$. On the other hand, it verifies the suffixes for every combination of $r_1$ with the objects in $CL'$. Due to a larger limit $\ell$ value, \limit now performs $8$ list intersections but has to verify only $10$ candidate object pairs by comparing their suffices, in total. \note{I think it is not necessary to describe both cases of $\ell=2$ and $\ell=3$. Describing just $\ell=3$ would be enough}\pnote{Minor, to allazoume argotera an theloume na kerdisoume xwro}\note{let's deal with this later}}
Finally, if $\ell = 3$ \limit traverses similarly prefix tree $\ell T_R$ in Figure~\ref{fig:running_example_limited_prefix_trees}(b) performing $8$ this time list intersections but verifying only $10$ candidate object pairs by comparing their suffixes.
\hfill $\blacksquare$
}
\end{example}

The advantage of \limit over \dasfaa and the partition-based join strategy of \cite{JampaniP05} is two-fold. First, building the prefix tree up to $\ell$ is \eat{naturally }faster than building the 
entire tree, but most importantly, with $\ell$, the space needed to store the tree in main memory is reduced. 
If the \emph{unlimited} $T_R$ does not fit in memory, \dasfaa would partition $R$ and construct a separate (memory-based)  $T_{R_i}$ for each partition $R_i$; therefore, two objects $r_i$, $r_j$ of $R$ that have the same $\ell$-prefix but belong to different partitions $R_i$ and $R_j$, would be considered separately, which increases\eat{affects negatively} the evaluation cost of the join. 
In other words, reducing the size of $T_R$ to fit in memory can have high impact on performance.
In contrast, \limit guarantees that, for every path of length up to $\ell$ on \emph{limited} $\ell T_R$, all redundant intersections are avoided similar to \eat{the case of}utilizing the \emph{unlimited}\eat{entire} prefix tree. Finally\eat{In addition}, an interesting aftermath of employing $\ell$ for set containment joins is related to the second shortcoming of \dasfaa. For instance, with $\ell = 3$ and \eat{the }prefix tree $\ell T_R$ in Figure~\ref{fig:running_example_indices}(b), \limit will verify object $r_1$ against $CL\! =\! \{s_2,\!s_5,\!s_9,\!s_{10},\!s_{12}\}$ and quickly determine that it is not part of the join result without performing two additional inverted list intersections. 

An issue still open involves how limit $\ell$ is defined 
and most importantly, whether there is an optimal value of $\ell$ that balances the benefits of using the \emph{limited} prefix tree over the cost of including a verification stage. Determining 
the optimal value for $\ell$ is a time-consuming task which involves more than\eat{ just an additional} an extra pass over the input collections. In specific, it requires computing expensive statistics with a process reminiscent to frequent itemsets mining; note that this process must take place online before building $\ell T_R$. Instead, in Section~\ref{sec:limit_effect} we\eat{ present and evaluate} discuss 
and evaluate four strategies for estimating a good $\ell$ value based on simple and cheap-to-compute statistics. 
Our analysis shows that typically these strategies tend to overestimate the optimal $\ell$. Besides, we also observe that the optimal $\ell$ value may in fact vary between different subtrees of $\ell T_R$ depending on the number of objects stored inside the nodes. In view of this, \eat{in the next section, }we next propose an {\em adaptive} extension to \limit which employs an ad-hoc limit $\ell$ for each path of $\ell T_R$ by dynamically choosing between list intersection and verification of the objects under the current subtree. 

\subsection{The \limita Algorithm}
\label{sec:limita}
As Example~\ref{ex:limit} shows\eat{demonstrates}, using\eat{employing} limit $\ell$ for\eat{to compute} set containment joins introduces an interesting trade-off between list intersection and candidates verification which is directly related to the second shortcoming of the \dasfaa algorithm. Specifically, as $\ell$ increases and \limit traverses longer paths of $\ell T_R$, candidates lists $CL$ shorten due to the additional list intersections performed. Consequently, the number of object pairs to be verified by accessing their suffixes also reduces. However, from some point on, the number of candidates in $CL$ no longer significantly reduces or, even worst, it remains unchanged; therefore, performing additional list intersections becomes a bottleneck. Similarly, if for a node $n$, $CL$ is already too short, verifying the candidate pairs between the contents of $CL$ and the objects contained under the subtree rooted at $n$ can be faster than performing additional list intersections.
 
The \limit algorithm \eat{may }addresses only a few of the cases when candidates verification is preferred over list intersection, for instance the case of object $r_1$ in Figure~\ref{fig:running_example_collections}(a) with limit $\ell = 3$. Due to \emph{global} limit $\ell$, the ``blind" approach of \limit processes every path of the prefix tree in the same manner. To tackle this problem, we devise an \emph{adaptive} strategy of processing $\ell T_R$ adopted by the \limita algorithm. Apart from \emph{global} limit $\ell$, \limita also employs a dynamically determined
 \emph{local} limit $\ell_p$ for each path $p$ of the prefix tree.
The basic idea behind this process is to decide on-the-fly for every node $n$ of the prefix tree between:
\begin{enumerate}[(A)]
\item performing the $CL' = CL \cap I_S[n.item]$ intersection, reporting the pairs  in $n.RL \times CL'$, and then, processing the descendant nodes of $n$ in a similar way, or
\item stopping the traversal of the current path and verifying the candidates between 
the objects of $R$ contained in the subtree rooted at $n$ 
denoted by $\ell T_R^n$ and those in $CL$, i.e., all candidate pairs in $\ell T_R^n \times CL$. 
\end{enumerate}  
In the first case, \limita  would operate exactly as \limit does for the \emph{internal} nodes of $\ell T_R$ while in the second case, it would treat node $n$ as a \emph{leaf} node but without performing the corresponding list intersection. Therefore, in practice, a \emph{local} limit for current path $n.path$ is employed by \limita.

\begin{algorithm}[!t]
\SetKwFunction{ContructPrefixTree}{ContructPrefixTree}
\SetKwFunction{ConstructInvertedIndex}{ConstructInvertedIndex}
\SetKwFunction{ProcessNode}{ProcessNode}
\SetKwFunction{MakeDecision}{ContinueAsLIMIT}
\SetKwFunction{Verify}{Verify}
\SetKwInOut{Input}{input}
\SetKwInOut{Output}{output}
\LinesNumbered
\Input{Collections $R$ and $S$, limit $\ell$; every object $r\! \in\! R$ and $s\! \in\! S$ is internally sorted such that the most frequent item in $R \cup S$ appears first}
\Output{the set $J$ of all object pairs $(r,s)$ such that $r \in R$, $s \in S$ and $r \subseteq s$}
\BlankLine
\nlset{1}$\ell  T_R \leftarrow \ContructPrefixTree(R,\ell)$\;
\nlset{2}$I_S \leftarrow \ConstructInvertedIndex(S)$\;
\ForEach{child node $c$ of the root in $T_R$}
{
	$CL \leftarrow \{s | s \in S\}$\tcp*[r]{Candidates list}
	$\ProcessNode(c, \ell, CL, I_S, J)$\;
}
\Return $J$;
%
\\
\vspace{3ex}
\bf{Function} $\ProcessNode(n, \ell, CL, I_S, J)$\\
\If{$\MakeDecision(n,CL,I_S)$}
{
	$CL' \leftarrow CL \cap I_S[n.item]$\tcp*[r]{List intersection}
	\ForEach{object $s \in CL'$}
	{
		\ForEach{object $r \in n.RL$}
		{
			\If{$|r| \leq \ell$}
			{
				$J \leftarrow J \cup (r,s)$\;
			}
			\Else
			{
				$\Verify(r,s,\ell,J)$\tcp*[r]{Compare object suffixes}
			}
		}
	}
	\ForEach{child node $c$ of $n$}
	{
		$\ProcessNode(c, \ell, CL',I_S, J)$\tcp*[r]{Recursion}
	}
}
\Else
{
	\ForEach{object $s \in CL$}
	{
		\ForEach(\tcp*[f]{$\ell T_R^n$:subtree under $n$}){object $r \in \ell T_R^n$\eat{$r \in n.RL \bigcup \ell T_R^n$}}
		{
			$\Verify(r,s,\ell\!-\!1,J)$\tcp*[r]{Compare object suffixes}
		}
	}
}
\caption{$\limita(R,S,\ell)$}
\label{algo:limita}
\end{algorithm}

Algorithm~\ref{algo:limita}  illustrates the pseudocode of \limita. Compared to \limit (Algorithm~\ref{algo:limit}), \limita only differs on how a node of $\ell T_R$ is processed. Specifically, given a node $n$, $\ProcessNode$ calls the $\MakeDecision$ function (Line 8) to determine whether the algorithm will continue processing $n$ similar to \limit (Lines 10--17), or it will stop traversing current path $n.path$ and start verifying 
all candidates in $\ell T_R^n\times CL$ invoking the \Verify function (Lines 18--21).
 In the latter case, notice that for every verifying pair $(r,s)$ with 
$r \in \ell T_R^n\times CL$  and $s \in CL$, the algorithm accesses the suffixes of $r$ and $s$ beyond length $\ell-1$ and not $\ell$ as the $CL \cap I_S[n.item]$ intersection has not taken place for current node $n$ (Line 21).

\begin{figure}
\begin{center}
\begin{tabular}{cc}
\begin{tabular}{l}
\includegraphics[width=0.32\linewidth]{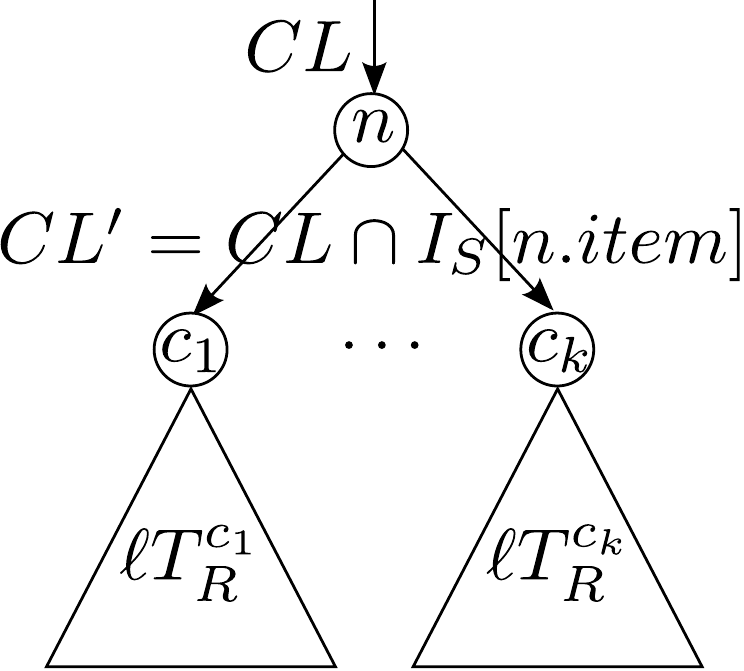}\\
\end{tabular}
&
\hspace*{-0.3cm}
\begin{tabular}{l}
\includegraphics[width=0.23\linewidth]{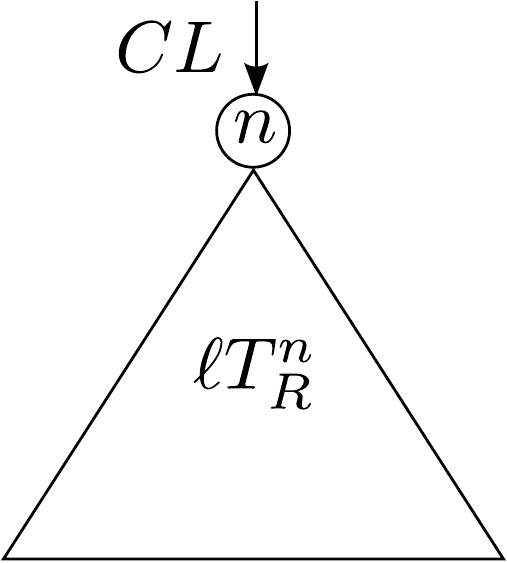}\\
\end{tabular}
\\
(a)  strategy for $\mathcal{C}_{\text{A}}$ &(b) strategy for $\mathcal{C}_{\text{B}}$
\end{tabular}
\end{center}
\caption{The two strategies considered by \limita}
\label{fig:alternatives}
\end{figure}

Next, we elaborate on $\MakeDecision$. Intuitively, in order to determine how \limita will process current node $n$ the function has to first estimate and then compare the computational costs $\mathcal{C}_\text{A}$ and $\mathcal{C}_\text{B}$ of the two alternative strategies: (A) processing current node and its descendants in the subtree $\ell T_R^n$ similar to \limit, or (B) verifying candidates in 
$\ell T_R^n \times CL$. In practice, 
it is not possible to estimate the cost of processing current node $n$ and its descendants in $\ell T_R^n$ similar to \limit since the involved intersections are not known in advance with the exception of $CL \cap I_S[n.item]$. 
Therefore, we estimate
$\mathcal{C}_\text{A}$
as the cost of computing 
the list intersection at current node $n$ and, 
verifying, for each child node $c_i$ of $n$,
the candidate pairs between \emph{all}\eat{the} objects under subtree $\ell T_R^{c_i}$\eat{ and $c_i.RL$,} and the objects in $CL'$.
Figure~\ref{fig:alternatives} illustrates the two alternative strategies, the costs of which are compared by \MakeDecision. 

We now discuss how costs $\mathcal{C}_\text{A}$ and $\mathcal{C}_\text{B}$ can be estimated. For this purpose, we first break \eat{down the contents of an} $n.RL$ set into two parts: $n.RL = n.RL^= \cup n.RL^{\supset}$, where $n.RL^=$ denotes the objects $r$ in $n.RL$ with $r = n.path$, while $n.RL^{\supset}$ the objects with $r \supset n.path$. Note that according to the definition of \emph{limited} prefix tree $\ell T_R$, $n.RL = n.RL^=$ holds for every \emph{internal} node $n$, as $n.RL^{\supset} = \emptyset$. Second, we introduce the following cost functions to capture the computational cost of the three tasks involved in strategies (A) and (B):
\begin{enumerate}[(i)]
\item \textbf{List intersection}. The cost of computing $CL' = CL \cap I_S[n.item]$ in current node $n$, denoted by $\mathcal{C}_{\cap}$, depends on the lengths of the involved lists and it is also related to the way list intersection is actually implemented. For instance, if list intersection is performed in a merge-sort manner, then $\mathcal{C}_{\cap}$ is linear to the sum of the lists' length, i.e., $\mathcal{C}_{\cap}\eat{(|CL|,|I_S[n.item]|)} = \alpha_1\cdot |CL| + \beta_1\cdot |I_S[n.item]| + \gamma_1$. On the other hand, if the intersection is based on a binary search over the $I_S[n.item]$ list then $\mathcal{C}_{\cap}\eat{(|CL|,|I_S[n.item]|)} = \alpha_2\cdot |CL| \cdot log_2(|I_S[n.item]|) + \beta_2$. 
Note that constants $\alpha_1$, $\alpha_2$, $\beta_1$, $\beta_2$ and $\gamma_1$ can be approximated by executing list intersection for several inputs and then, employing regression analysis over the collected measurements.
\item \textbf{Direct output of results.} \eat{Recall that for current node $n$ the $n.RL$ list consists of every object $r$ in $R$ with $r \supseteq n.path$. }Similar to \dasfaa and \limit, after list intersection $CL' = CL \cap I_S[n.item]$, every pair $(r,s)$ with $r \in n.RL$ and $s \in CL'$ such that  $r = n.path$, i.e., $r \in n.RL^=$, is guaranteed to be among the join results and it would be directly reported. The cost of this task, denoted by $\mathcal{C}_\text{d}$, is linear to the number of object pairs to be reported, and thus, $\mathcal{C}_\text{d}\eat{(|CL'|,|n.RL^=|)} = \alpha_3\cdot|CL'|\cdot|n.RL^=| + \beta_3$. 
Constants $\alpha_3$ and $\beta_3$ can be approximated by regression analysis.
\item \textbf{Verification.} To determine whether an $(r,s)$ \eat{object }pair is part of the join result \Verify would compare their suffixes in a merge-sort manner. Under this, the verification cost for each candidate pair is linear to the sum of their suffixes' length. Both alternative strategies considered by \MakeDecision involve verifying all candidate pairs between a subset of objects in $R$ and a subset in $S$ (candidates list $CL$ or $CL'$). Without loss of generality consider the case of strategy (A). In total, 
$|\ell T_R^n \smallsetminus n.RL^=|\cdot |CL'|$ candidates would be verified. 
Considering the length sum of the objects in 
$\ell T_R^n$ and of the objects in $CL'$, the total verification cost for (A) is 
\begin{align*}
\mathcal{C}_\text{v}\eat{(|n.RL^{\supset} \cup \ell T_R^n|,|CL'|,\ell)} &= \alpha_4\cdot |CL'| \cdot \sum_{r \in \{\ell T_R^n \smallsetminus n.RL^=\}}{(|r|-\ell)}\\
 &+ \beta_4 \cdot |\ell T_R^n \smallsetminus n.RL^=| \cdot \sum_{s \in CL'}{(|s|-\ell)} + \gamma_4
\end{align*}
where $|r|-\ell$ ($|s|-\ell$) equals the length of the suffix for a object $r$ ($s$) with respect to limit $\ell$. 
Similar to the previous tasks, constants $\alpha_4$, $\beta_4$ and $\gamma_4$ can be approximated by regression analysis. 
On the other hand, to approximate $|CL'| = |CL \cap I_S[n.item]|$ and $\sum_{s \in CL'}{(|s|-\ell)}$, we adopt an independent assumption approach based on the frequency of the item contained in current node $n$. Under this, $|CL'|  \approx |CL|\cdot \frac{|I_S[n.item]|}{|S|}$ while the length sum of the objects in $CL'$ can be estimated with respect to the $\frac{|CL'|}{|CL|} \approx \frac{|I_S[n.item]|}{|S|}$ decrease ratio, hence, we have $\sum_{s \in CL'}{(|s|-\ell)} \approx \frac{|I_S[n.item]|}{|S|}\cdot\sum_{s \in CL}{(|s|-\ell)}$. Finally, note that $\sum_{r \in \{\ell T_R^n \smallsetminus n.RL^=\}}{(|r|-\ell)}$ can be computed using statistics gathered while building prefix tree $\ell T_R$ and that $\sum_{s \in CL}{(|s|-\ell)}$ can be computed while performing the list intersection at the parent of current node $n$.
\end{enumerate}
With $\mathcal{C}_{\cap}$, $\mathcal{C}_\text{d}$, and $\mathcal{C}_\text{v}$, the computational costs of the (A) and (B) strategies considered by \MakeDecision are estimated by: \eat{the following formulas:}
\begin{align*}
&\mathcal{C}_\text{A} = \mathcal{C}_{\cap}(CL,I_S[n.item]) + \mathcal{C}_\text{d}(n.RL^=,CL') +\mathcal{C}_\text{v}(\{\ell T_R^n \smallsetminus n.RL^=\},CL',\ell)\\
&\mathcal{C}_\text{B} = \mathcal{C}_\text{v}(\ell T_R^n,CL,\ell-1)
\end{align*}
As intersection $CL' = CL \cap I_S[n.item]$  is not computed in (B), candidates list $CL$ and object suffixes beyond $\ell-1$ are considered by $\mathcal{C}_\text{B}$ in place of  $CL'$ and\eat{, while} suffixes beyond $\ell$ considered by $\mathcal{C}_\text{A}$.


\begin{example}
\label{ex:limita}
{\em
We illustrate the functionality of \limita using Example~\ref{ex:limit}. Assuming \eat{that }$\ell = 3$, \limita constructs prefix tree $\ell T_R$ of Figure~\ref{fig:running_example_limited_prefix_trees}(b) and inverted index $I_S$ of Figure~\ref{fig:running_example_indices}(b). 
First, the algorithm traverses the subtree of $\ell T_R$ under the node labeled by item $G$. The computational cost of the alternative strategies \eat{considered }for this node are\eat{ defined} as follows. $\mathcal{C}_\text{A}$ involves the cost of computing \eat{the }$CL' = \{s_1,\ldots,s_{12}\} \cap I_S[G] = \{s_2,s_4,s_5,s_7,s_8,$ $s_9,s_{10},s_{11},s_{12}\}$ \eat{intersection }and  based on the two child nodes, the cost of verifying all candidates in $\{r_1,r_2,r_5,r_7\}\! \times\! CL'$ and\eat{ in} $\{r_3\}\! \times\! CL'$\eat{, respectively}; note that no direct join results exist \eat{are reported }as $RL$ for current node is empty. On the other hand, $\mathcal{C}_\text{B}$ captures the cost of verifying all candidates in $\{r_1,r_2,r_3,r_5,r_7\} \times CL$. Without loss of generality assume \eat{that }$\mathcal{C}_\text{A} < \mathcal{C}_\text{B}$. Hence, \limita processes current node ($G$) similar to \limit: \eat{i.e.,}\eat{ meaning that} path $\</,G,F\>$ and the node labeled by $F$ are next considered. 
Assuming $\mathcal{C}_\text{A}\! >\! \mathcal{C}_\text{B}$ for this node, \limita \eat{decides to }imposes a local limit equal to $2$ and verifies all candidates in $\{r_1,r_2,r_5,r_7\}\! \times\! CL$ with $CL\!=\! \{s_2,s_4,s_5,s_7,s_8,s_9,s_{10},s_{11},s_{12}\}$\eat{, i.e.,} (objects in $S$ containing item $G$). Notice the resemblance to Example~\ref{ex:limit} for $\ell\! =\! 2$ with the exception that \eat{intersection }$\{s_2,s_4,s_5,s_7,s_8,s_9,s_{10},s_{11},$ $s_{12}\} \cap I_S[F]$ is not computed.
\hfill $\blacksquare$
}
\end{example}

\eat{
\note{...}
With respect to how computational costs $\mathcal{C}_\text{A}$ and $\mathcal{C}_\text{B}$ are determined, we introduce two variations of the \limita algorithm, termed $\limita\mathtt{A}$ and $\limita\mathtt{S}$. Intuitively, $\limita\mathtt{A}$ attempts to estimate the total cost for processing the object pairs involved in alternatives (A) and (B). In this spirit, it distinguishes between the pairs needed to be verified and those that are directly reported as results. Specifically, costs $\mathcal{C}_\text{A}$ and $\mathcal{C}_\text{B}$ are defined as follows:
\vspace*{1ex}
\begin{itemize}
\item $\mathcal{C}_\text{A} = \mathcal{C}_{\cap} + |CL'|\times |n.RL^=| + c\times|CL'|\times(|T_R^n|-|n.RL^=|)$ \hfill (1) 
\item $\mathcal{C}_\text{B} = c\times|CL|\times|T_R^n|$ \hfill (2)
\end{itemize}
\vspace*{1ex}
where $\mathcal{C}_{\cap}$ captures the cost of performing the $CL' = CL \cap I_S[n.item]$ intersection, $n.RL^=$ denotes the subset of objects $r$ in $n.RL$ with $r = n.path$, $|T_R^n|$ equals the number of objects contained in the subtree of prefix tree $T_R$,  rooted at current node $n$, and constant $c$ captures the cost of comparing the suffixes of two objects.

Cost $\mathcal{C}_{\cap}$ is related to the way the intersection is implemented. For instance, if the intersection is performed in a merge-sort way, then $\mathcal{C}_{\cap}$ is linear to the sum of the lists' length, i.e., $\mathcal{C}_{\cap} = |CL| + |I_S[n.item]|$. Next, $|CL'|\times |n.RL^=|$ and  $c\times|CL'|\times(|T_R^n|-|n.RL^=|)$ capture the cost of reporting the join results that involve the objects in $n.RL^=$ (similar to \dasfaa no verification takes place for $n.RL^=$), and the cost of verifying the pairs between the contents of $CL'$ and the objects under the subtrees rooted at the children of current node $n$, respectively. For the latter, the total number of objects under these subtrees equals the number of objects under the subtree $T_R^n$ rooted at $n$ excluding the objects in $n.RL^=$, and thus, $|CL'|\times(|T_R^n|-|n.RL^=|)$ of object pairs need to be verified, in total. 
Finally, $\mathcal{C}_\text{B}$ is defined with respect to the total number of object pairs between candidates list $CL$ and subtree $T_R^n$ to be verified. Note that in case of alternative (B) pairs that involve objects from $n.RL^=$ also need to be verified as list intersection in current node has not take place.

On the other hand, $\limita\mathtt{S}$ takes on a more agnostic approach where the costs for alternatives (A) and (B) are dominated entirely by the number of objects to be verified. Particularly, $\limita\mathtt{S}$ ignores the cost of comparing the suffixes of the objects, i.e., constant $c = 1$, and the cost of reporting join results that involve the objects in $n.RL^=$, and thus, $\mathcal{C}_\text{A}$ and $\mathcal{C}_\text{B}$ are defined as follows:
\vspace*{1ex}
\begin{itemize}
\item $\mathcal{C}_\text{A} = \mathcal{C}_{\cap} + |CL'|\times(|T_R^n|-|n.RL^=|)$ \hfill (3)
\item $\mathcal{C}_\text{B} = |CL|\times|T_R^n|$ \hfill (4)
\end{itemize}
\vspace*{1ex}
As our experiments show, due to its agnostic approach, $\limita\mathtt{S}$ always performs a smaller number of list intersections compared to $\limita\mathtt{A}$, which, on the other hand, results in a larger number of candidate object pairs to be verified.

Finally, one issue that remains still undiscussed is how $|CL'|$ and constant $c$ are computed. Since $CL'$ is not known in advanced, \MakeDecision computes an estimation $|CL'_e|$ of \eat{its actual size }$|CL'|$. For this task, in Section~\ref{sec:inter_size} we discuss a number of \eat{five }strategies that differ on what kind of pre-join statistics are computed on the right-hand collection $S$ to better estimate $|CL'|$. Note that using estimation $|CL'_e|$ in place of $|CL'|$ does not affect the correctness of the $\limita\mathtt{A}$ and $\limita\mathtt{S}$ algorithms. As far as constant $c$, we consider that the suffixes $suff(r)$ and $suff(s)$ of two objects $r$ and $s$, respectively, are compared in a merge-sort manner with a cost linear to $|suff(r)|+|suff(s)|$. However, since the objects to be compared are not known in advanced, \MakeDecision estimates the comparison cost with respect to the weighted average length of the objects in collections $R$ and $S$ and limit $\ell$ which captures the part of the objects already compared while traversing the prefix tree. Recall that in case of alternative (B) this part involves the first $\ell-1$ items in each object as the list intersection at current node has not taken place.

\begin{example}
{\em
We demonstrate \limita using Example~\ref{ex:dasfaa}. Similar to Example~\ref{ex:limit} we consider the \emph{limited} prefix trees $T_R$ shown in Figure~\ref{fig:running_example_limited_prefix_trees} for the collection $R$ in Figure~\ref{fig:running_example_collections}(a), and the inverted index $I_S$ in Figure~\ref{fig:running_example_indices}(a). Further, without loss of generality, we assume that the result of a $CL' = CL \cap I_S[n.item]$ intersection is estimated by the length of the shortest list, i.e., $|CL'_e| = min(|CL|,|I_S[n.item]|)$, and that the cost $\mathcal{C}_{\cap}$ of such an intersection is linear to the sum of the lists' length, i.e., $\mathcal{C}_{\cap} = |CL| + |I_S[n.item]|$.

First, we discuss $\limita\mathtt{S}$ and $\ell = 2$; the case of $\ell = 3$ involves exactly the same steps and therefore, it is omitted. Similar to \dasfaa and \limit, the algorithm traverses the subtrees of $T_R$ in Figure~\ref{fig:running_example_limited_prefix_trees}(a) starting from the left most subtree under the node labeled by item $G$. After performing the intersection for this node, consider path $\{/,G,F\}$. In this case we have candidates list $|CL| = |\{s_2,\!s_4,\!s_5,\!s_7,\!s_8,\!s_9,\!s_{10},\!s_{11},\!s_{12}\}| = 9$ and $|I_S[F]| = 8$, which means that the cost of performing the intersection $CL' = CL \cap I_S[F]$ is $\mathcal{C}_{\cap} = 17$, and that the estimation of $|CL'|$ is $|CL'_e| = 8$. Further, for current node labeled by item $F$, $n.RL^=$ is empty while its subtree contains $4$ objects. Using Equations (3) and (4), we get $\mathcal{C}_\text{A} = 17 + 8\times(4-0) = 49$ and $\mathcal{C}_\text{B} = 9\times 4 = 36$. With $\mathcal{C}_\text{A} > \mathcal{C}_\text{B}$, $\limita\mathtt{S}$ decides to stop performing list intersections and verifies object pairs between $CL$ and $\{r_1,\!r_2,\!r_5,\!r_7\}$. In a similar manner, at the next steps, the algorithm decides to stop performing any additional intersections and verifies all involved object pairs also for paths $\{/,G,D\}$, $\{/,F,D\}$ and $\{/,E,C\}$. To sum up, $\limita\mathtt{S}$ performs $3$ list intersections and verifies $61$ candidates pairs by comparing their suffices, in total.

Next, we discuss $\limita\mathtt{A}$. For simplicity, in the context of this example, we define constant $c$ only in terms of the weighted average length of the objects in $R$ and $S$, i.e., $c = 8$ for both $\mathcal{C}_\text{A}$ and $\mathcal{C}_\text{B}$ costs. For $\ell = 2$, consider again the case of path $\{/,G,F\}$. Recall that for current node $n$ labeled by item $F$, $|CL| = 9$, $|CL'_e| = 8$, $\mathcal{C}_{\cap} = 17$, $|n.R^=| = 0$ and $|T_R^n| = 4$ as discussed for $\limita\mathtt{S}$. Using Equations (1) and (2), we get $\mathcal{C}_\text{A} = 17 + 8\times0 + 8\times 8\times(4-0) = 273$ and $\mathcal{C}_\text{B} = 8\times 9\times 4 = 288$, and as a result, $\limita\mathtt{A}$ decides to proceed similar to \limit. Therefore, it performs the $CL \cap I_S[F]$ intersection, computes the $CL' = \{s_2,\!s_4,\!s_5,\!s_9,\!s_{10},\!s_{11},\!s_{12}\}$ candidates list and finally, verifies object pairs between $CL'$ and $T_R^n = \{r_1,\!r_2,\!r_5,\!r_7\}$. Note the difference with $\limita\mathtt{S}$, where candidates list $CL$ is used for the same task, as the list intersection did not take place. $\limita\mathtt{A}$ proceeds in a similar way to examine the rest of the prefix tree nodes. In total, it performs $5$ list intersections and verifies $45$ object pairs by comparing their suffices. Finally, for $\ell = 3$, $\limita\mathtt{A}$ performs $10$ list intersections and verifies $12$ object pairs by comparing their suffices.
 \hfill $\blacksquare$
}
\end{example}

}
\section{A Novel Join Paradigm}
\label{sec:new}
\eat{
As discussed in Section~\ref{sec:dasfaa}, the join paradigm of \dasfaa
\cite{JampaniP05}, in order to
construct the prefix tree $T_R$ for the left-hand collection
$R$, considers a global ordering of the items. Specifically, 
each object of $R$ is internally sorted in order to have its items in
decreasing order of their global frequencies in Objects of $R$.
Further, the join process is driven solely by the traversal of the
prefix tree $T_R$; the objects in collection $R$ are examined with
respect to their prefix (i.e., paths of the prefix tree), while at the
same time lists of objects from collection $S$ are intersected to
produce join results. 
The focus of this examination order is on batch processing; i.e., by grouping
the Objects under $T_R$ \dasfaa performs the corresponding list
intersections only once for groups of objects corresponding to each prefix.
\note{de se pianw: IMHO, the object ordering is meant only for maximizing tree
  compression, by avoiding empty tree nodes as much as possible. 
Regardless the ordering of items, the grouping of
  objects is identical.}\pnote{To ordering xreiazetai mono gia na
  ftiaxtei to dentro kai kata sunepeia gia na exoume batch
  processing. Den kovoume empty tree nodes kai den kanoume kapoiou
  eidous compression, oute to \dasfaa ta kanei}
\note{allo lew egw... ``The focus of this examination order...'',
  dhladh ena allo order tha eixe allo focus? ti ennoeis by
  ``examination order''. pio katw milas gia Objects, edw gia items,
  gamise ta.}

On the other hand, in the context of the set-similarity join, which is
another set join operator, works that employ the prefix filtering
\cite{BayardoMS07,XiaoWLS09,XiaoWLY08}, consider both the global
ordering of the items and the examination order of the Objects to
minimize the number of candidate Object pairs. The idea is to examine
the internally ordered Objects from either input collection in ascending
order of their lengths and at the same time progressively build the
inverted indices employed to perform the join. Consequently, a Object,
e.g., $r$ of the left-hand collection $R$, is probed against an
incomplete inverted index $I_S$ and joined only with Objects of the
right-hand collection $S$ which can be similar to $r$.
\note{This is unreadable to someone unfamiliar with set-similarity
  joins. Not to mention that it challenges the novelty of this
  section. Also, we do not review set-similarity joins in the related
  work. Normally, we should and discuss the differences between these
  methods and set-containment join algorithms.}\pnote{As syzitisoume tote ta set similarity joins. Vasiki diafora pou mporw na skeftw einai oti to containment den einai symmetric kai oti methods opws to ppjoin pou xrisimopoioun prefix filtering ousiastika de doulevoun giati to prefix tha perieixe 1 item opote exeis poly mikro pruning, peftei olo to varos sto verification.}

In this section, we propose a novel paradigm for set containment joins
termed \new that combines the advantages of the aforementioned
approaches \note{which approaches?}. 
Intuitively, \new comes as a hybrid that primarily accesses Objects
from either input collection in order of their most frequent item
\note{unclear} while secondarily, it examines the Objects in collection
$R$ with respect to their prefix \note{unclear again}.\pnote{Na suzitisoume auta ta unclear sxolia kai to "approaches" avrio twra den mporw na leitourgisw...}
Specifically, the paradigm has the following key features:
\begin{enumerate}[(i)]
\item The Objects of \emph{both} input collections are internally
  ordered according to a global ordering based on the \emph{union} of
  their domains \note{unclear: what does it mean to order based on the
    union of the domains? make it explicit.}. Similar to the \dasfaa
  framework, this ordering brings the most frequent item inside each
  Object first. Further, the Objects in collection $S$ are grouped in
  partitions based on their first item.
%
\item Similar to the framework of \dasfaa, \new employs a prefix tree $T_R$ on the left-hand collection $R$ and an inverted index $I_S$ on the right hand $S$, but only $T_R$ is built before the actual join process.
\item During the join process, \new considers Objects from $R$ and $S$
  in decreasing frequency order of their first item.
Consider an item $i$. The paradigm first considers collection $S$. It
iterates through the Objects that start by $i$ accessing in practice
the contents of partition $P_i$, to incrementally build inverted index
$I_S$. After processing all Objects in $P_i$, \new traverses the
subtree of $T_R$ that contains all Objects in $R$ also starting by $i$
and joins them with the version of $I_S$ built so far. \note{++
  explain what's next.}\pnote{Ti ennoeis? Proxwraei sto epomeno item
  kai epanalamvanei ta idia}\note{tespa, den yparxei periptwsh na
  katalavei kaneis tipota apo ta parapanw, mexri na diavasei ton
  algorithmo meta.}
\end{enumerate}
}

\eat{
Due to grouping the Objects of collection $R$ under prefix tree $T_R$,
similar to the \dasfaa join paradigm, \new is able to avoid redundant
processing, i.e., list intersections are computed only once for each
common prefix. However, due to incrementally building inverted index
$I_S$ \new exhibits a two-fold advantage over \dasfaa paradigm. First,
a Object in $R$ is joined only with Objects in $S$ that may contain it. As the Objects are primarily examined in decreasing frequency order of their first item when the Objects in $R$ starting by an item $i$ are joined against $I_S$, the inverted index contains only Objects starting either by $i$ or a more frequent item; in the latter case the Objects may still contain $i$. On the other hand, the Objects of $S$ that cannot contain $i$, i.e., those that start by a less frequent item, are not yet examined and therefore, they are not contained in $I_S$. Second, with the inverted lists being shorter compared to the ones in the full inverted index, list intersections are faster; note though that the total number of intersections performed does not change.
}

As discussed in Section~\ref{sec:dasfaa}, the join paradigm of \dasfaa
\cite{JampaniP05}, which is also followed by \limit and \limita, 
constructs the entire prefix tree $T_R$  (or $\ell T_R$) and the entire inverted index $I_S$ 
before joining them.
However, we observe that the construction of $T_R$ and $I_S$ can be
interleaved with the join process since for joining a set of objects from
$R$ that lie in a subtree of $T_R$ it is not necessary to have
constructed the entire $I_S$.
For example, consider again the $T_R$ and $I_S$ indices of Figure
\ref{fig:running_example_indices}. When performing the join for the
nodes in the subtree rooted at node $G$, obviously, we need not have
constructed the subtrees rooted at nodes $F$ and $E$ already.
At the same time, only the objects from $S$ that contain item $G$ can be
joined with each object in that subtree. Therefore, we only need a
partially built $I_S$ which includes just these objects.
In this section, we propose a new paradigm, termed 
Order and Partition Join (\new),
which is based on this observation. \new operates as follows:

\begin{enumerate}[(i)]
\item Assume that for each object (in  either $R$ or $S$), the items
are considered in a certain order (i.e., 
in decreasing order of their
  frequency in $R\cup S$). 
\new partitions the objects of each collection into groups based on their
first item.%
\footnote{This is different than the external-memory partitioning of
  the \dasfaa paradigm, discussed at the end of Section \ref{sec:dasfaa}.}
Thus, for each item $i$, there is a partition $R_i$
($S_i$) of $R$ ($S$) that includes all objects $r\in R$ ($s\in S$),
for which the {\em first} item is $i$.
For example, partition $R_G$ of collection $R$ \eat{shown }in Figure~\ref{fig:running_example_collections}(a) 
includes
$\{r_1,r_2,r_3,r_5,r_7\}$, while partition $R_E$ includes just $r_6$.
Due to the internal sorting of the objects, an object in $R_i$ or $S_i$
includes $i$ but does not include any item $j$, which comes before $i$
in the order (e.g., $r_6\in R_E$ cannot contain $G$ or $F$). Then, \new initializes an empty inverted index $I_S$ for $S$.

\item For each item $i$ in order, \new creates a prefix tree
  $T_{R_i}$ for partition $R_i$ and updates $I_S$ to include all objects from partition
  $S_i$. Then,  $T_{R_i}$ is joined with $I_S$ using \dasfaa (or our
  algorithms \limit and \limita). After
  the join, $T_{R_i}$ is dumped from the memory and \new proceeds with
  the next item $i+1$ in order to construct $T_{R_{i+1}}$ using
  $R_{i+1}$, update $I_S$ using $S_{i+1}$ and 
  join $T_{R_{i+1}}$ with $I_S$.
\end{enumerate}

\new has several advantages over the \dasfaa join paradigm. First, the
entire $T_R$ needs not be constructed and held in memory. For each
item $i$ the subtree of $T_R$ rooted at $i$ (i.e., $T_{R_i}$) is
built\eat{constructed}, joined, and then removed from memory. Second, the
inverted index $I_S$ is incrementally constructed, therefore\eat{ the}
$T_{R_i}$ for each item $i$ in order is joined with a
smaller $I_S$ which (correctly) excludes objects of $S$ having only items that
come after $i$.
Thus, the inverted lists of the partially constructed $I_S$ are
shorter and the join is faster.%
\footnote{Note that \new and
\dasfaa perform the same number of list intersections; i.e., \new does
not save list intersections, but makes them cheaper.}
Finally, the overall memory requirements of \new are much lower
compared to\eat{ those of} \dasfaa join paradigm as 
\new only
keeps one $T_{R_i}$ in memory at a time (instead of the entire $T_R$).
\eat{, and
(ii) at the latter stages of \new, when $I_S$ is almost complete (and
thus large), the $T_{R_i}$'s are expected to be quite small
(because the $R_i$'s are expected to contain small combinations of items).
\pnote{Proteinw na kopsoume to (ii), den isxyei gia increasing sta BMS, FLICKR, KOSARAK. Genika xtypane asxima arguments pou vasizontai se sygkekrimeno order to opoio sta peiramata diexnoume oti den einai kalo!}
}
\begin{algorithm}[t]
\SetKwFunction{ContructPrefixTree}{ContructPrefixTree}
\SetKwFunction{Partition}{Partition}
\SetKwFunction{Sort}{Sort}
\SetKwFunction{UpdateInvertedIndex}{UpdateInvertedIndex}
\SetKwFunction{ProcessNode}{ProcessNode}
\SetKwFunction{Verify}{Verify}
\SetKwInOut{Input}{input}
\SetKwInOut{Output}{output}
\LinesNumbered
\Input{Collections $R$ and $S$, limit $\ell$; every Object $r\! \in\! R$ and $s\! \in\! S$ is internally sorted such that the most frequent item in $R \cup S$ appears first}
\Output{the set $J$ of all Object pairs $(r,s)$ such that $r \in R$, $s \in S$ and $r \subseteq s$}
\BlankLine
\nlset{1}$\Partition(S)$; $\Partition(R)$\tcp*[r]{w.r.t. the first item in each Object}
\nlset{2}$I_S \leftarrow \emptyset$\;
\eat{
\ForEach{partition $P_i$ of $S$ in lexicographical order}
{
	$i \gets$ the first item of the Objects in $P_i$\;
	$c \gets$ child node of the root in $T_R$ with $c.item = i$\;
	$CL \gets \{s | s \in S\}$\;
	$I_S \gets \UpdateInvertedIndex(I_S,P_i)$\;
	$\ProcessNode(c, CL, I_S,J,\ell)$\tcp*[r]{\dasfaa,\limit,\limita}
}
}
\ForEach{item $i$ in decreasing
  frequency order}
{
  $\ell T_{R_i} \leftarrow \ContructPrefixTree(R_i,\ell)$\;
$I_S \gets \UpdateInvertedIndex(I_S,S_i)$\;
$c \gets$ child node of $\ell T_{R_i}$'s root\tcp*[r]{$\ell T_{R_i}$'s root
  has a single child $c$ with $c.item =i$}
	$CL \gets$ Objects in $S$ seen so far\tcp*[r]{Candidates list}
	$\ProcessNode(c, CL, I_S,J,\ell)$\tcp*[r]{\dasfaa,\limit,\limita}
\textbf{delete} $\ell T_{R_i}$\;
}
\Return $J$;
%
\eat{
\\
\vspace{3ex}
\bf{Function} $\ProcessNode(n, CL, I_S, J)$\\
$CL' \leftarrow CL \cap I_S[n.item]$\tcp*[r]{List intersection}
\ForEach{Object $s \in CL'$}
{
	\ForEach{Object $r \in n.RL$}
	{
		$J \leftarrow J \cup (r,s)$\;
	}
}
\ForEach{child node $c$ of $n$}
{
	$\ProcessNode(c,CL',I_S,J)$\tcp*[r]{Recursion}
}
}
\caption{$\new(R,S,\ell)$}
\label{algo:new}
\end{algorithm}

Algorithm~\ref{algo:new} illustrates a high-level sketch of the
\new\eat{ join} paradigm.
\new
receives as input collections $R$ and $S$, and limit $\ell$; for\eat{in case
of} \dasfaa $\ell = \infty$\eat{ parameter is set to infinity} (i.e., $\ell T_{R_i}$
becomes $T_{R_i}$\eat{ of \dasfaa}). 
Initially, collections $R$ and $S$ are partitioned \eat{in order }to put all
objects having \eat{item }$i$ as their first item inside partitions $R_i$ and
$S_i$, respectively (Line 1). Also, $I_S$ (the inverted index of $S$) is
initialized (Line 2). 
Then, for each item $i$, \new computes the join results between
objects from $R$ having $i$ as their first item and objects from $S$
having $i$ or a previous item in order as their first item (Lines~3--9).
Specifically, for each item $i$ in order, \new builds a (limited) prefix tree
$\ell T_{R_i}$ using partition $R_i$, adds all objects of partition
$S_i$ into $I_S$, and finally joins $\ell T_{R_i}$ with $I_S$ using
the methodology of \dasfaa, \limit, or \limita.
Note that for each $\ell T_{R_i}$ the root has a single child $c$ with
$c.item =i$, because all objects in $R_i$ have $i$ as their first item. 
Thus, \new has to invoke the \ProcessNode function (of either \dasfaa,
\limit or \limita) only for $c$. In addition, note that 
candidates list $CL$ is initialized with only the objects in $S$
accessed so far instead of all objects in $S$ according to the \dasfaa
join paradigm; the examination order guarantees that the rest of the
objects in $S$ cannot be joined with the objects in $R$ under node
$c$.

\eat{
Then, the Objects in collection $S$ are grouped into partitions based on
their first item and an empty inverted index $I_S$ is initialized. The
second phase of the \new paradigm involves the computation of the join
result set $J$ (Lines~4--11). As discussed earlier, \new examines the
Objects from the input collections in decreasing frequency order of
their first item. In practice this is done by iterating through the
items inside $domain(R)\cup domain(S)$ following the global item
ordering. 
Let  $i$ be the current item. \eat{First,}If $domain(S)$ contains $i$
\eat{(Lines~5--7)},  \new accesses the Objects in $S$ that start by $i$,
i.e., partition $P_i$, to update inverted index $I_S$ (Lines~6--7). Then, if
$domain(R)$ also contains item $i$ (Lines~8-11), the paradigm considers the
Objects in $R$ that start by $i$ accessing the child node $c$ of the
prefix tree $\ell T_R$ root with $c.item = i$ to perform the actual join.
\note{I suggest to drop the if statements and consider a common domain
  for $R$ and $S$, which is the typical case. Otherwise, there are
  special cases to consider. For example if $i$ is in R but not in S,
  then all the sets in $R$ that contain $i$ should be pruned.}
  \pnote{De tha doulevei tote sti geniki periptwsi. Enallaktika ekana mia allagi sto pseudocode kai allaksa ligo tin perigrafi. Des twra}
\eat{The final step is to perform the actual join and }For this purpose, \new invokes the \ProcessNode function of either \dasfaa, \limit or \limita to join the Objects inside the subtree of $T_R$ under node $c$ with the version of inverted index $I_S$ built so far. Note that candidates list $CL$ is initialized with only the Objects in $S$ accessed so far instead of all Objects in $S$ according to the \dasfaa join paradigm; the examination order guarantees that the rest of the Objects in $S$ cannot be joined with the Objects in $R$ under node $c$.
}

\eat{
\begin{algorithm}[t]
\SetKwFunction{ContructPrefixTree}{ContructPrefixTree}
\SetKwFunction{Partition}{Partition}
\SetKwFunction{Sort}{Sort}
\SetKwFunction{UpdateInvertedIndex}{UpdateInvertedIndex}
\SetKwFunction{ProcessNode}{ProcessNode}
\SetKwFunction{Verify}{Verify}
\SetKwInOut{Input}{input}
\SetKwInOut{Output}{output}
\LinesNumbered
\scriptsize
\Input{Collections $R$ and $S$ and limit $\ell$; every Object $r \in R$ and $s \in S$ is internally ordered such that the most frequent item with respect to $domain(R)\cup domain(S)$ appears first}
\Output{the set $J$ of all Object pairs $(r,s)$ such that $r \in R$, $s \in S$ and $r \subseteq s$}
\BlankLine
\nlset{1}$T_R \leftarrow \ContructPrefixTree(R,\ell)$\;
\nlset{2}$\Partition(S)$\tcp*[r]{w.r.t. the first item on each Object}
\nlset{3}$I_S \leftarrow \emptyset$\;
\eat{
\ForEach{partition $P_i$ of $S$ in lexicographical order}
{
	$i \gets$ the first item of the Objects in $P_i$\;
	$c \gets$ child node of the root in $T_R$ with $c.item = i$\;
	$CL \gets \{s | s \in S\}$\;
	$I_S \gets \UpdateInvertedIndex(I_S,P_i)$\;
	$\ProcessNode(c, CL, I_S,J,\ell)$\tcp*[r]{\dasfaa,\limit,\limita}
}
}
\ForEach{item $i \in\! domain(R)\!\cup\!domain(S)$ in decreasing frequency order}
{
	\If{$i \in domain(S)$}
	{
		$P_i$ $\gets$ partition of Objects in $S$ starting with $i$\;
		$I_S \gets \UpdateInvertedIndex(I_S,P_i)$\;
	\If{$i \in domain(R)$}
	{
		$c \gets$ child node of the root in $T_R$ with $c.item = i$\;
		$CL \gets$ Objects in $S$ examined so far\;
		$\ProcessNode(c, CL, I_S,J,\ell)$\tcp*[r]{\dasfaa,\limit,\limita}
	}
}
}
\Return $J$;
%
\eat{
\\
\vspace{3ex}
\bf{Function} $\ProcessNode(n, CL, I_S, J)$\\
$CL' \leftarrow CL \cap I_S[n.item]$\tcp*[r]{List intersection}
\ForEach{Object $s \in CL'$}
{
	\ForEach{Object $r \in n.RL$}
	{
		$J \leftarrow J \cup (r,s)$\;
	}
}
\ForEach{child node $c$ of $n$}
{
	$\ProcessNode(c,CL',I_S,J)$\tcp*[r]{Recursion}
}
}
\caption{$\new(R,S,\ell)$}
\label{algo:new}
\end{algorithm}
}

\begin{figure}
\begin{center}
\begin{tabular}{clcl}
&$s_2\!\!: \{G,F,E,D,C,A\}$ & &$A\!\!: \{s_2\}$\\
&$s_4\!\!: \{G,F,C,B\}$ & &$B\!\!: \{s_4,\!s_5,\!s_7,\!s_8\}$\\
&$s_5\!\!: \{G,F,E,B\}$ & &$C\!\!: \{s_2,\!s_4,\!s_7,\!s_8\}$\\
&$s_7\!\!: \{G,E,D,C,B\}$ & &$D\!\!: \{s_2,\!s_7,\!s_8,\!s_9,\!s_{10}\}$\\
$S_G$ &$s_8\!\!: \{G,E,D,C,B\}$ & &$E\!\!: \{s_2,\!s_5,\!s_7,\!s_8,\!s_9,\!s_{10},\!s_{12}\}$\\
&$s_9\!\!: \{G,F,E,D\}$ & &$F\!\!: \{s_2,\!s_4,\!s_5,\!s_9,\!s_{10},\!s_{11},\!s_{12}\}$\\
&$s_{10}\!\!: \{G,F,E,D\}$ & &$G\!\!: \{s_2,\!s_4,\!s_5,\!s_7,\!s_8,\!s_9,\!s_{10},\!s_{11},\!s_{12}\}$\\
&$s_{11}\!\!: \{G,F\}$\\
&$s_{12}\!\!: \{G,F,E\}$\\
\cline{1-2}
& & &$B\!\!: \{s_4,\!s_5,\!s_6,\!s_7,\!s_8\}$\\
& & &$C\!\!: \{s_2,\!s_4,\!s_6,\!s_7,\!s_8\}$\\
$S_F$ &$s_6\!\!: \{F,E,D,C,B\}$ & &$D\!\!: \{s_2,\!s_6,\!s_7,\!s_8,\!s_9,\!s_{10}\}$\\
& & &$E\!\!: \{s_2,\!s_5,\!s_6,\!s_7,\!s_8,\!s_9,\!s_{10},\!s_{12}\}$\\
& & &$F\!\!: \{s_2,\!s_4,\!s_5,\!s_6,\!s_9,\!s_{10},\!s_{11},\!s_{12}\}$\\
\cline{1-2}
\\
&$s_1\!\!: \{D,C,A\}$ & &$A\!\!: \{s_1,\!s_2\}$\\
$S_D$ &$s_3\!\!: \{D,B\}$ & &$B\!\!: \{s_3,\!s_4,\!s_5,\!s_6,\!s_7,\!s_8\}$\\
& & &$C\!\!: \{s_1,\!s_2,\!s_4,\!s_6,\!s_7,\!s_8\}$\\
& & &$D\!\!: \{s_1,\!s_2,\!s_3,\!s_6,\!s_7,\!s_8,\!s_9,\!s_{10}\}$\\
\\
\multicolumn{2}{c}{\normalsize{(a) Partitions of $S$}} &\multicolumn{2}{c}{
\normalsize{(b) Updates in $I_S$}}\\
\end{tabular}
\end{center}
\eat{
\begin{center}
\hspace*{-0.3cm}
{\scriptsize
\begin{tabular}{cc}
\begin{tabular}{cl}
&$s_2\!\!: \{G,F,E,D,C,A\}$\\
&$s_4\!\!: \{G,F,C,B\}$\\
&$s_5\!\!: \{G,F,E,B\}$\\
&$s_7\!\!: \{G,E,D,C,B\}$\\
$P_G$ &$s_8\!\!: \{G,E,D,C,B\}$\\
&$s_9\!\!: \{G,F,E,D\}$\\
&$s_{10}\!\!: \{G,F,E,D\}$\\
&$s_{11}\!\!: \{G,F\}$\\
&$s_{12}\!\!: \{G,F,E\}$\\
\hline
\\\\
$P_F$ &$s_{10}\!\!: \{F,E,D,C,B\}$\\
\\\\
\hline
\\$P_D$ &$s_{11}\!\!: \{D,C,A\}$\\
&$s_{12}\!\!: \{D,B\}$\\
\\
\end{tabular}
&
\begin{tabular}{l}
$A\!\!: \{s_1\}$\\
$B\!\!: \{s_2,\!s_3,\!s_4,\!s_5\}$\\
$C\!\!: \{s_1,\!s_2,\!s_4,\!s_5\}$\\
$D\!\!: \{s_1,\!s_4,\!s_5,\!s_6,\!s_7\}$\\
$E\!\!: \{s_1,\!s_3,\!s_4,\!s_5,\!s_6,\!s_7,\!s_9\}$\\
$F\!\!: \{s_1,\!s_2,\!s_3,\!s_6,\!s_7,\!s_8,\!s_9\}$\\
$G\!\!: \{s_1,\!s_2,\!s_3,\!s_4,\!s_5,\!s_6,\!s_7,\!s_8,\!s_9\}$\\
\\
\\
$B\!\!: \{s_2,\!s_3,\!s_4,\!s_5,\!s_{10}\}$\\
$C\!\!: \{s_1,\!s_2,\!s_4,\!s_5,\!s_{10}\}$\\
$D\!\!: \{s_1,\!s_4,\!s_5,\!s_6,\!s_7,\!s_{10}\}$\\
$E\!\!: \{s_1,\!s_3,\!s_4,\!s_5,\!s_6,\!s_7,\!s_9,\!s_{10}\}$\\
$F\!\!: \{s_1,\!s_2,\!s_3,\!s_6,\!s_7,\!s_8,\!s_9,\!s_{10}\}$\\
\\
$A\!\!: \{s_1,\!s_{11}\}$\\
$B\!\!: \{s_2,\!s_3,\!s_4,\!s_5,\!s_{10},\!s_{12s}\}$\\
$C\!\!: \{s_1,\!s_2,\!s_4,\!s_5,\!s_{10},\!s_{11}\}$\\
$D\!\!: \{s_1,\!s_4,\!s_5,\!s_6,\!s_7,\!s_{10},\!s_{11},\!s_{12}\}$\\
\end{tabular}
\\\\
\normalsize{(a) Partitions of $S$} &\normalsize{(b) Updates in $I_S$}\\
\end{tabular}
}
\end{center}
}
\caption{Employing the \new join paradigm}
\label{fig:running_example_new}
\end{figure}

\begin{example}
\label{ex:new}
{\em
We demonstrate \new on collections $R$ and $S$ in
Figure~\ref{fig:running_example_collections}.
The items in decreasing frequency order over $R\cup S$ are
$G(14),F(13),E(12),D(11),C(9),$ $B(9),$ $A(3)$, resulting in the internally
sorted objects shown in the figure. 
Without loss of generality, assume that the \dasfaa algorithm is used
to perform the join between each $\ell T_{R_i}$ and $I_S$ (i.e.,
\eat{assume }$\ell = \infty$ and $\ell T_{R_i}=T_{R_i}$).
Initially, the objects are partitioned according to their first item. 
The partitions for $R$ are $R_G\!=\!\{r_1,r_2,r_3,r_5,r_7\}$,
$R_F\!=\!\{r_4\}$, and $R_E\!=\!\{r_6\}$; the partitions for $S$ are shown in
Figure~\ref{fig:running_example_new}(a).
\new first accesses partition $R_G$ and builds $T_{R_{G}}$, which is
identical to the leftmost subtree of the \emph{unlimited} $T_{R}$ in
Figure~\ref{fig:running_example_indices}(a). Then, \new updates the
(initially empty) inverted index $I_S$ to include the objects of
$S_G$; the resulting $I_S$ is shown on the right of $S_G$, at the top
of Figure~\ref{fig:running_example_new}(b).
After joining $T_{R_{G}}$ with $I_S$, $T_{R_{G}}$ is deleted from
memory, and the next item $F$ in order is processed.
\new builds $T_{R_{F}}$ (which is identical to the 2nd subtree of
$T_R$ in Figure~\ref{fig:running_example_indices}(a)) and updates $I_S$
to include the objects in $S_F$; these updates are shown on the right
of $S_F$ in Figure~\ref{fig:running_example_new}(b). Then,  $T_{R_{F}}$
is joined with $I_S$, and \new proceeds to the next item $E$. In this
case, $T_{R_{E}}$ is built (the rightmost subtree of $T_R$
in Figure~\ref{fig:running_example_indices}(a)), but $I_S$ is not
updated as $S_E$ is empty. Still, $T_{R_{E}}$ is joined with
current $I_S$. In the next round (item $D$), there is no join to be
performed, because $R_D$ is empty. If there were additional partitions $R_i$
to be processed, $I_S$ would have to be updated to include the objects in
$S_D$, as shown on the right of $S_D$ in
Figure~\ref{fig:running_example_new}(b). However, since all objects from
$R$ have been processed, \new can terminate without processing $S_D$.
\eat{
Figure~\ref{fig:running_example_collections}(a) and Figure~\ref{fig:running_example_new}(a) show collections $R$ and $S$, respectively,  input to \new; their objects are internally ordered according to the $\{G,F,E,D,C,B,A\}$ global ordering. Notice the difference of collection $S$ compared to Figure~\ref{fig:running_example_collections}(b) where no internal ordering was imposed. Initially, the prefix tree $T_R$ in Figure~\ref{fig:running_example_indices}(a) is built while the Objects of $S$ are grouped into partitions $P_G$, $P_F$ and $P_D$ pictured in Figure~\ref{fig:running_example_new}(a).
In what follows, \new will iterate through the items in $domain(R)\cup
domain(S)$ following the global ordering. With item $G$, the objects
inside partition $P_G$ are examined to build the inverted index $I_S$
shown at the top part of
Figure~\ref{fig:running_example_new}(b). Next, the leftmost subtree of
$T_R$ under the node labeled by $G$ is traversed. When the paradigm
considers path $\{/,G,F\}$ the $CL \cap I_S[F]$ intersection is
computed. However, compared to Example~\ref{ex:dasfaa} and the
inverted index in Figure~\ref{fig:running_example_indices}(b), the
$I_S[F]$ postings list is in this case shorter as it does not contain
$s_6$. Object $s_6$ is grouped inside partition $P_F$ and thus, it is
not accessed yet. After traversing entirely the leftmost subtree of
$T_R$ and utilizing \dasfaa to compute join results, \new considers
items $F$ and $D$ in the same manner;
Figure~\ref{fig:running_example_new}(b) demonstrates how the postings
lists of $I_S$ are updated when accessing partitions $P_F$ and
$P_D$. The join procedure terminates after performing 14 list
intersections similar to Example~\ref{ex:dasfaa}.
}
\hfill $\blacksquare$
}
\end{example}

\eat{
Similar to the framework of \dasfaa,
the \new paradigm employs a prefix tree $T_R$ on the left hand
collection $R$ and an inverted index $I_S$ on the right hand $S$ but
only $T_R$ is built before the actual join process. In addition, apart
from the order of the \emph{items} inside the Objects of $R$ imposed
by \dasfaa for building the prefix tree $T_R$, \new also groups the
\emph{Objects} of $S$ based on their first item and considers these
groups in lexicographical order of their designated item. The basic
idea of the \new join paradigm is to iterate through the Objects of the
right hand collection $S$ and incrementally build the $I_S$ inverted
index. After all Objects starting by an item $i$ are examined,
\new joins the subtree of $T_R$ that contains all Objects in the left
hand collection $R$ also starting by $i$ with the version of $I_S$ built
so far. The advantage of the \new paradigm is three-fold. First, due
to incrementally building the inverted index $I_S$ a Object in $R$ is
joined only with Objects in $S$ that may contain it as they share at least one
item. Second, with the inverted lists being shorter compared to the
ones in the full inverted index, intersections are faster. Finally,
\new can be easily adapted both by \dasfaa and our algorithms \limit and \limita.

Algorithm~\ref{algo:new} illustrates a high-level sketch of the \new join paradigm. The algorithm receives as input collections $R$ and $S$, and limit $\ell$; \note{note that in case of \dasfaa the limit $\ell$ parameter is set equal to infinity}. The Objects of both collections $R$ and $S$ are internally ordered according to a global ordering of the union of their domains which brings the most frequent items inside each Object first. Similar to the framework discussed in the previous sections, the computation of the join involves two phases. First, during the initialization phase (Lines 1--3), the \emph{limited} prefix tree $T_R$ on collection $R$ is constructed; note that if $\ell = \infty$ then the full prefix tree as 
}
\eat{
The right hand collection $S$ is partitioned such that Objects beginning with the same item $i$ fall in the same partition $P_i$. The number of partitions created at most equal to the domain of $S$. (Line 2)

The partitions of $S$ and therefore, also the Objects of $S$, are examined following the global ordering of the items such that the first partition considered would contain Objects starting with the most frequent item. (Lines 4--9)

For each partition $P_i$ of $S$, first, the inverted index $I_S$ on the right hand collection $S$ is updated considering the Objects in $P_i$ (Line 8). Then, the subtree $T_R^{c_i}$ rooted at the child $c_i$ of the prefix tree $T_R$ root that corresponds to item $i$ is traversed in a depth first manner to perform the set containment join (Line 9).

For traversing every subtree $T_R^{c_i}$ and performing the join the \ProcessNode function of either the \dasfaa, \limit or \limita algorithm is employed. Note that in case \ProcessNode of \dasfaa is used the limit $\ell$ parameter is set equal to infinity, and thus, the full prefix tree of the left hand collection $R$ is built in Line 1.
}
\section{Experimental Evaluation}
\label{sec:exps}
In this section, we present an experimental evaluation of our methodology for set containment joins. Section~\ref{sec:setup} details the setup of our analysis. Section~\ref{sec:order} investigates the preferred global ordering of the items\eat{order of the items inside the objects \eat{when constructing the prefix tree }of the left-hand collection}, while Section~\ref{sec:employ_new} demonstrates the advantage of the \new join paradigm. \eat{Next, }Section~\ref{sec:limit_effect} shows how limit $\ell$ affects the efficiency of our methodology and presents four strategies for estimating its optimal value.
Finally, Section~\ref{sec:comparison} conducts a\eat{n extensive} performance analysis of our\eat{ proposed} methods against the state-of-the-art \dasfaa \cite{JampaniP05}.


\subsection{Setup}
\label{sec:setup}
\begin{table}[t]
\caption{Characteristics of real datasets}
\begin{center}
\begin{tabular}{l@{~~~}cccc}\hline
{\bf characteristic} &BMS &FLICKR &KOSARAK &NETFLIX\\\hline\hline
Cardinality &$515K$ &$1.7M$ &$990K$ &$480K$\\
Domain size &$1.6K$ &$810K$ &$41K$ &$18K$\\
Avg object length &$63$ &$52$ &$398$ &$1,\!557$\\
Weighted avg &\multirow{2}{*}{$7$} &\multirow{2}{*}{$10$} &\multirow{2}{*}{$9$} &\multirow{2}{*}{$210$}\\
object length &&&&\\
Max object length &$164$ &$102$ &$2497$ &$17,\!653$\\
File size (Mb) &11 &76 &31 &407\\
\eat{                      &                         &                            &\bf{object length}     &\bf{object length}    &\bf{object length}\\\hline\hline
BMS\eat{-POS}\eat{ (dup)} &$515,\!597$ &$1,\!657$ &$63$ &$7$ &$164$\\
FLICKR\eat{-LONDON}\eat{ (dup)} &$1,\!680,\!490$ &$810,\!660$ &$52$ &$10$ &$102$\\
KOSARAK\eat{ (dup)} &$990,\!002$ &$41,\!270$ &$398$ &$9$ &$2497$\\
NETFLIX\eat{ (dup)} &$480,\!189$ &$17,\!770$ &$1,\!557$ &$210$ &$17,\!653$\\
}\hline
\end{tabular}
\end{center}
\label{tab:real}
\end{table}
\eat{
\begin{table*}
\caption{Datasets' characteristics}
\begin{center}
\scriptsize
\begin{tabular}{|l|c|c|c|c|c|}\hline
\bf{dataset}   &\bf{cardinality} &\bf{domain size} &\bf{average}           &\bf{weighted average} &\bf{maximum}\\
                      &                         &                            &\bf{object length}     &\bf{object length}    &\bf{object length}\\\hline\hline
BMS\eat{-POS}\eat{ (dup)} &$515,\!597$ &$1,\!657$ &$63$ &$7$ &$164$\\
FLICKR\eat{-LONDON}\eat{ (dup)} &$1,\!680,\!490$ &$810,\!660$ &$52$ &$10$ &$102$\\
KOSARAK\eat{ (dup)} &$990,\!002$ &$41,\!270$ &$398$ &$9$ &$2497$\\
NETFLIX\eat{ (dup)} &$480,\!189$ &$17,\!770$ &$1,\!557$ &$210$ &$17,\!653$\\
\hline
\end{tabular}
\end{center}
\label{tab:real}
\end{table*}
}

\eat{
\begin{table}
\caption{Parameters for synthetic datasets generation}
\begin{center}
\scriptsize
\begin{tabular}{|c|c|}\hline
\bf{Parameter} &\bf{Values}\\\hline\hline
$|R|$ &$100,\!000$, $300,\!000$, $\mathbf{500,\!000}$, $1,\!000,\!000$, $3,\!000,\!000$\\
$|D|$ &$1,\!000$, $5,\!000$, $\mathbf{10,\!000}$, $50,\!000$, $100,\!000$\\
$L_{avg}$ &$5$, $10$, $\mathbf{30}$, $50$, $100$\\
$Zipf$ &$0.2$, $0.4$, $\mathbf{0.6}$, $0.8$, $1$\\
\hline
\end{tabular}
\end{center}
\label{tab:synthetic}
\end{table}
}

\begin{table}[t]
\caption{Characteristics of synthetic datasets}
\begin{center}
\begin{tabular}{lcccc|}\hline
{\bf characteristic} &{\bf values} &{\bf default value} &{\bf file size (Gb)}\\\hline\hline
Cardinality &1$M$, 3$M$, 5$M$, 7$M$, 10$M$ &5$M$ &0.3, 0.8, 1.4, 1.9, 2.7\\
Domain size &10$K$, 50$K$, 100$K$, 500$K$, 1$M$ &100$K$ &1.1, 1.3, 1.4, 1.6, 1.6\\
Weighted avg &\multirow{2}{*}{10, 30, 50, 70, 100} &\multirow{2}{*}{50} &\multirow{2}{*}{0.3, 0.8, 1.4, 1.9, 2.7}\\
object length &&\\
Zipfian &\multirow{2}{*}{0, 0.3, 0.5, 0.7, 1} &\multirow{2}{*}{0.5} &\multirow{2}{*}{1.4, 1.4, 1.4, 1.3, 1.1}\\
distribution & & &\\
\hline
\end{tabular}
\end{center}
\label{tab:synthetic}
\end{table}
Our experimental analysis involves both real and synthetic collections. Particularly, we use the following real datasets:
\begin{itemize}
\item BMS\eat{-POS} is a collection of click-stream data from Blue Martini Software and KDD 2000 cup \cite{ZhengKM01}.
\item FLICKR is a collection of photographs from Flickr website for the city of London\eat{ taken over a period of 2 years, previously used in} \cite{BourosGM12}. 
Each object contains the union of ``tags'' and ``title'' elements.
\item KOSARAK is a collection of click-stream data from a hungarian on-line news portal \eat{publicly }available at http://fimi.ua.ac.be/data/.
\item NETFLIX is a collection of user ratings on movie titles over a period of 7 years from the Netflix Prize and KDD 2007 cup.
\end{itemize}
Table~\ref{tab:real} summarizes the characteristics of the real datasets. BMS\eat{-POS} covers the case of small domain collections while FLICKR the case of datasets with very large domains. NETFLIX is a collection of extremely long objects.
In addition, to study the scalability of the methods, we generated synthetic datasets with respect to (i) the collection cardinality, \eat{$|R|$ }(ii) the domain size, \eat{$|D|$, }(iii) the weighted average object length\eat{ $L_{avg}$} and (iv) the order of the Zipfian distribution \eat{$Zipf$ }for the item frequency. Table~\ref{tab:synthetic} summarizes the characteristics of the synthetic collections. On each test, we vary one of the above parameters while the rest are set to their default values.

Similar to \eat{the work in }\cite{JampaniP05} for set containment joins 
(and other works on
set similarity joins \cite{BayardoMS07,XiaoWLY08\eat{,XiaoWLS09,XiaoWLYW11}}),
our experiments\eat{ in our analysis} involve only self-joins, i.e.,
$R = S$ (note, however, that our methods operate exactly as in case of
non self-joins, i.e., they take as input two copies of the same dataset).
\eat{Note that t}The collections and the indexing structures used by all join
methods are stored entirely in main memory; as discussed in the introduction we focus on the main module of the evaluation methods which joins two in-memory partitions, but our proposed methodology is easily integrated in the block-based approaches of \cite{JampaniP05,Mamoulis03}. Further, we do not consider
any compression techniques, as they are orthogonal to our methodology. 

To assess the performance of each method, we measure its response
time, the total number of intersections performed and the total number
of candidates; note that the response time includes both the indexing
and joining cost of the method, and in case of the \new paradigm, also
the cost of sorting and partitioning the inputs. 
Finally, all tested methods are written in C++\eat{ and compiled with gcc,} and the evaluation is carried out on an 3.6Ghz Intel Core i7 CPU with 64GB RAM running Debian Linux.

\subsection{Items Global Ordering}
The goal of the first experiment is to determine the most appropriate ordering for the items inside an object\eat{ when constructing the prefix tree on the left-hand collection}.
In practice, only the characteristics of prefix tree $T_R$ and how it is utilized are affected by how we order the items inside each object (neither the size of inverted index $I_S$ nor the number of objects accessed from $S$ depend on this ordering).
Therefore, in this experiment, we only focus on
the \dasfaa join paradigm. In \cite{JampaniP05}, to construct a compact prefix tree $T_R$ the items inside an object are arranged in decreasing order of their frequency.
On the other hand, arranging the items in increasing frequency order allows for faster candidate pruning as the candidates list $CL$ rapidly shrinks after a small number of list intersections. In other words, the ordering of the items affects not only the building cost and the storage requirements of $T_R$, but most importantly, the response time of the join method. In practice, we observe that the best\eat{preferred} ordering is also related to how the $CL \cap I_S[n.item]$ list intersection is implemented.  Although the problem\eat{implementation details} of list intersection is out of scope of this paper per se, we implemented\eat{ two approaches for computing $CL \cap I_S[n.item]$}: (i) a merge-sort based approach, and (ii) a hybrid approach based on \cite{Baeza-Yates04} that either adopts the merge-sort approach or binary searches every object of $CL$ inside the $I_S[n.item]$ postings list. Table~\ref{tab:vary-order}\eat{Figure~\ref{fig:vary-order}} confirms our claim regarding the correlation between the global ordering of the items and the response time of the \dasfaa join algorithm (note that the reported time involves both the indexing and the join phase of the method). Arranging the items in decreasing order of their frequency is generally better only if the merge-sort based approach is adopted for the list intersections, while in case of the hybrid approach, the objects should be arranged in increasing order; an exception arises for  NETFLIX where adopting the increasing ordering is always more beneficial because of its extremely long objects. In summary, the combination of the hybrid approach and the increasing frequency global ordering minimizes the response time of the \dasfaa algorithm in all cases. Thus, for the rest of this analysis, we employ the hybrid approach for list intersection and arrange the items inside an object in the increasing order of their frequency. Note that for matters of reference and completion we also include the original version of \cite{JampaniP05} denoted by $\mathtt{org}\dasfaa$ corresponding to the Decreasing-Hybrid combination of Table~\ref{tab:vary-order}.
\label{sec:order}
\begin{table}
\caption{Determining items global ordering, response time (sec) of the \dasfaa algorithm}
\begin{center}
\begin{tabular}{lcccc}\hline
\multirow{2}{*}{{\bf Dataset}} &\multicolumn{2}{c}{{\bf Increasing}} &\multicolumn{2}{c}{{\bf Decreasing}}\\\cline{2-5}
 &\bf{Merge-sort} &\bf{Hybrid} &\bf{Merge-sort} &\bf{Hybrid}\\\hline\hline
BMS &$407$ &$42$ &$106$ &$71$\\
FLICKR &$1606$ &$30$ &$187$ &$108$\\
KOSARAK &$1606$ &$73$ &$282$ &$136$\\
NETFLIX &$18,\!399\eat{18\!398.9}$ &$504$ &$35,\!169\eat{32,\!959}$ &$14,\!051$\\
%
\hline
\end{tabular}
\end{center}
\label{tab:vary-order}
\end{table}
\eat{
\begin{figure}[!ht]
\begin{center}
\begin{tabular}{cccc}
\hspace*{-0.7cm}
\includegraphics[width=0.235\linewidth]{plots/potemkin/lol.pdf}
&\hspace*{-0.1cm}\includegraphics[width=0.235\linewidth]{plots/potemkin/lol.pdf}
&\hspace*{-0.1cm}\includegraphics[width=0.235\linewidth]{plots/potemkin/lol.pdf}
&\hspace*{-0.1cm}\includegraphics[width=0.235\linewidth]{plots/potemkin/lol.pdf}\\
(a) BMS\eat{-POS}
&(b) FLICKR-LONDON
&(c) KOSARAK
&(d) NETFLIX\\
\end{tabular}
\caption{Vary the order of the items inside the objects}
\label{fig:vary-order}
\end{center}
\end{figure}
}

\subsection{Employing the {\large\new} Join Paradigm}
\label{sec:employ_new}
\eat{
\begin{table}
\caption{Employing the \new join paradigm, response time (sec)\eat{ of the \dasfaa method}}
\begin{center}
\begin{tabular}{|l|c|c|c|}\hline
\bf{Dataset} &\bf{\dasfaa} &$\dasfaa*$ &\bf{Improvement}\\\hline\hline
BMS &$42$ &$28$ &$33\%$\\
FLICKR &$30$ &$20$ &$33\%$\\
KOSARAK &$73$ &$54$ &$26\%$\\
NETFLIX &$504$ &$391$ &$23\%$\\
\hline
\end{tabular}
\end{center}
\label{tab:new}
\end{table}
}
\eat{
\begin{table}
\caption{Employing the \new join paradigm, response time (sec)\eat{ of the \dasfaa method}}
\begin{center}
\begin{tabular}{|l|c|c|c|c|c|}\hline
\multirow{2}{*}{\bf{Dataset}} &\multirow{2}{*}{$\mathtt{org}\dasfaa$} &\multirow{2}{*}{\bf{\dasfaa}} &\multirow{2}{*}{$\dasfaa*$} &\multicolumn{2}{c|}{\bf{Rel. improvement over}}\\\cline{5-6}
& & & &$\mathtt{org}\dasfaa$ &\dasfaa\\\hline\hline
BMS &$71$ &$42$ &$28$ &$60\%$ &$33\%$\\
FLICKR &$108$ &$30$ &$20$ &$81\%$ &$33\%$\\
KOSARAK &$136$ &$73$ &$54$ &$60\%$ &$26\%$\\
NETFLIX &$14,\!051$ &$504$ &$391$  &$97\%$ &$23\%$\\
\hline
\end{tabular}
\end{center}
\label{tab:new}
\end{table}
}
\begin{table}
\caption{Employing the \new join paradigm, response time (sec)\eat{ of the \dasfaa method}}
\begin{center}
\begin{tabular}{lccccc}\hline
\multirow{2}{*}{\bf{Dataset}} &\multirow{2}{*}{$\mathtt{org}\dasfaa$} &\multirow{2}{*}{\bf{\dasfaa}} &\multirow{2}{*}{$\dasfaa*$} &\multicolumn{2}{c}{\bf{Improvement ratio over}}\\\cline{5-6}
& & & &$\mathtt{org}\dasfaa$ &\dasfaa\\\hline\hline
BMS &$71$ &$42$ &$28$ &$2.5\times$ &$1.5\times$\\
FLICKR &$108$ &$30$ &$20$ &$5.4\times$ &$1.5\times$\\
KOSARAK &$136$ &$73$ &$54$ &$2.5\times$ &$1.4\times$\\
NETFLIX &$14,\!051$ &$504$ &$391$  &$38.5\times$ &$1.3\times$\\
\hline
\end{tabular}
\end{center}
\label{tab:new}
\end{table}
Next, we investigate the advantage of \new (Section \ref{sec:new}) over the \dasfaa join paradigm of \cite{JampaniP05}. For this purpose we devise an extension to the \dasfaa algorithm that follows \new, denoted by $\dasfaa*$.
Table~\ref{tab:new} reports the response time of the\eat{ two} algorithms. The results experimentally prove the superiority of the \new paradigm; 
$\dasfaa*$ is from $1.3$ to $1.5$ times faster than \dasfaa.
Recall at this point that compared to the algorithm discussed in
\cite{JampaniP05}, our version of \dasfaa arranges the items in
increasing order of their frequency as discussed in
Section~\ref{sec:order}; thus, the overall \eat{relative }improvement of
$\dasfaa*$ (which follows \new) over the original method of \cite{JampaniP05} 
$\mathtt{org}\dasfaa$ is even greater: 
$2.5\times$ for BMS-POS, $5.4\times$ for FLICKR, $2.5\times$ for KOSARAK and $38.5\times$ for NETFLIX.
For the rest of our analysis we adopt the \new\eat{ join} paradigm for all tested methods.

\eat{
\begin{figure}[!ht]
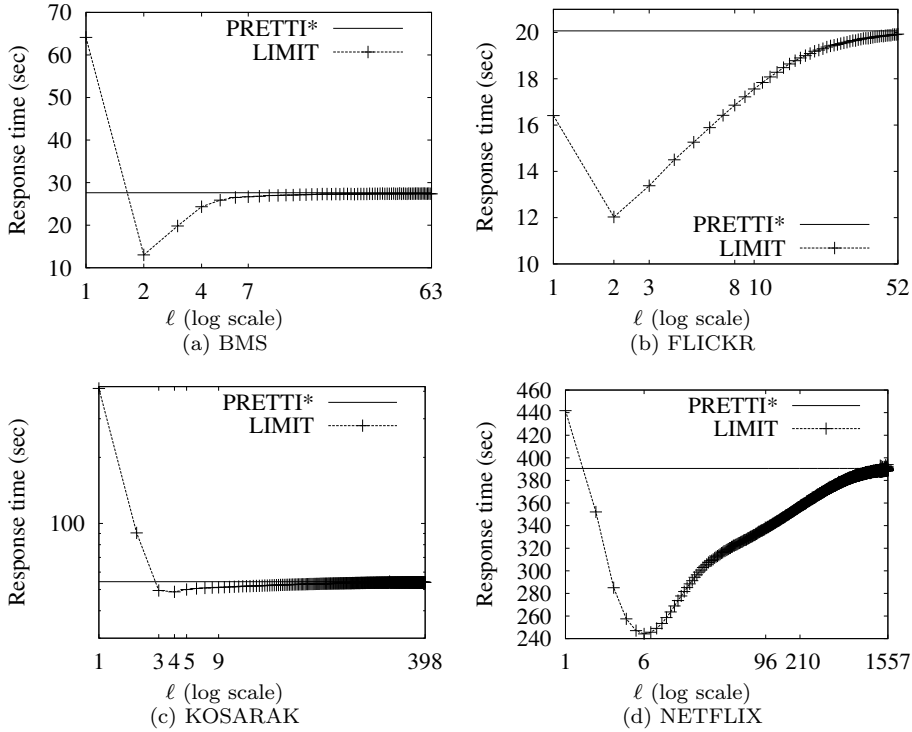

\begin{center}
\begin{tabular}{cccc}
\hspace*{-0.7cm}
\includegraphics[width=0.235\linewidth]{plots/BMS-POS_dr_mem-idx_vary-partitions.pdf}
&\hspace*{-0.1cm}\includegraphics[width=0.235\linewidth]{plots/BMS-POS_dr_mem-idx_vary-items.pdf}
&\hspace*{-0.1cm}\includegraphics[width=0.235\linewidth]{plots/BMS-POS_dr_mem-idx_vary-items.pdf}
&\hspace*{-0.1cm}\includegraphics[width=0.235\linewidth]{plots/NETFLIX_dr_mem-idx_vary-items.pdf}\\
$\ell$
&$\ell$
&$\ell$
&$\ell$\\
(a) BMS\eat{-POS}\eat{ (no dup)}
&(b) FLICKR\eat{-LONDON (no dup)}
&(c) KOSARAK\eat{ (no dup)}
&(d) NETFLIX\eat{ (no dup)}
\end{tabular}
\caption{Vary limit $\ell$, response time}
\label{fig:vary-lm_time}
\end{center}
\end{figure}
}
\eat{
\begin{figure}[!ht]
\begin{center}
\begin{tabular}{cc}
\hspace*{-0.3cm}
\includegraphics[width=0.5\linewidth]{plots/all_mem-idx_vary-partitions.pdf}
&\hspace*{-0.2cm}\includegraphics[width=0.5\linewidth]{plots/all_mem-idx_vary-partitions.pdf}\\
{\small Partitions examined ($\%$)} &{\small Partitions examined ($\%$)}\\
(a) &(b)
\end{tabular}
\caption{Storage requirements for join indices: (a) $\dasfaa*$ over \dasfaa, (b) \limit/\limita over \dasfaa}
\label{fig:mem-imprint}
\end{center}
\end{figure}
}

\subsection{The Effect of Limit {\large$\ell$}}
\label{sec:limit_effect}
\begin{table}
\caption{Limit $\ell$ determined by each estimation strategy} \label{tab:limits}
\begin{center}
\begin{tabular}{lccccc}\hline
\bf{Dataset} &{\bf Optimal} &$AVG$ &$W$--$AVG$  &$MDN$ &$FRQ$\\\hline\hline
BMS\eat{-POS}\eat{ (dup)} &$2$ &$63$  &$7$ &$4$ &$4$\\
FLICKR\eat{-LONDON}\eat{ (dup)} &$2$ &$52$ &$10$ &$8$ &$3$\\
KOSARAK\eat{ (dup)} &$4$ &$398$ &$9$ &$3$ &$5$\\
NETFLIX\eat{ (dup)} &$6$ &$1,\!557$ &$210$ &$96$ &$6$\\
\hline
\end{tabular}
\end{center}
\end{table}

\begin{figure}
\begin{center}
\begin{tabular}{cc}
\includegraphics[width=0.47\linewidth]{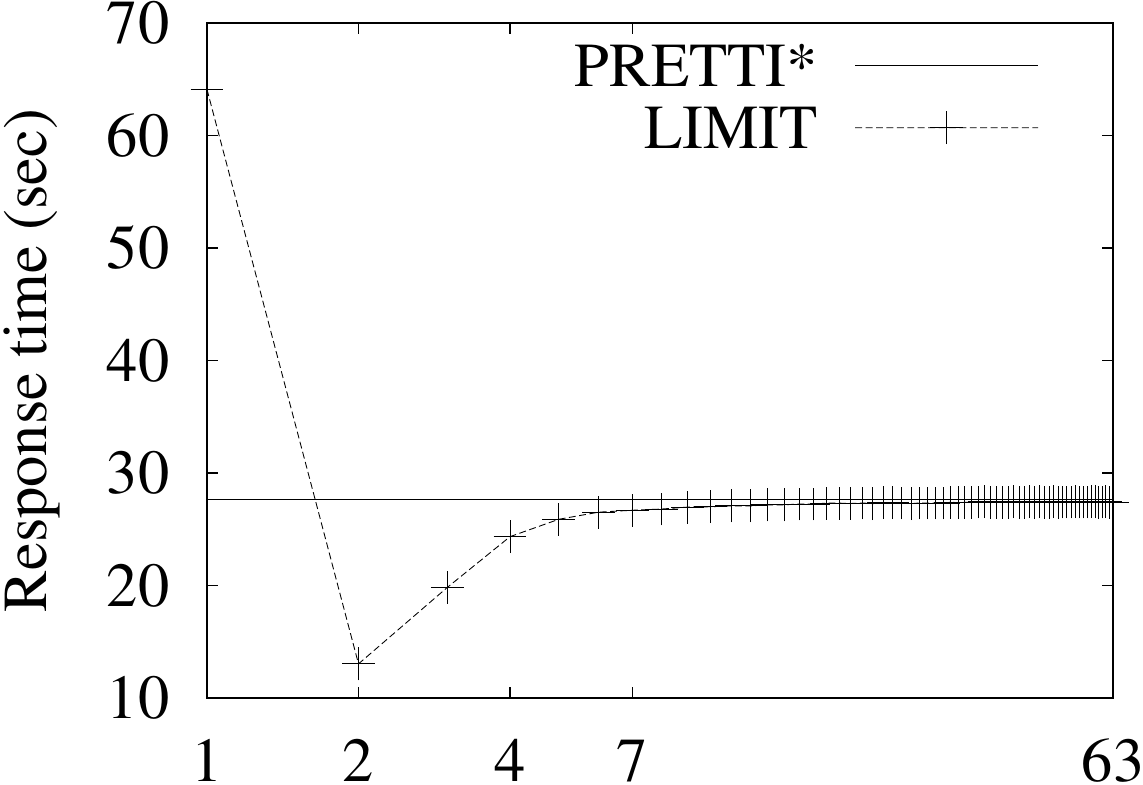}
&\includegraphics[width=0.47\linewidth]{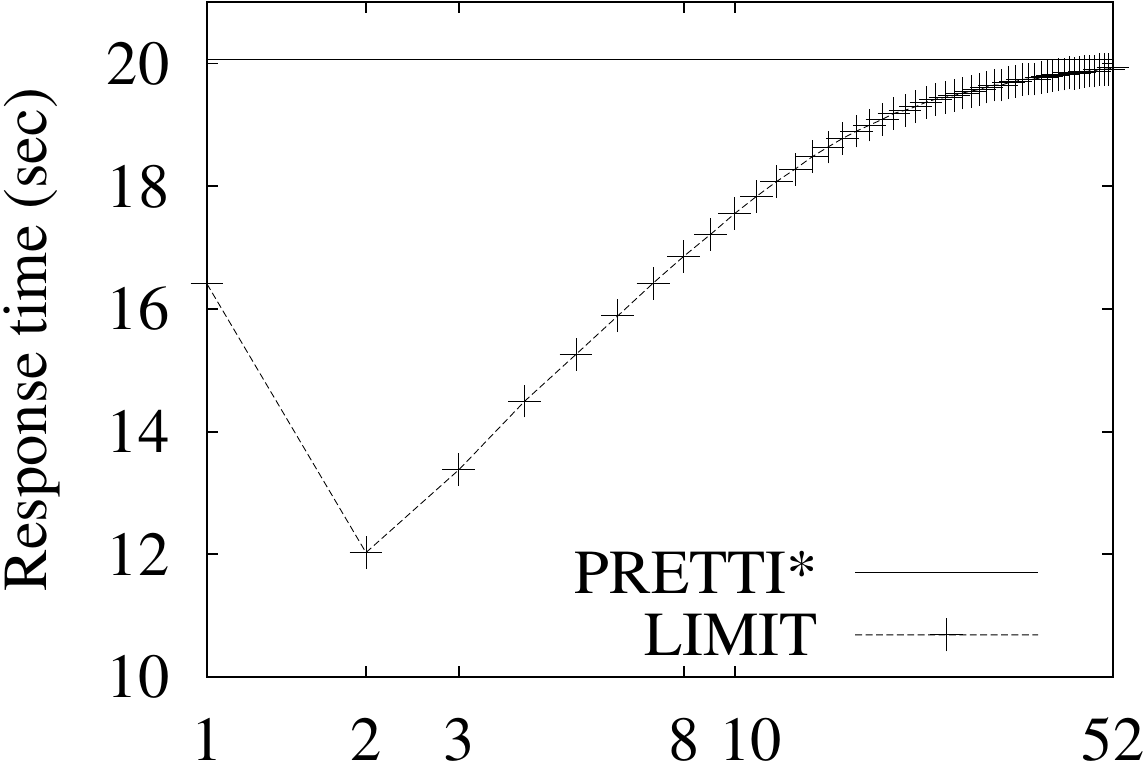}\\
$\ell$ (log scale)
&$\ell$ (log scale)\\
(a) BMS\eat{-POS}\eat{ (no dup)}
&(b) FLICKR\eat{-LONDON (no dup)}\\\\
\includegraphics[width=0.47\linewidth]{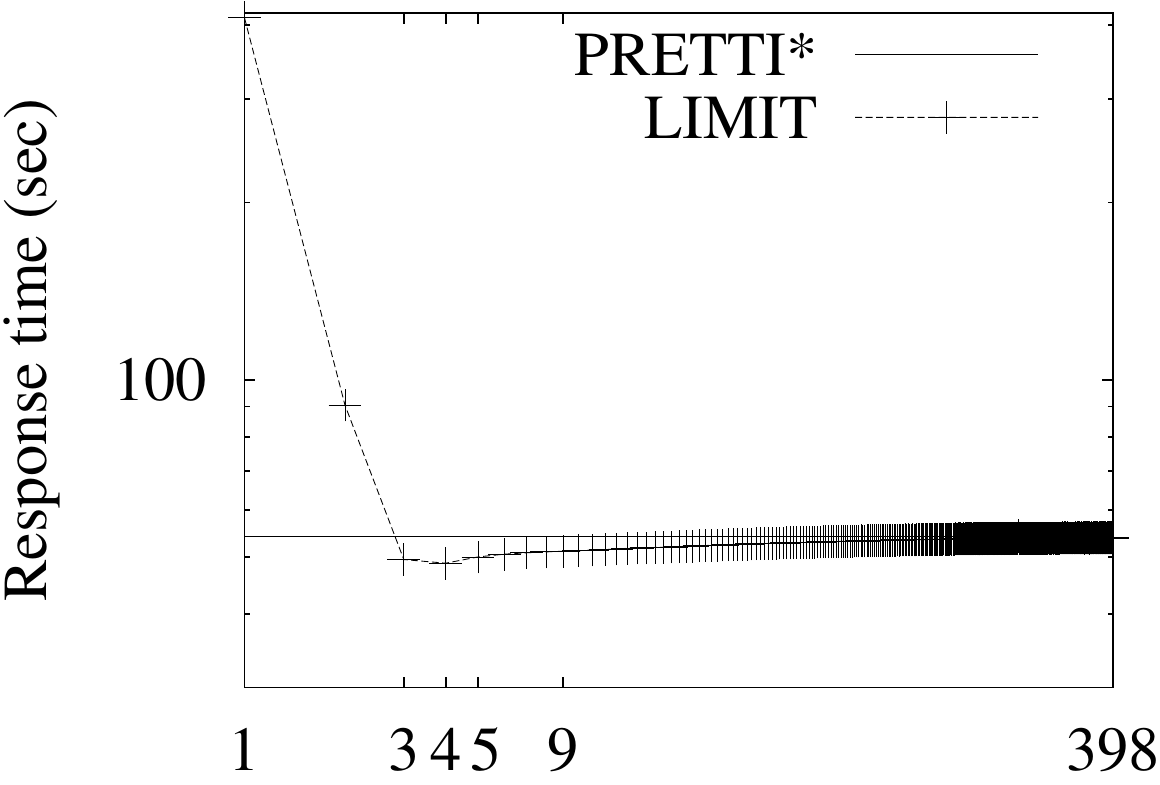}
&\includegraphics[width=0.47\linewidth]{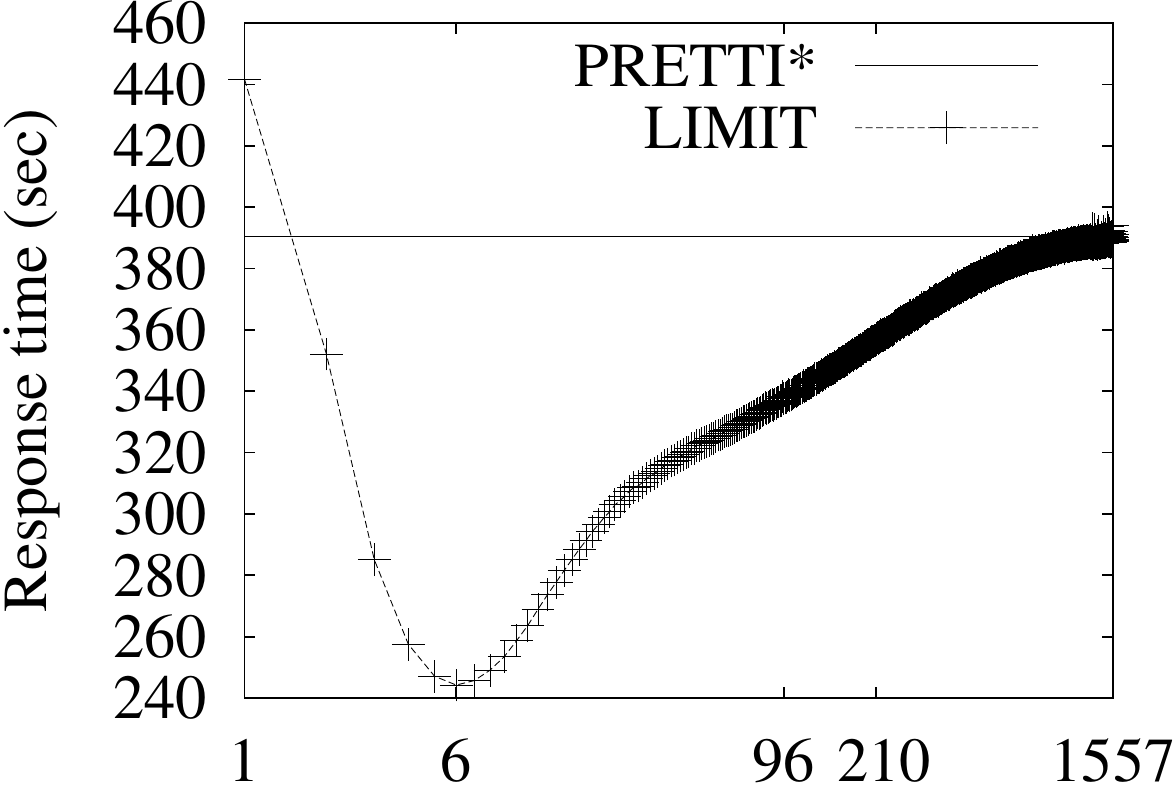}\\
$\ell$ (log scale)
&$\ell$ (log scale)\\
(c) KOSARAK\eat{ (no dup)}
&(d) NETFLIX\eat{ (no dup)}
\end{tabular}
\caption{Vary limit $\ell$, response time}
\label{fig:vary-lm_time}
\end{center}
\end{figure}

\begin{figure}
\begin{center}
\begin{tabular}{cc}
\includegraphics[width=0.47\linewidth]{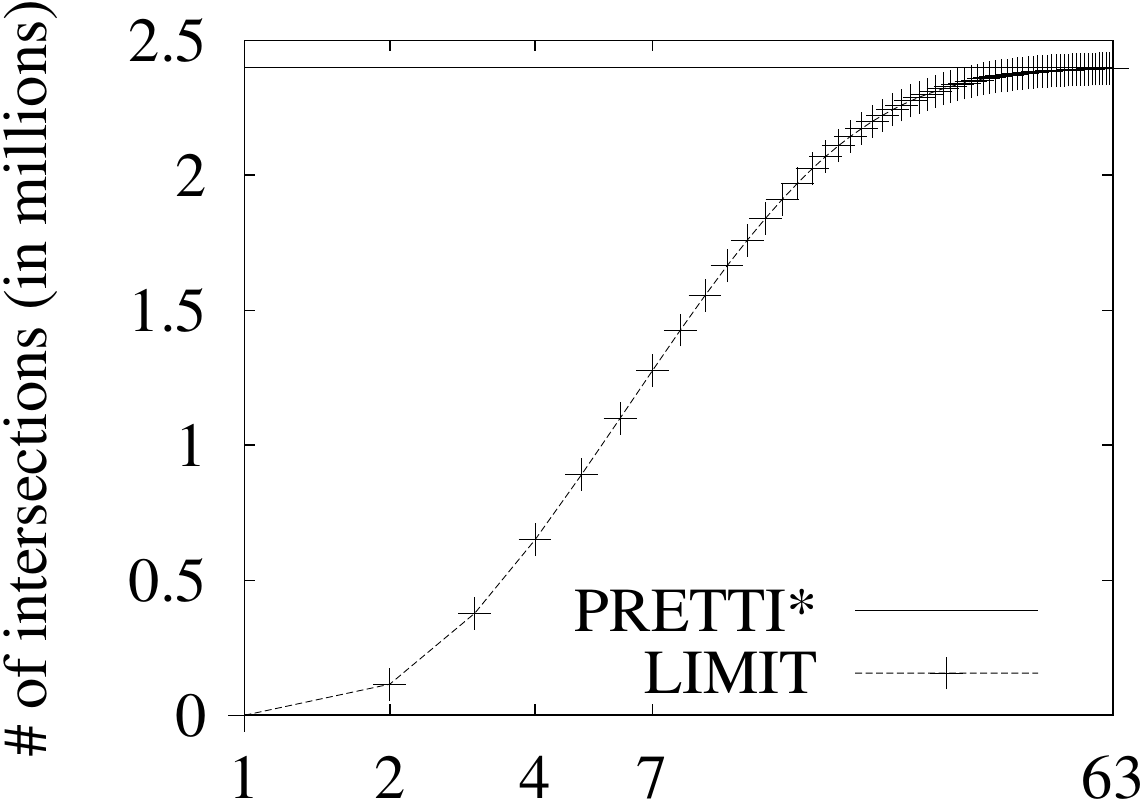}
&\includegraphics[width=0.47\linewidth]{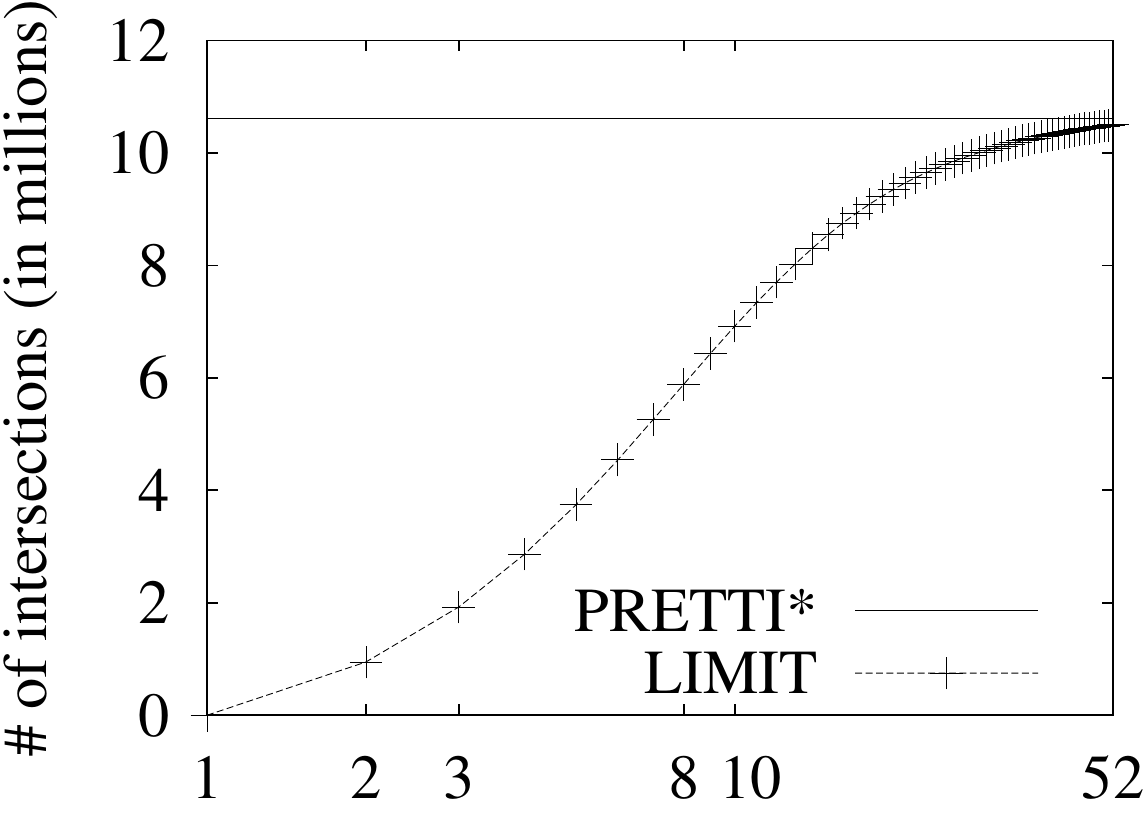}\\
$\ell$ (log scale)
&$\ell$ (log scale)\\
(a) BMS\eat{-POS}\eat{ (no dup)}
&(b) FLICKR\eat{-LONDON (no dup)}\\\\
\includegraphics[width=0.47\linewidth]{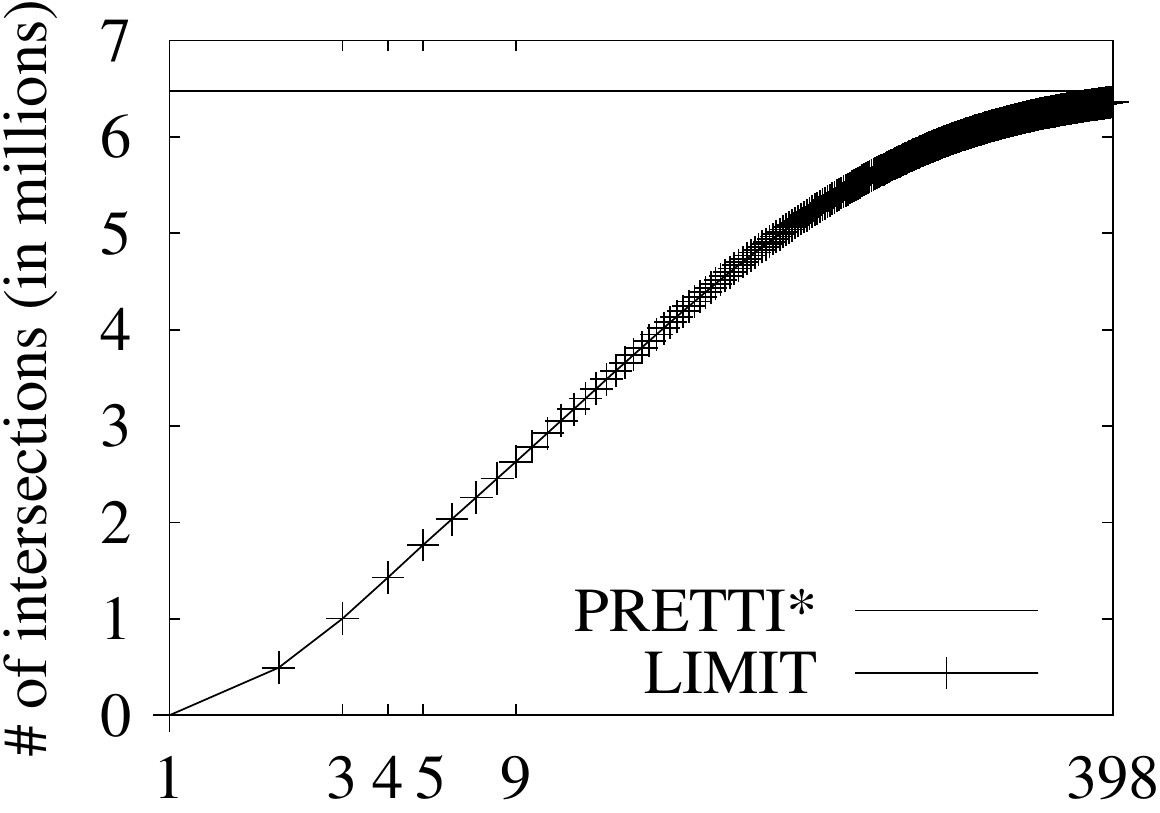}
&\includegraphics[width=0.47\linewidth]{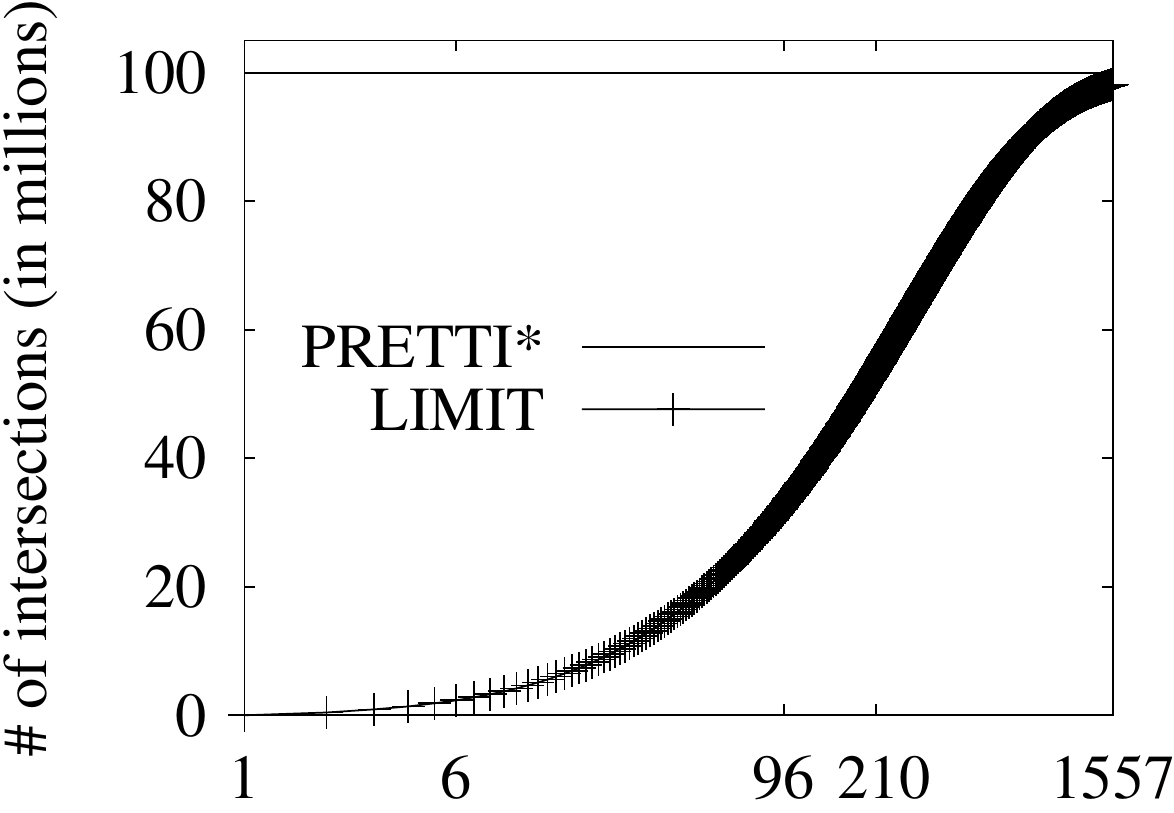}\\
$\ell$ (log scale)
&$\ell$ (log scale)\\
(c) KOSARAK\eat{ (no dup)}
&(d) NETFLIX\eat{ (no dup)}
\end{tabular}
\caption{Vary limit $\ell$, number of intersections.}
\label{fig:vary-Im_inters}
\end{center}
\end{figure}

\begin{figure}
\begin{center}
\begin{tabular}{cc}
\includegraphics[width=0.47\linewidth]{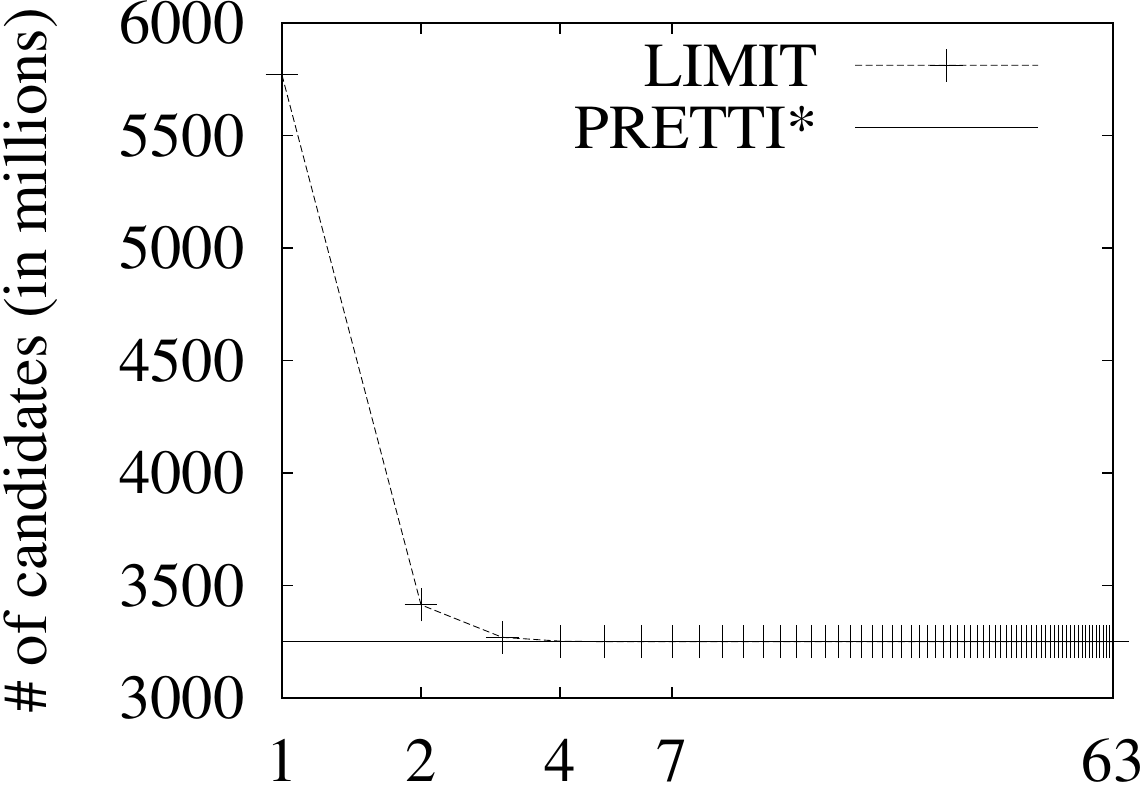}
&\includegraphics[width=0.47\linewidth]{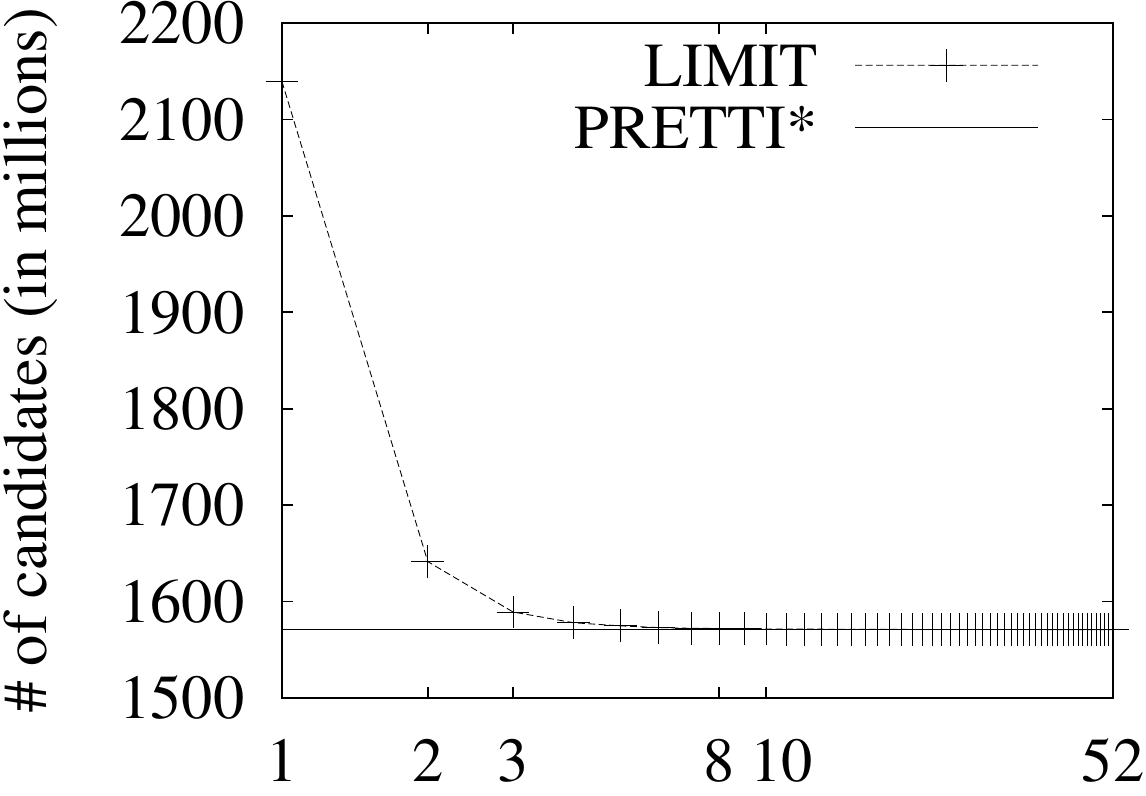}\\
$\ell$ (log scale)
&$\ell$ (log scale)\\
(a) BMS\eat{-POS}\eat{ (no dup)}
&(b) FLICKR\eat{-LONDON (no dup)}\\\\
\includegraphics[width=0.47\linewidth]{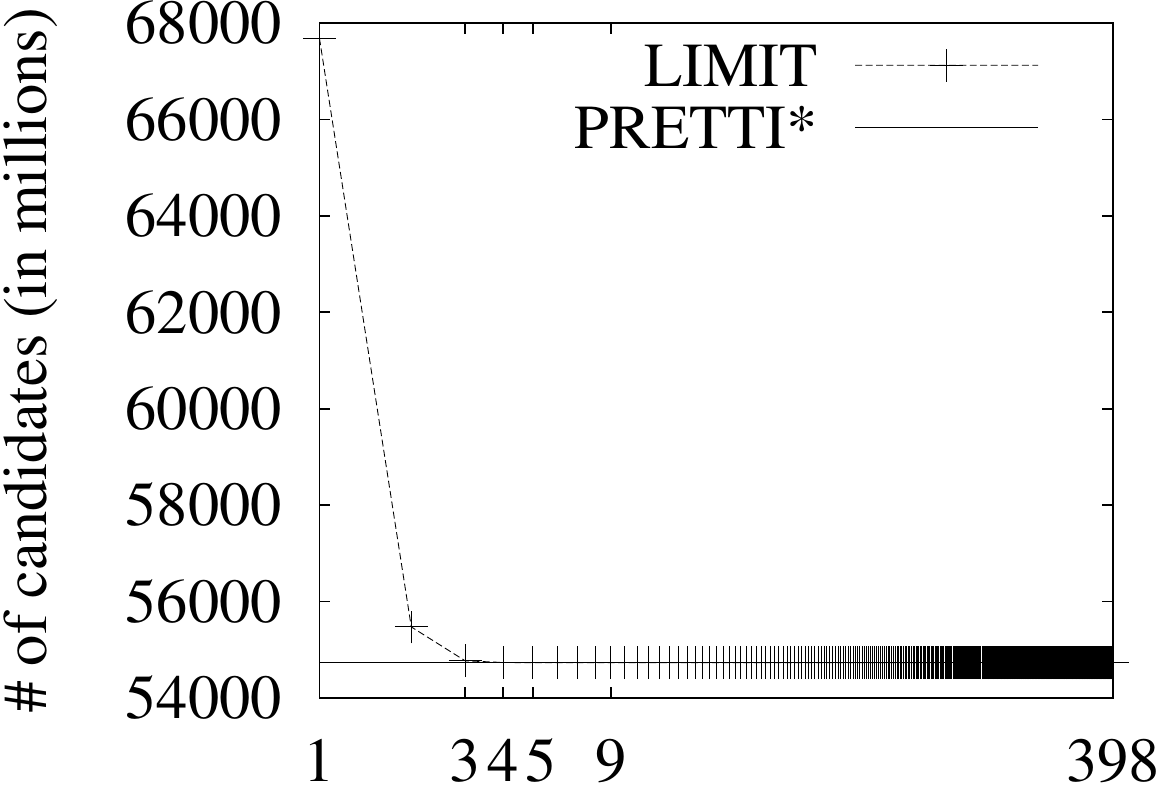}
&\includegraphics[width=0.47\linewidth]{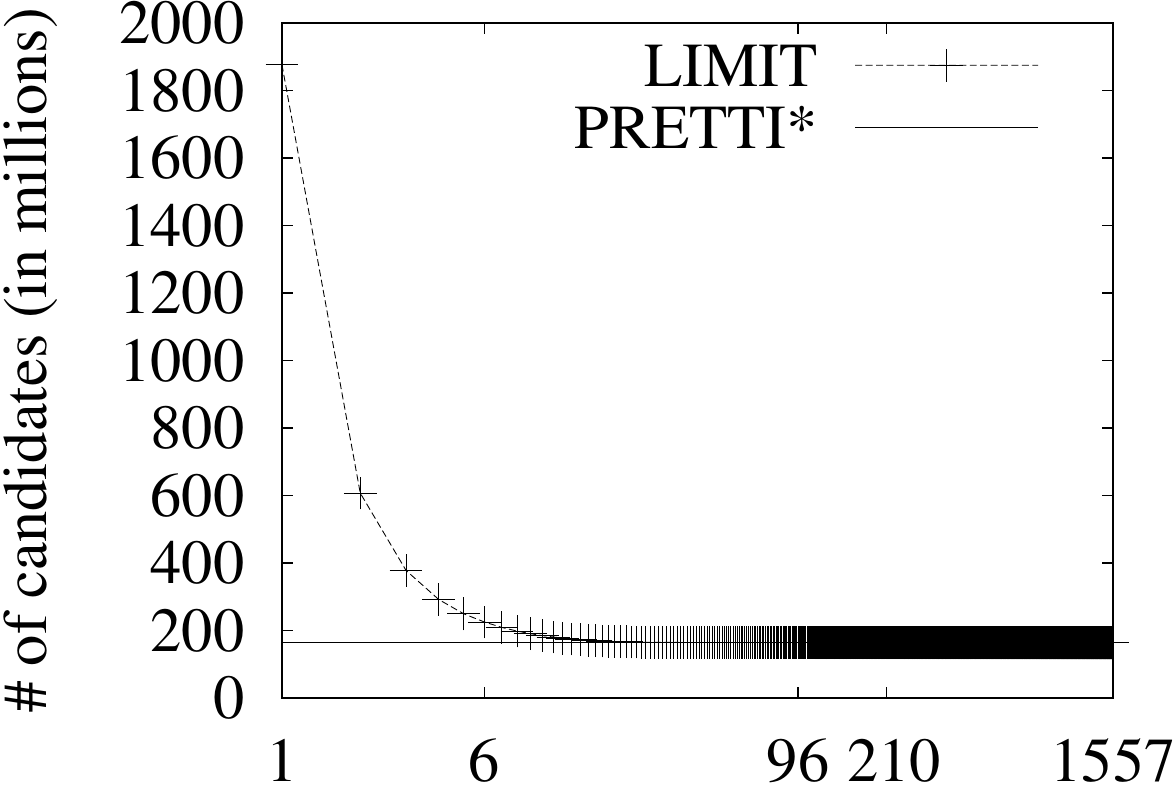}\\
$\ell$ (log scale)
&$\ell$ (log scale)\\
(c) KOSARAK\eat{ (no dup)}
&(d) NETFLIX\eat{ (no dup)}
\end{tabular}
\caption{Vary limit $\ell$, number of candidates (for $\dasfaa*$ equals the number of results)}
\label{fig:vary-lm_cands}
\end{center}
\end{figure}
As discussed in Section~\ref{sec:limits}, employing limit $\ell$ for
set containment joins introduces a trade-off between list intersection
and candidates verification. To demonstrate this effect, we run the
\limit algorithm (adopting \eat{the }\new\eat{ paradigm}) while varying limit
$\ell$ from $1$ to the average object length in $R$, and then plot its
response time (Figure~\ref{fig:vary-lm_time}), the number of list
intersections performed (Figure~\ref{fig:vary-Im_inters}) and the
total number of candidates (Figure~\ref{fig:vary-lm_cands}). The total
number of candidates includes both\eat{ the} $(r,s)$ pairs which are
directly reported as results, i.e., with $|r| \leq \ell$, and those
that are verified by comparing their prefixes beyond $\ell$, i.e., with $|r| > \ell$. To have a
better understanding of this experiment we also include the
measurements for $\dasfaa*$ which uses an \emph{unlimited}
$T_R$.
The figures clearly show the trade-off introduced by limit $\ell$ and confirm the existence of an optimal value that balances the benefits of using the \emph{limited} prefix tree over the cost of including a verification stage\eat{, compared to $\dasfaa*$ that utilizes the \emph{full} prefix tree}. According to Figures~\ref{fig:vary-Im_inters} and \ref{fig:vary-lm_cands}, as $\ell$ increases, \limit naturally performs more list intersections, and thus, the number of candidate pairs decreases until it becomes equal to the join results, i.e., the number of candidates \eat{considered by }for $\dasfaa*$. However, regarding its performance shown in Figure~\ref{fig:vary-lm_time}, although \limit initially benefits from having to verify fewer candidate pairs, when $\ell$ increases beyond a specific value, performing additional list intersections becomes a bottleneck and the algorithm slows down until its response time becomes almost equal to the time of $\dasfaa*$.

Apart from the trade-off introduced by limit $\ell$,
Figures~\ref{fig:vary-lm_time}, \ref{fig:vary-Im_inters} and
\ref{fig:vary-lm_cands} also show that the \limit algorithm can be
faster than $\dasfaa*$ as long as \eat{the value of }$\ell$ is properly set,
i.e., close to its optimal value\eat{: $2$ for BMS and FLICKR, $4$ for
  KOSARAK and $6$ for NETFLIX}. However\eat{On the other hand}, as discussed in
Section~\ref{sec:limits}, determining the optimal $\ell$ value is a
time-consuming procedure, reminiscent to frequent itemsets mining which
cannot be employed in practice; 
recall that $\ell$ must be determined online. 
For this purpose, we propose the following simple
strategies to select a good $\ell$ value based on cheap-to-compute
statistics that require no more than a pass over the input collection
$R$. First, strategies $AVG$ and $W$--$AVG$ set $\ell$ equal to the
average and the weighted average object length in $R$,
respectively. Similarly, strategy $MDN$ sets $\ell$ to the median
value of the object length in $R$. Last, we also devise a
frequency-based strategy termed $FRQ$. 
The idea behind $FRQ$ is to estimate when paths greater than $\ell$ would
only be contained in very few objects. 
We start with a path $p$ that contains the most frequent item in $R$
and progressively add the next items in decreasing frequency order. We
estimate the probability that this path appears in a object by
considering only the support of the items. When this probability falls
under a threshold, which makes the expected cost of list intersection
greater than the cost of verification (according to our analysis in
Section \ref{sec:limita}),
we stop adding items in $p$ and set
$\ell\!=\!|p|$.
Note that this probability serves as an upper bound for all paths of
length $\ell$ (assuming item independence), since $p$ includes the
most frequent items.
Table~\ref{tab:limits} summarizes the values of $\ell$ determined by each strategy for the experimental datasets. 
Overall $FRQ$ provides the best estimation of optimal $\ell$; in fact for NETFLIX it identifies\eat{manages to find} the actual optimal value\eat{ of $\ell$}.
Figures~\ref{fig:vary-lm_time}, \ref{fig:vary-Im_inters} and \ref{fig:vary-lm_cands} confirm this observation as the performance of \limit with a limit set by $FRQ$ is very close to its performance for the optimal $\ell$\eat{ value}.
Thus, for the rest of our analysis we adopt $FRQ$ to set limit $\ell$ value.
\eat{
Finally, in Table~\ref{tab:vary-lm_pt_size} we highlight another
advantage of $FRQ$ as the \emph{limited} prefix tree $T_R$ w.r.t. to
the $\ell$ value determined by $FRQ$ always occupies less than one
third of the space employed by \dasfaa and $\dasfaa*$ to store the
\emph{entire} prefix tree; note that for NETFLIX the required space is
one order of magnitude less. Thus, for the rest of our analysis we
adopt $FRQ$ to determine $\ell$.
}


\eat{
\begin{table}[h]
\caption{Vary limit $\ell$, prefix Tree $T_R$ size}
\begin{center}
\scriptsize
\begin{tabular}{|l|c|c|c|c|c|}\hline
\multirow{3}{*}{\bf{Dataset}} &\multicolumn{5}{c|}{\bf{Prefix tree size}}\\\cline{2-6}
                     &\dasfaa/           &\multicolumn{4}{c|}{\limit/\limita}\\\cline{3-6}
                     &$\dasfaa*$       &$AVG$ &$W$--$AVG$ &$MDN$ &$FRQ$\\\hline\hline
BMS\eat{-POS}\eat{ (dup)} &$860$ MB &$860$ MB	&$500$ MB	&$300$ MB	&$300$ MB\\
FLICKR\eat{-LONDON}\eat{ (dup)} &$3.8$ GB &$3.8$ GB &$2.7$ GB &$2.4$ GB &$1.1$ GB	\\
KOSARAK\eat{ (dup)} &$2.2$ GB &$2.2$ GB &$1$ GB	&$520$ MB &$770$ MB\\
NETFLIX\eat{ (dup)} &$32$ GB &$31$ GB &$17$ GB &$11$ GB &$1.5$ GB\\
\hline
\end{tabular}
\end{center}
\label{tab:vary-lm_pt_size}
\end{table}
}

\eat{
\subsection{Estimating Intersection Size}
\label{sec:inter_size}
\note{
$CL' = CL \cap IL$
\\Four methods:
\begin{itemize}
\item Naive: $|CL'| = |CL|$, upper bound
\item Independent approach: $|CL'| = \frac{|CL|\times|IL|}{|S|}$ (Terrovitis)
\item Selectivity-based: consider prefix tree path $p$ from root to current node with keyword $k$, $p(k_1,\ldots,k_n)$. Select $k_i \in p$ such that the combination of $(k_i,k)$ is the most selective one, selectivity $\sigma(k_i,k) = \frac{freq(k_i,k)}{freq(k_i)\times freq(k)}$. Then $|CL'| = freq(k_1,k)$ if $|p| = 1$ or $|CL'| = |CL|\times\frac{freq(k_i,k)}{freq(k_i)}$ if $|p| > 1$ (Nikos)
\item Selectivity-based naive: consider only $(k_1,k)$ combination since $k_1$ is the less frequent keyword, $|CL'| = freq(k_1,k)$ if $|p| = 1$ or $|CL'| = |CL|\times\frac{freq(k_1,k)}{freq(k_1)}$ if $|p| > 1$ (Nikos naive)
\item Hybrid: independent approach when $|p| > 1$, selectivity-based otherwise. Thus $|CL'| = freq(k_1,k)$ if $|p| = 1$ or $|CL'| = \frac{|CL|\times|IL|}{|S|}$ if $|p| > 1$.
\end{itemize}
See Table~\ref{tab:relerror} and \ref{tab:cost}.
}

\begin{table*}
\caption{Relative error}
\begin{center}
\scriptsize
\begin{tabular}{|l|c|c|c|c|c|c|}\hline
\bf{Dataset} &\bf{Upper bound} &\bf{Independent} &\bf{Selectivity-based} &\bf{Selectivity-based naive} &\bf{Hybrid} &\bf{OFFLINE}\\\hline\hline
BMS\eat{-POS} &$1280\%$ &$52\%$ &$42\%$ &$42\%$ &$41\%$ &$0$\\
FLICKR &$2951\%$ &$20\%$ &$10\%$ &$51\%$ &$12\%$\\
KOSARAK &$2892\%$ &$43\%$ &$43\%$ &$72\%$ &$32\%$\\
NETFLIX &$1207\%$ &$85\%$ &$49\%$ &$81\%$ &$69\%$\\
\hline
\end{tabular}
\end{center}
\label{tab:relerror}
\end{table*}

\begin{table*}
\caption{Cost for computing pre-join statistics (sec)}
\begin{center}
\scriptsize
\begin{tabular}{|l|c|c|c|c|c|c|}\hline
\bf{Dataset} &\bf{Upper bound} &\bf{Independent} &\bf{Selectivity-based} &\bf{Selectivity-based naive} &\bf{Hybrid} &\bf{OFFLINE}\\\hline\hline
BMS\eat{-POS} &$0$ &$0$ &$1$ &$-$ &$1$\\ 
FLICKR\eat{-LONDON} &$0$ &$0$ &$12$ &$-$ &$12$\\ 
KOSARAK &$0$ &$0$ &$3$ &$-$ &$3$\\ 
NETFLIX &$0$ &$0$ &$188$ &$-$ &$188$\\  
\hline
\end{tabular}
\end{center}
\label{tab:cost}
\end{table*}
}

\subsection{Comparison of the Join Methods}
\label{sec:comparison}
\begin{figure}
\eat{
\begin{center}
\fbox{\includegraphics[width=0.4\linewidth]{Fig10leg.pdf}}
\end{center}
\vspace*{-0.84cm}
}
\begin{center}
\begin{tabular}{cc}
\includegraphics[width=0.47\linewidth]{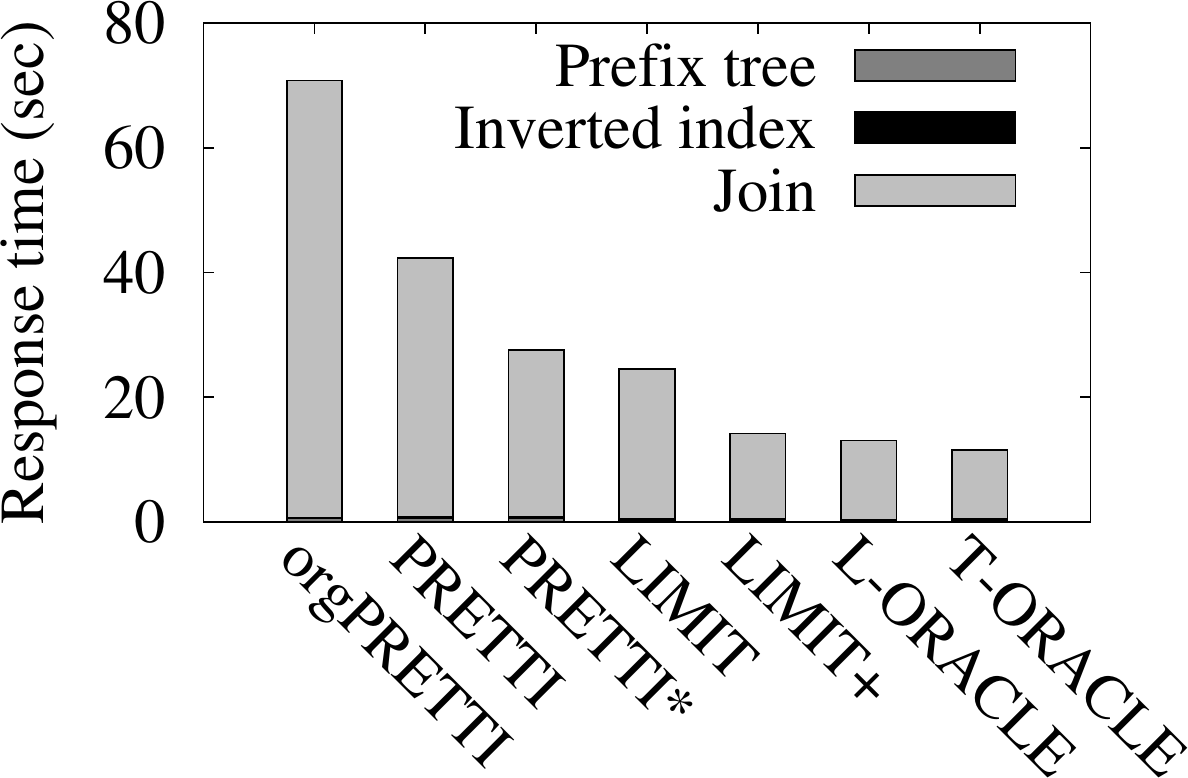}
&\includegraphics[width=0.47\linewidth]{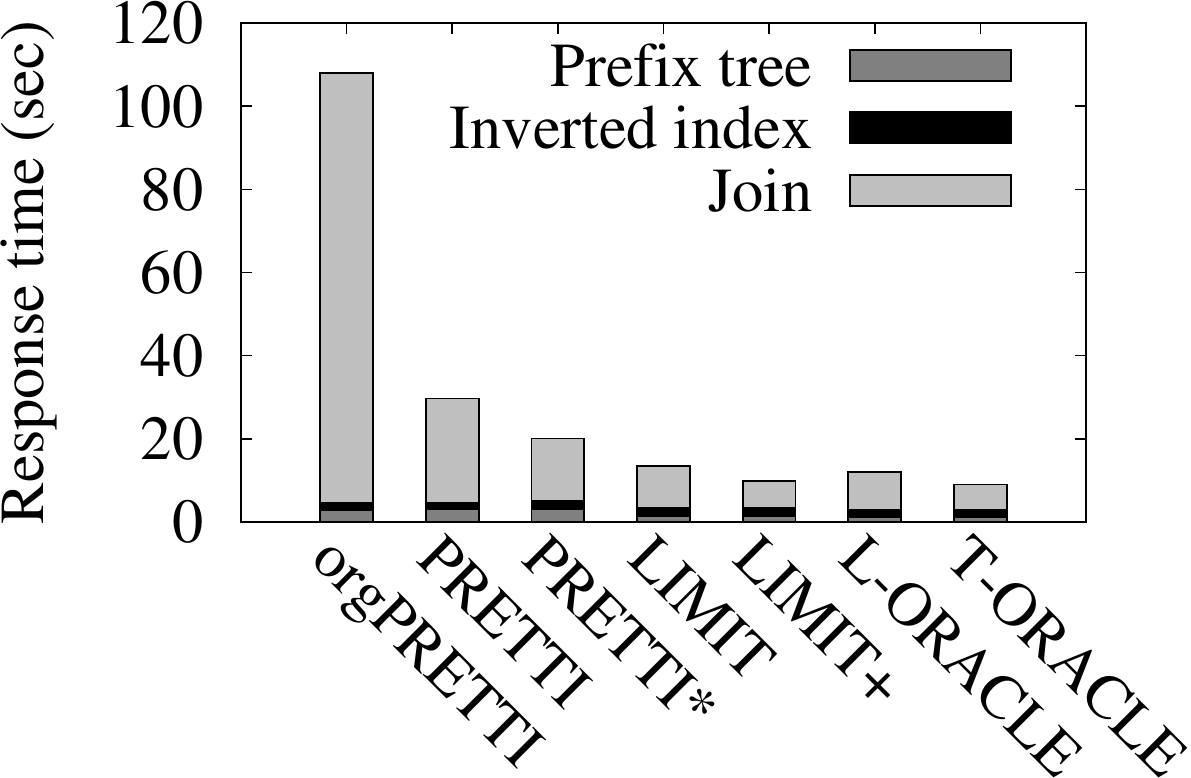}\\
(a) BMS\eat{-POS}
&(b) FLICKR\eat{-LONDON}\\\\
\includegraphics[width=0.47\linewidth]{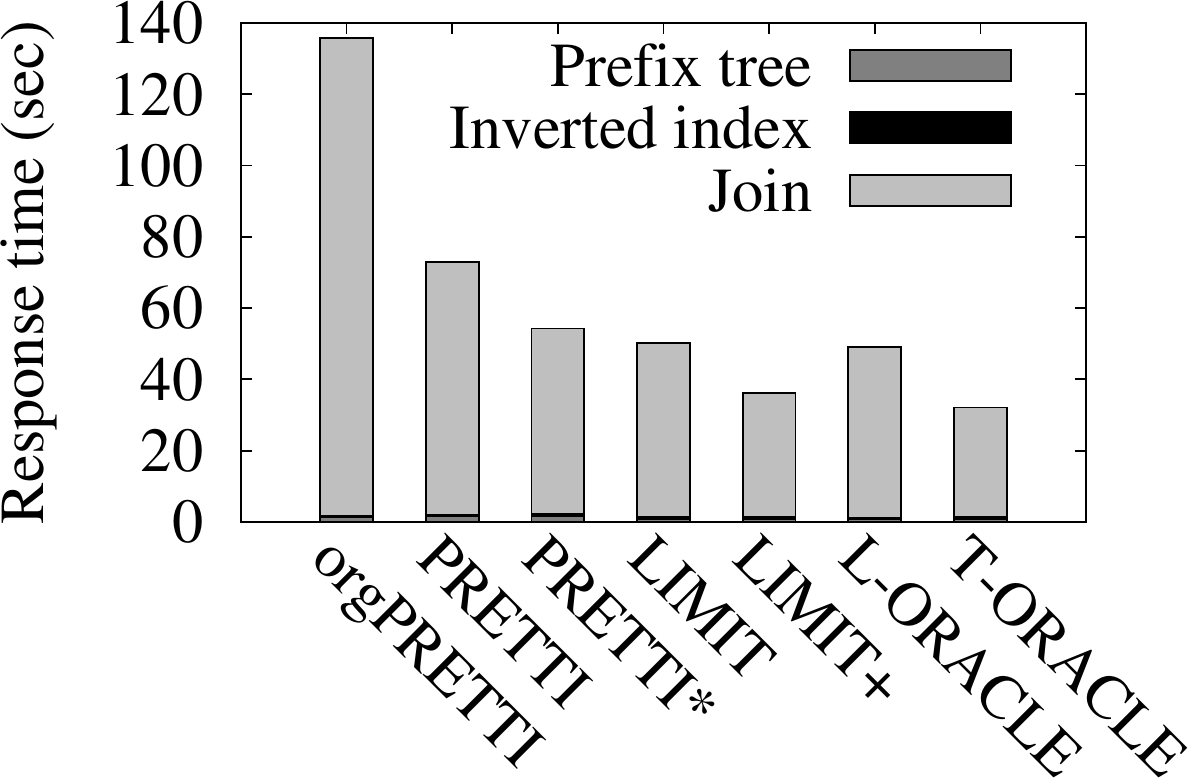}
&\includegraphics[width=0.47\linewidth]{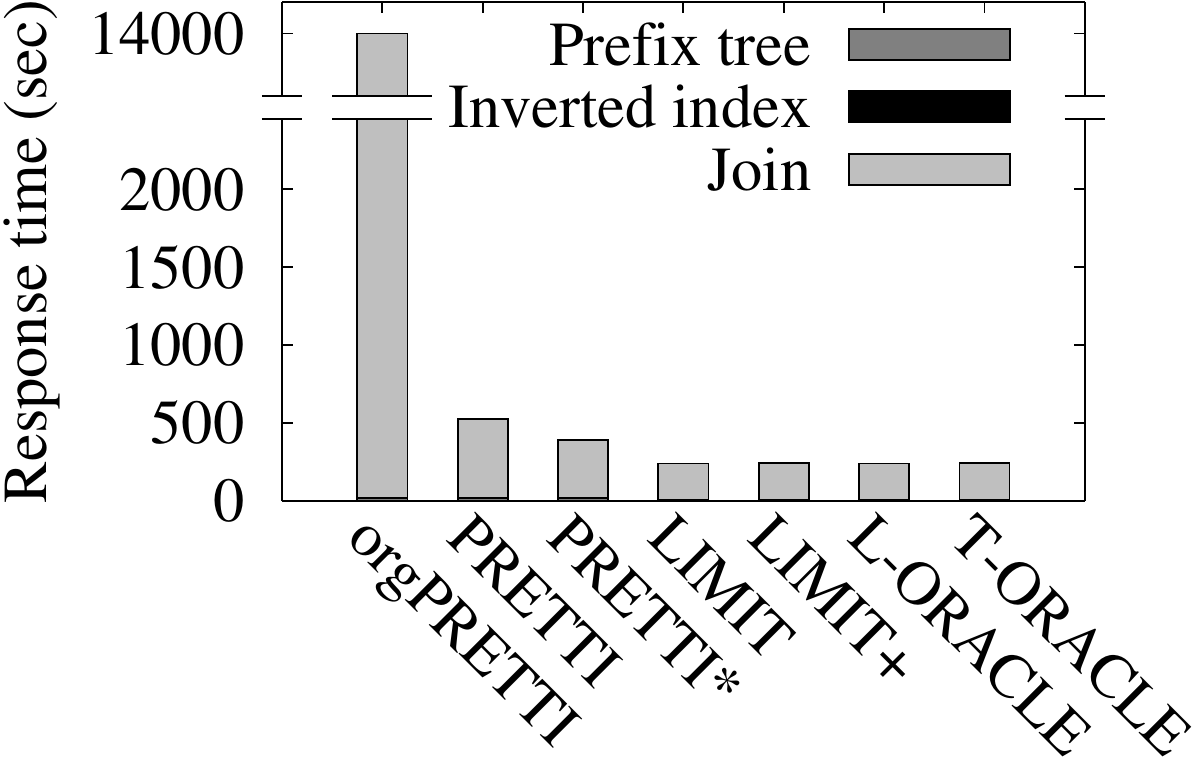}
\\
(c) KOSARAK
&(d) NETFLIX
\end{tabular}
\caption{Comparison of the set containment join methods on real datasets (limit $\ell$ set by FRQ according to Table~\ref{tab:limits})}
\label{fig:algos}
\end{center}
\end{figure}
In Section~\ref{sec:limit_effect}, we showed that by properly selecting limit $\ell$ ($FRQ$ strategy), \limit outperforms $\dasfaa*$ and, based on Sections~\ref{sec:employ_new} and \ref{sec:order}, also \dasfaa and  $\mathtt{org}\dasfaa$.
Next, we experiment with \limita which (like \limit) employs $FRQ$. Figure~\ref{fig:algos} reports the response time of $\mathtt{org}\dasfaa$, \dasfaa, $\dasfaa*$, \limit and \limita on all four real datasets. To further investigate the properties of \limita, we also include the response time of two oracle methods%
\footnote{These are {\em infeasible} methods using\eat{utilizing} apriori knowledge which is not known at runtime and it is extremely expensive to compute before the join.}: (i) \loracle corresponds to \limit with $\ell$ set to its optimal value (see Table~\ref{tab:limits}), (ii) \toracle is a version of \limita which compares the actual execution time of the two alternative strategies for current prefix tree node instead of utilizing the cost model of Section~\ref{sec:limita}; note that for this purpose we run offline both alternative strategies for every prefix tree node and store their execution time.
With the exception of $\mathtt{org}\dasfaa$ and \dasfaa the rest of
the algorithms follow the \new join paradigm. We break the response
time of all methods into three parts, (i) building prefix tree $T_R$,
(ii) building inverted index $I_S$ and (iii) computing the join
results. 
Note that for $\dasfaa+$, \limit, \limita and the oracles, the
indexing time additionally includes the sorting and partitioning cost
of the input objects. 
As expected the total indexing time\eat{, i.e., building $T_R$ and $I_S$,} is negligible compared to the joining time; an exception arises for FLICKR due its large number of objects.\eat{ contained in the input datasets. }

Figure~\ref{fig:algos} shows that \limita is \eat{clearly }the most efficient method for set containment joins. 
It is at least two times faster than \dasfaa.
\limita also outperforms \limit for the BMS, FLICKR and KOSARAK
datasets while for NETFLIX, both algorithms perform similarly as
(i) the $FRQ$ strategy sets limit $\ell$ to
its optimal value and (ii) the $T_R$ prefix tree for NETFLIX is quite balanced.
The adaptive approach of \limita that dynamically chooses between list
intersection and candidates verification, copes better with
(i) overestimated $\ell$ values and (ii) cases where $T_R$ is unbalanced.
Specifically, due to employing an ad-hoc limit for
each path of the prefix tree, \limita 
can be faster than \limit even 
with optimal $\ell$,
i.e., faster than \loracle (see
Figures~\ref{fig:algos}(b) and (c)). 
For these datasets, $T_R$ is quite unbalanced and thus, 
there is no fixed value of $\ell$ to outperform
the adaptive strategy. 
Note that even if $\ell$ is 
overestimated, e.g., \eat{by }using strategy  $W$--$AVG$, 
the
performance of \limita is almost the same as when 
an optimal (or close to optimal) $\ell$ is used.
Note also that the response time of \limita is very close to that of \toracle which proves the accuracy of our cost model proposed in Section~\ref{sec:limita}.
\eat{
We would like to stress that, as pointed out in\eat{ the previous section} Section~\ref{sec:employ_new}, our version of
\dasfaa is actually an improved version of the algorithm in
\cite{JampaniP05} due to arranging the items inside a object in
increasing order of their frequency (see
Section~\ref{sec:order}). Therefore, \eat{regarding the join efficiency }the
overall performance improvement achieved by \limita is in fact larger;
\limita is 5 times for BMS, 11 times for FLICKR, 3.5 times for KOSARAK
and 70 times for NETFLIX faster than the original 
\dasfaa 
(Decreasing-Hybrid in Table~\ref{tab:vary-order}).
}
We would like to stress at this point that the
overall performance improvement achieved by \limita over the original method of \cite{JampaniP05} which arranges the items inside an object in decreasing frequency order is as expected even larger compared to our version of \dasfaa;
\limita is 5 times faster than $\mathtt{org}\dasfaa$ for BMS, 11 times for FLICKR, 3.5 times for KOSARAK
and 70 times for NETFLIX.

\eat{
First, Figure~\ref{fig:memory}(a) shows that by just
using a \emph{limited} prefix tree $\ell T_R$ instead of the
\emph{entire} $T_R$,
we save at least 50\% of space \note{to 40\% apo pou prokyptei? egw
  vlepw 50\%. pes ti
  methodous sygkrineis sto figure (a) -- p.x. PRETTI* vs LIMIT+ but no
  OPJ-- kai
  vale captions sta (a) kai (b)};
note that for NETFLIX, for which $T_R$ has he highest storing cost,
due to its extremely long objects, the savings are over 90\%.
Then, Figure~\ref{fig:memory}(b) shows that, with the use of \new,
\limita uses at least $50\%$ less space in total to store indices
$\ell T_R$ and $I_S$ and fully compute the join.
As expected, due to progressively building $I_S$ the amount of space
used by \limita increases while examining the collection partitions but
it is always lower than the space for the original method of
\cite{JampaniP05} because of never actually building and storing the
entire prefix tree; only one subtree of $T_R$ is kept in memory at a
time. \pnote{Ta noumera allazoun ...}
\note{kala, kane mia sovarh perigrafh gi auth thn paragrafo. prwta apo
  ola prepei na eksigiseis ti deixnei to kathe figure kai meta na
  kaneis comment.}
  }
\eat{
\begin{figure}
\begin{center}
\begin{tabular}{|l|c|}
\hline
\multirow{2}{*}{\textbf{Dataset}} &\textbf{memory ratio}\\
 &$\ell T_R/T_R$\\\hline\hline
BMS &$50\%$\\
FLICKR &$44\%$\\
KOSARAK &$46\%$\\
NETFLIX &$3\%$\\
\hline
\end{tabular}
\\
(a) \limita  (not \new) Vs $\mathtt{org}\dasfaa$ 
\\\vspace{2ex}
\begin{tabular}{c}
\includegraphics[width=0.47\linewidth]{plots/all_mem-idx_vary-partitions.pdf}
\end{tabular}\\
(b)  \limita  (\new) Vs $\mathtt{org}\dasfaa$\\
\caption{Memory requirements (\limita using $FRQ$)}
\label{fig:memory}
\end{center}
\end{figure}
}
\begin{figure}
\begin{center}
\begin{tabular}{cc}
\begin{tabular}{lc}
\hline
\multirow{2}{*}{\textbf{Dataset}} &\textbf{memory ratio}\\
 &$\ell T_R/T_R$\\\hline\hline
BMS &$50\%$\\
FLICKR &$44\%$\\
KOSARAK &$46\%$\\
NETFLIX &$3\%$\\
\hline
\end{tabular}
&
\begin{tabular}{c}
\includegraphics[width=0.47\linewidth]{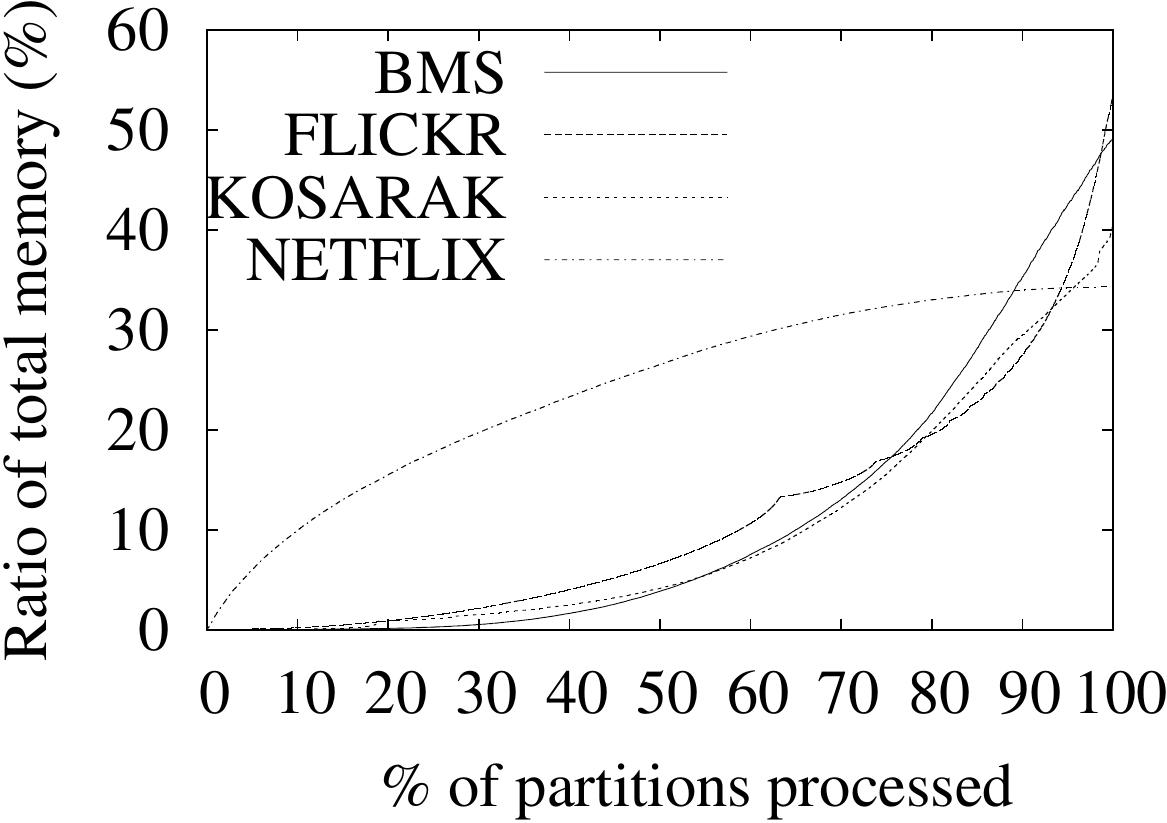}
\end{tabular}\\
(a) \limita  (not \new) Vs $\mathtt{org}\dasfaa$ &(b)  \limita  (\new) Vs $\mathtt{org}\dasfaa$\\
\end{tabular}
\caption{Memory requirements (\limita using $FRQ$)}
\label{fig:memory}
\end{center}
\end{figure}
Next,\eat{In our last experiment,} we analyze the advantage of \limita (using $FRQ$) over 
$\mathtt{org}\dasfaa$ of \cite{JampaniP05}
that arranges the items in decreasing frequency order, with respect to their memory
requirements. Figure~\ref{fig:memory}(a) shows the space
for indexing only the left-hand collection $R$ when neither method
follows the \new paradigm. We observe that by constructing
\emph{limited} prefix tree $\ell T_R$ instead of \emph{unlimited}
$T_R$, \limita saves at least 50\% of space compared to 
$\mathtt{org}\dasfaa$;
for NETFLIX, where $T_R$ has the highest storing cost due to
its extremely long objects, the savings are over 90\%. Then, in
Figure~\ref{fig:memory}(b) we consider \limita adopting \new and
report the space for indexing both input collections  while evaluating
the join, compared to 
$\mathtt{org}\dasfaa$
which does not follow the \new paradigm. We observe that by incrementally\eat{progressively} building $\ell T_R$ and $I_S$, \limita uses at least $50\%$ less space than\eat{compared to} 
$\mathtt{org}\dasfaa$. Naturally, the amount of space
used by \limita increases while examining the collection partitions, but it is always lower than the space for 
$\mathtt{org}\dasfaa$ due to never actually building and storing the
entire prefix tree; only one subtree of $\ell T_R$ is kept in memory
at a time. Finally,  notice 
the different trend 
for NETFLIX
as  its partitions have balanced sizes; in contrast for BMS,
FLICKR and KOSARAK, the first partitions contain very few objects
while the last ones are very large.

\eat{
\begin{figure}
\begin{center}
\end{center}
\begin{center}
\begin{tabular}{cc}
\includegraphics[width=0.47\linewidth]{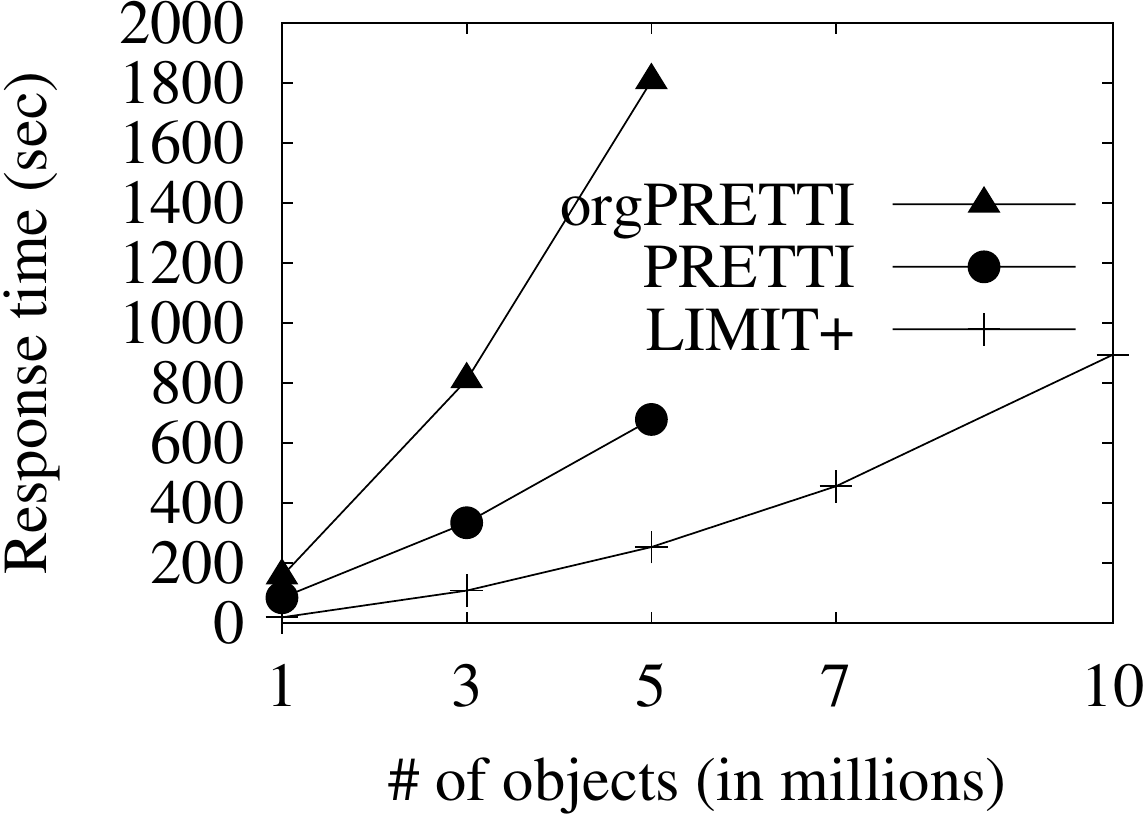}
&\includegraphics[width=0.47\linewidth]{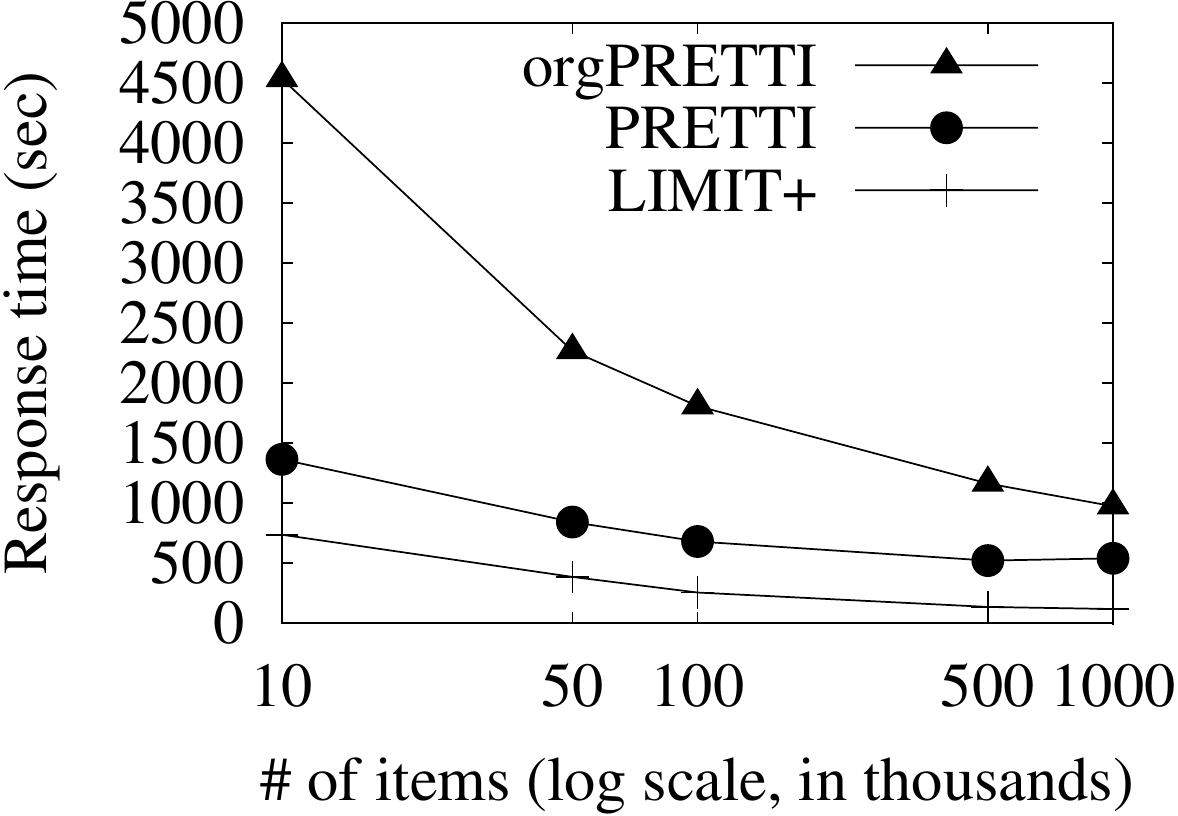}\\
(a) Cardinality
&(b) Weighted avg object length\\\\
\multicolumn{2}{c}{\includegraphics[width=0.47\linewidth]{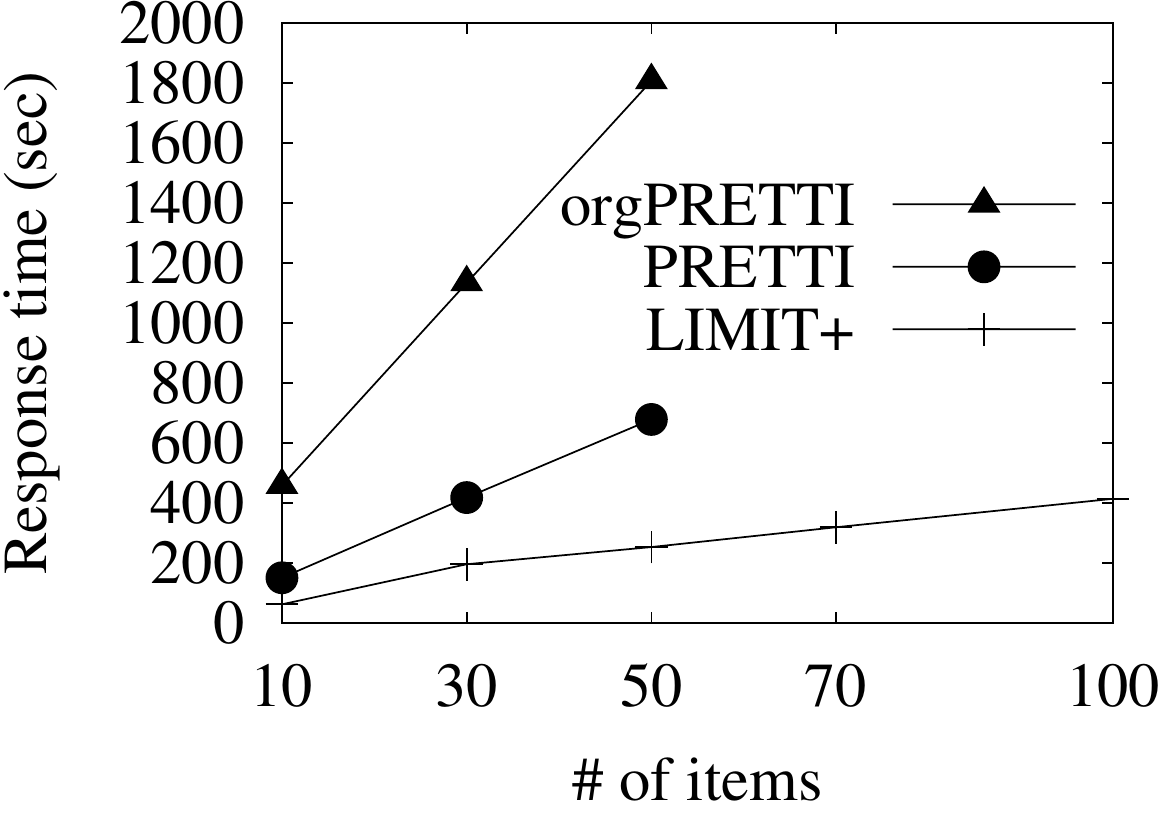}}\\
\multicolumn{2}{c}{(c) Domain size}
\end{tabular}
\caption{Scalability tests (FLICKR) \eat{, response time}}
\label{fig:scalability}
\end{center}
\end{figure}
}

\begin{figure}
\begin{center}
\begin{tabular}{cc}
\includegraphics[width=0.47\linewidth]{Fig11a.pdf}
&\includegraphics[width=0.47\linewidth]{Fig11b.pdf}\\
(a) Cardinality
&(b) Domain size\\\\
\includegraphics[width=0.47\linewidth]{Fig11c.pdf}
&\includegraphics[width=0.47\linewidth]{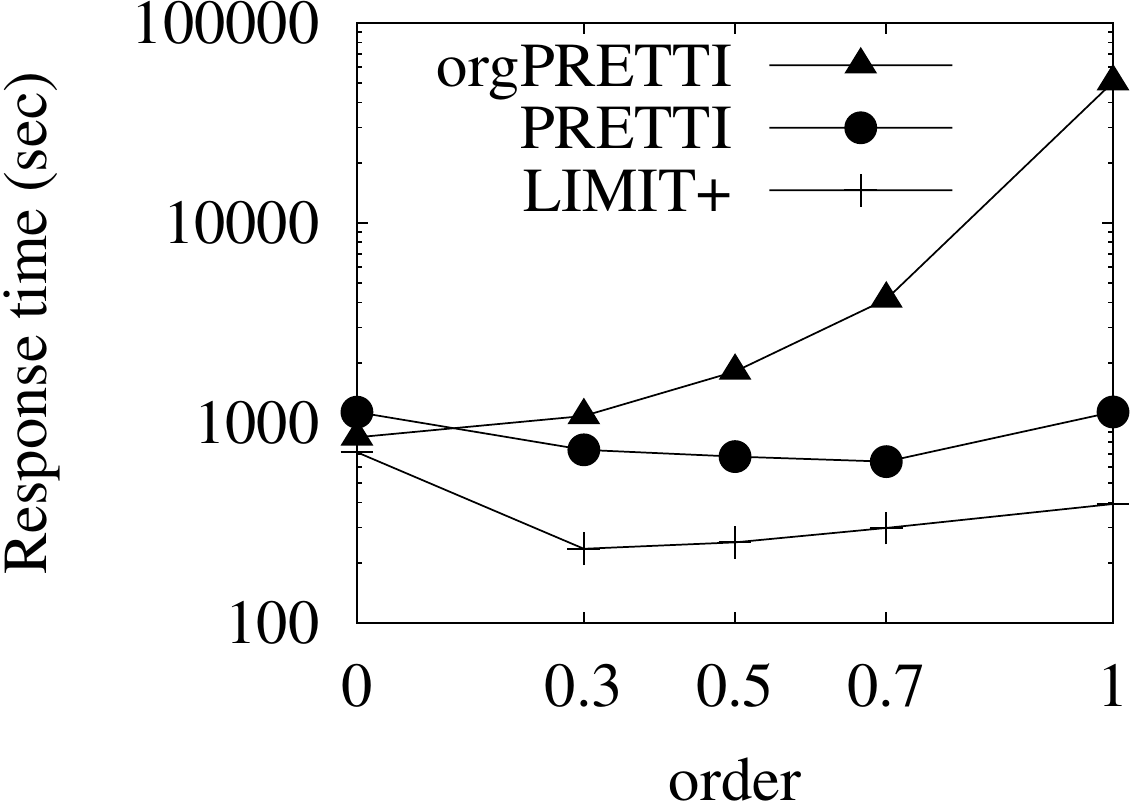}
\\
(c) Weighted avg object length
&(d) Zipfian distribution
\end{tabular}
\caption{Scalability tests on synthetic datasets (limit $\ell$ set by FRQ), default parameter values: candinality $5M$ objects, domain size $100K$ items, weighted avg object length 50 items, order of Zipfian distribution 0.5}
\label{fig:algos_synthetic}
\end{center}
\end{figure}
\eat{
Finally, we also perform scalability tests by populating FLICKR \note{what do you mean by ``populating''?}. We plot the response time of our best join method \limita against \dasfaa while varying the cardinality of the input, the weight average object length and its domain size \note{what is the ``weight average object length''? give more details on how you change these parameters.}. Figure~\ref{fig:scalability} reports on the results. As expected, the response time of both algorithms increases as the input contains more or longer objects and decreases while the domain size becomes larger. Yet, \limita is in all cases less affected compared to \dasfaa and always faster.
}

\eat{
\eat{Finally, w}We also perform scalability tests on FLICKR\eat{ dataset}
increasing its cardinality, \eat{the }weighted average object length and\eat{ its}
domain size. Note that on each test we vary \eat{only }one of the
dataset characteristics while the rest are set 
according to Table~\ref{tab:real}. To double, for instance, the
cardinality of the dataset, we generated a copy of it\eat{ by}
redistributing its items to the objects, assuring\eat{ making sure} that the
frequency of each item does not change compared to the initial input.
The new dataset produced by merging the initial \eat{dataset }and its copy has exactly the same domain and weighted average object length, but twice the number of objects. Figure~\ref{fig:scalability} reports the response time of our best \eat{join }method \limita, \dasfaa and $\mathtt{org}\dasfaa$. Naturally\eat{As expected}, the \eat{response }time of all\eat{both} algorithms increases as the input contains more or longer objects and decreases while the domain size becomes larger. Yet, \limita is in all cases less affected than\eat{compared to} \dasfaa and $\mathtt{org}\dasfaa$ and, it is always faster.
}

Finally, we present the results of our scalability tests on the synthetic datasets of Table~\ref{tab:synthetic}. Figure~\ref{fig:algos_synthetic} reports the response time of our best method \limita and the $\mathtt{org}\dasfaa$ and \dasfaa competitors. The purpose of these tests is twofold: (i) to demonstrate how the characteristics of a dataset affect the performance of the methods, and (ii) to determine their ``breaking point''. First, we notice that all methods are affected in a similar manner; their response time increases as the input contains more or longer objects and decreases while the domain size becomes larger. An exception arises in Figure~\ref{fig:algos_synthetic}(d). The performance of $\mathtt{org}\dasfaa$ is severely affected when increasing the order of the Zipfian distribution; recall that $\mathtt{org}\dasfaa$ arranges the items inside an object, in decreasing frequency order. As expected, \limita outperforms $\mathtt{org}\dasfaa$ and \dasfaa under all setups, similar to the case of real datasets. Second, we also observe that both $\mathtt{org}\dasfaa$ and \dasfaa are unable to cope with the increase of the cardinality and weighted average object length of the datasets. These two factors directly affect the size of the $T_R$ prefix tree and the memory requirements. In practice, $\mathtt{org}\dasfaa$ and \dasfaa failed to run for inputs with more than 5$M$ objects and/or when their weighted average length is larger than 50, because the \emph{unlimited} prefix tree cannot fit inside the available memory; in these cases the methods would have to adopt a block-based evaluation approach similar \cite{JampaniP05,Mamoulis03}. In contrast, \limita is able to index left-hand relation $R$ due to employing limit $\ell$ and following \new, and hence, compute the join results.

\eat{
\subsection{Scalability}
\label{sec:scalability}
\begin{figure}[!ht]
\begin{center}
\begin{tabular}{cccc}
\hspace*{-0.7cm}
vary $|R|$
&vary $|D|$
&vary $L_{avg}$
&vary $Zipf$
\\
(a)
&(b)
&(c)
&(d)\\
\end{tabular}
\caption{Execution time varying dataset characteristics}
\label{fig:scalability}
\end{center}
\end{figure}
See Figure~\ref{fig:scalability}.
}

\eat{
\subsection{The Case of NETFLIX}
\note{
\begin{itemize}
\item Extremely long objects
\item \dasfaa cannot work, failed to construct the prefix tree
\item Our methods work fine as they built only a small part of the prefix tree
\item Extremely correlated dataset, \limit outperforms
\end{itemize}
}
} 
\section{Related Work}
\label{sec:related}
Our work is related to query operators on sets\eat{-valued data}.
In this section, we 
summarize previous work done for set containment queries, set containment joins, and 
set similarity joins. 
In addition, we review previous work on efficient computation of list intersection, which is 
a core module of our algorithms.

\subsection{Set Containment Queries}
Signatures and inverted files are two alternative indexing structures for set-valued data.
Signatures are bitmaps used to exactly or approximately represent sets. With $|D|$ being the cardinality of the items domain, a set $x$ is represented by a $|D|$-length signature $sig(x)$. The $i$-th bit of $sig(x)$ is set to $1$ iff the $i$-th item of domain $D$ is present in $x$. If the sets are very small compared to $|D|$, exact signatures are expensive to store, and therefore, approximations of fixed length $l < |D|$ are typically used. 
Experimental studies \cite{HelmerM03,ZobelMR98} showed that inverted files outperform signature-based indices for set containment queries on datasets with low cardinality set objects, e.g., typical text databases. 

In \cite{TerrovitisBVSM11,TerrovitisPVS06}, the authors proposed extensions of the classic inverted file data structure, which optimize the indexing set-valued data with 
skewed item distributions. 
In \cite{ChaudhuriCKS07}, the authors proposed an indexing scheme for text documents, 
which includes inverted lists for frequent word combinations.
A main-memory method for addressing  error-tolerant set containment queries was proposed in \cite{AgrawalAK10}. 
In \cite{ZhangCSCGT12}, Zhang et al. addressed the problem of probabilistic set containment, where the contents of the sets are uncertain. The proposed solution relies on an inverted file where postings are populated with the item's probability of belonging to a certain object. The study in \cite{IbrahimF13} focused on containment queries on nested sets, and proposes an evaluation mechanism that relies on an inverted file which is populated with information for the placement of an element in the tree of nested sets. The above methods use classic inverted files or extend them either by trading update and creation costs for response time \cite{AgrawalAK10,ChaudhuriCKS07,TerrovitisBVSM11,TerrovitisPVS06} or by adding information that is needed for more complex queries \cite{IbrahimF13,ZhangCSCGT12}.
Employing these extended inverted files for set containment joins (i.e., in place of our $I_S$) is orthogonal to our work.

\subsection{Set Containment Joins}
In \cite{HelmerM97}, the \emph{Signature Nested Loops} ($\mathtt{SNL}$) Join and the \emph{Signature Hash} Join ($\mathtt{SHJ}$) algorithm for set containment joins were proposed, with $\mathtt{SHJ}$ shown to be the fastest. 
For each set object $r$ in the left-hand collection $R$, both algorithms compare signatures to identify every object $s$ in the right-hand collection $S$ with
$sig(r)~\&~\neg sig(s)=0$
and $|r| \leq |s|$ (filter phase), and then, perform explicit set comparison to discard false drops (verification phase). Later, the hash-based algorithms \emph{Partitioned Set Join} ($\mathtt{PSJ}$) in \cite{RamasamyPNK00} and \emph{Divide-and-Conquer Set Join} ($\mathtt{DCJ}$) in \cite{MelnikG02} aimed at reducing the quadratic cost of the algorithms in \cite{HelmerM97}. In these approaches, the input collections are partitioned based on hash functions such that object pairs of the join result fall in the same partition. Finally, Melnik and Molina \cite{MelnikG03} proposed adaptive extensions to $\mathtt{PSJ}$ and $\mathtt{DCJ}$, termed $\mathtt{APSJ}$ and $\mathtt{ADCJ}$, respectively, to overcome the problem of a potentially poor partitioning quality.

Inverted files were employed by \cite{JampaniP05,Mamoulis03} for set containment joins. Specifically, in \cite{Mamoulis03}, Mamoulis proposed a \emph{Block Nested Loops} ($\mathtt{BNL}$) Join algorithm that indexes the right-hand collection $S$ by an inverted file \eat{$S_{IL}$}$I_S$. The algorithm iterates through each object $r$ in the left-hand collection $R$ and intersects the corresponding postings lists of \eat{$S_{IL}$}$I_S$ to identify the objects in $S$ that contain $r$. The experimental analysis in \cite{Mamoulis03} showed that $\mathtt{BNL}$ is significantly faster than previous signature-based methods \cite{HelmerM97,RamasamyPNK00}. In \cite{JampaniP05}, Jampani and Pudi targeted the major weakness of $\mathtt{BNL}$; the fact that the overlaps between set objects are not taken into account. T\eat{o this end, t}he proposed algorithm \dasfaa, employs a prefix tree on the left-hand collection, allowing 
list intersections for multiple objects with a common prefix to be performed just once. Experiments in \cite{JampaniP05} showed that \dasfaa outperforms $\mathtt{BNL}$ and previous signature-based methods of \cite{MelnikG03,RamasamyPNK00}. Our work first identifies and tackles the shortcomings of the \dasfaa algorithm and then, proposes a new join paradigm.

\subsection{Set Similarity Joins}
The set similarity join
finds object pairs $(r,s)$ from \eat{two }input collections $R$ and $S$, such that $sim(r, s) \geq \theta$, where $sim(\cdot,\cdot)$ is a similarity function (e.g., Jaccard coefficient) and $\theta$ is a given threshold. Computing set similarity joins based on inverted files was first proposed in \cite{SarawagiK04}: for each object in one input, e.g., $r \in R$, the inverted lists that correspond to $r$'s elements on the other collection are scanned to accumulate the overlap between $r$ and all objects $s \in S$. Among the optimization techniques on top of this baseline, Chaudhuri et al. \cite{ChaudhuriGK06} proposed a filter-refinement framework based on \emph{prefix filtering}; for two internally sorted set objects $r$ and $s$ to satisfy $sim(r, s) \geq \theta$ their prefixes should have at least some minimum overlap. Later, \cite{ArasuGK06,BayardoMS07,RibeiroH09,XiaoWLY08} built upon\eat{ and extended} prefix filtering to reduce the number of candidates generated.  Recently, Bouros et al. \cite{BourosGM12} proposed a grouping optimization technique to boost the performance of the method in \cite{XiaoWLY08}, and Wang et al. \cite{WangLF12} devised a cost model to judiciously select the appropriate prefix for a set object. An experimental comparison of set similarity join methods can be found in \cite{JiangLFL14}.
%
In theory, the above methods can be employed for set containment joins, considering for instance the asymmetric containment Jaccard measure, $sim(r,s) = \frac{|r \cap s|}{|r|}$ and \eat{a }threshold $\theta\! =\! 1$. In practice, however, this approach is not efficient\eat{ because} as it generates a large number of candidates. For each object $r \in R$\eat{ the} prefix filtering can only prune objects in $S$ that do not contain $r$'s first item while the rest of the candidates need to be verified by comparing the actual set objects. Therefore, the ideas proposed in previous work on set similarity joins are not applicable to set containment joins.

\subsection{List Intersection}
In \cite{DemaineLM00,DemaineLM01}, Demaine et al. presented an adaptive algorithm for computing set intersections, unions and differences. Specifically, the algorithm in \cite{DemaineLM00} (ameliorated in \cite{DemaineLM01} and extended in \cite{BarbayK02}) polls each list in a round robin fashion. Baeza-Yates \cite{Baeza-Yates04} proposed an algorithm that adapts to the input values and performs quite well in average. It can be seen as a natural hybrid of the binary search and the merge-sort approach. Experimental comparison of the above, among others, methods of list intersection, with respect to their CPU cost can be found in \cite{Baeza-YatesS05,Baeza-YatesS10,BarbayLLS09}. The trade-off between the way sets\eat{-valued data} are stored and the way they are accessed in the context of the intersection operator was studied in \cite{CulpepperM10}. Finally, recent work \cite{TatikondaCJ11,TatikondaJCP09,TsirogiannisGK09} considered list intersection with respect to the characteristics of modern hardware and focused on balancing the load between multiple cores. In \cite{TatikondaCJ11,TatikondaJCP09}, Tatikonda et al. proposed inter-query parallelism and intra-query parallelism. The former exploits parallelism between different queries, while the latter parallelizes the processing within a single query. On the other hand, the algorithm in \cite{TsirogiannisGK09} probes the lists in order to gather statistics that would allow efficient exploration of the multi-level cache hierarchy.
Efficient 
list intersection is orthogonal to our set containment join problem.
Yet, in Section~\ref{sec:order}, we employ a hybrid list intersection method based on 
\cite{Baeza-Yates04} to determine the preferred 
ordering of the items inside the objects.

\subsection{Estimating Set Intersection Size}
Estimating the intersection size of two sets has received a lot of attention in the area of information retrieval \cite{Broder97,Broder00,ChenKKM00,ChenKKM03,Kohler10}\eat{ in order}, to determine the similarity between two documents modelled as sets of terms. Given \eat{two }sets $A$ and $B$, the basic idea is to compute via sampling small {\em sketches}\eat{ of the sets,}
$\mathcal{S}(A)$ and $\mathcal{S}(B)$, respectively. Then, $|\mathcal{S}(A) \cap \mathcal{S}(B)|$  is used as an estimation of $|A \cap B|$.
Our adaptive methodology for set containment joins (Section \ref{sec:limita}) involves estimating the size of a list intersection. Yet\eat{However}, the methods discussed above are not applicable as \eat{because }they require an expensive preprocessing step, i.e., precomputing and indexing the sketches for every list of the inverted index at the right-hand collection.
In addition, one of the two lists at each intersection (i.e., candidates list $CL$) is the result of previous intersections.
Thus, computing the sketch of $CL$ should be done on-the-fly, i.e., the overall cost of the sketch-based intersection would exceed the cost of performing the exact list intersection 
(especially since $CL$ becomes shorter every time it is intersected with a inverted list of the right-hand collection).



\section{Conclusion}
\label{sec:concl}
In this paper we revisited the set containment join $R \bowtie_{\subseteq} S$ between two collections $R$ and $S$ of set objects $r$ and $s$, respectively.  We presented a framework which improves the state-the-art method \dasfaa, greatly reducing the space requirements and time cost of the join.
Particularly, we first proposed an adaptive methodology (algorithms \limit and \limita) that limits the prefix tree constructed for the left-hand collection $R$. Second, we proposed a novel join paradigm termed \new that partitions the objects of each collection based on their first contained item, and then examines these partitions to evaluate the join while progressively building the indices on $R$ and $S$.
Finally, we conducted extensive experiments on real datasets to demonstrate the advantage of our methodology.

Besides the fact that the \new\eat{ join} paradigm significantly reduces both the join cost and the maximum memory requirements, it can be applied in a parallel processing environment. For instance, by assigning each partition $R_i$ of the left-hand collection to a single computer node $v_i$ while replicating the partitions of the right-hand collection such that node $v_i$ gets every object in $S$ which starts either by item $i$ or an item before $i$ according to the global item ordering, our method 
runs at each node and there is no need for communication among the nodes, since join results are independent and there are no duplicates.  In the future, we plan to investigate the potential of such an implementation.
\eat{
\pnote{
Future work:
\begin{itemize}
\item Distributed environments
\item GPU processing
\end{itemize}
}
}

\eat{
\begin{acknowledgements}
This work was supported by the HKU 714212E grant from Hong Kong RGC and the MEDA project within GSRT’s KRIPIS action, funded by Greece and the European Regional Development Fund of the European Union under the O.P. Competitiveness and Entrepreneurship, NSRF 2007-2013 and the Regional Operational Program of ATTIKI.
\end{acknowledgements}
}

%
%
\bibliographystyle{abbrv}
\bibliography{references}

\eat{
\section*{Author Biographies}
\leavevmode

\vbox{%
\begin{wrapfigure}{l}{80pt}\vspace*{-21pt}
 {\includegraphics[height=1.5in]{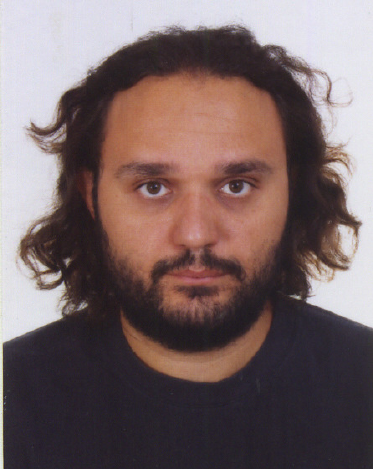}}
\vspace*{140pt}
\end{wrapfigure}
\noindent\small 
\textbf{Panagiotis Bouros} received his diploma and PhD degree from the School of Electrical and Computer Engineering at the National Technical University of Athens, Greece, in 2003 and 2011, respectively. He is currently a post-doctoral researcher for the Department of Computer Science at Aarhus University, Denmark. Prior to that he was with Humboldt-Universit\"{a}t zu Berlin, Germany and the University of Hong Kong, Hong Kong SAR, China. His research focuses on managing and querying complex data types including spatial, temporal and text, and on routing optimization problems.
\vadjust{\vspace{40pt}} 
}

\vbox{%
\begin{wrapfigure}{l}{80pt}\vspace*{-21pt}
{\includegraphics[height=1.62in]{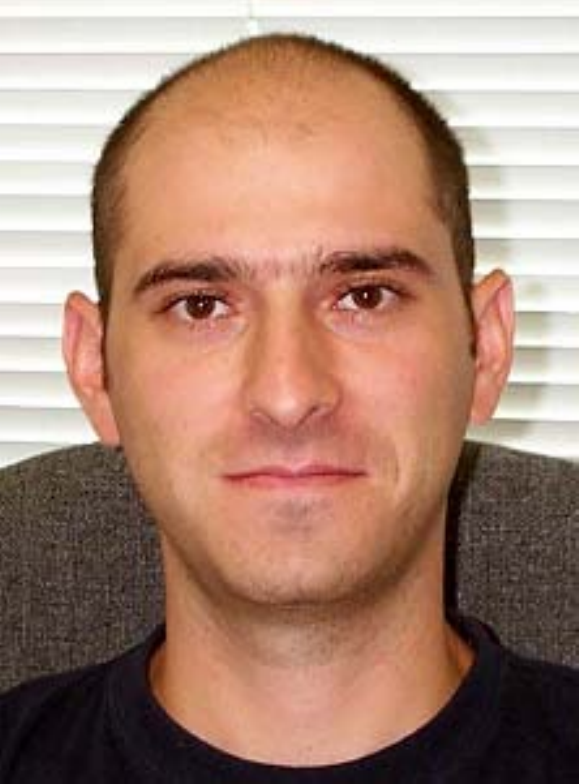}}
\vspace*{140pt}
\end{wrapfigure}
\noindent\small 
\textbf{Nikos Mamoulis} received a diploma in Computer Engineering and Informatics in 1995 from the University of Patras, Greece, and a PhD in Computer Science in 2000 from the Hong Kong University of Science and Technology. He is currently a professor at the Department of Computer Science, University of Hong Kong, which he joined in 2001. His research focuses on management and mining of complex data types, including spatial, spatio-temporal, object-relational, multimedia, text and semi-structured data. He has served on the program committee of over 80 international conferences on data management and mining. He is an associate editor for IEEE TKDE and the VLDB Journal.
\vadjust{\vspace{40pt}} 
}

\vbox{%
\begin{wrapfigure}{l}{80pt}\vspace*{-21pt}
{\includegraphics[height=1.6in]{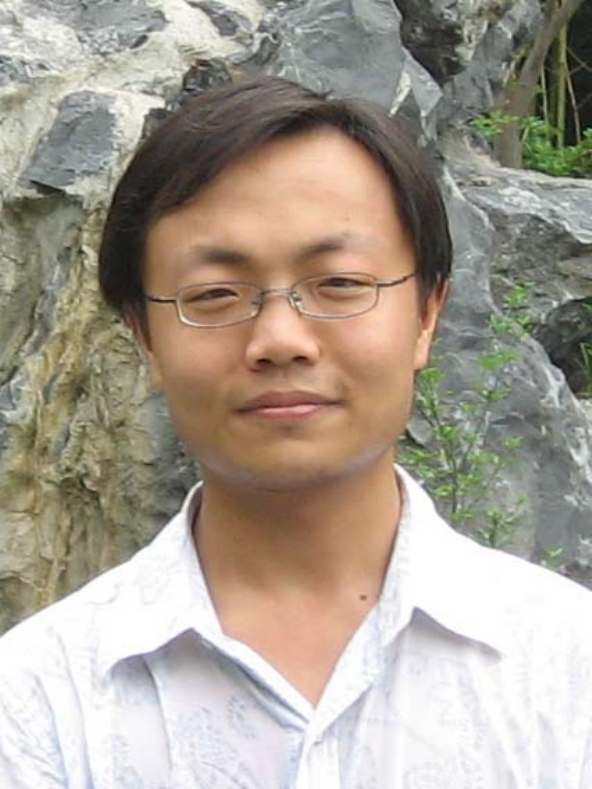}}
\vspace*{140pt}
\end{wrapfigure}
\noindent\small 
\textbf{Shen Ge} received his PhD degree from the Department of Computer Science, University of Hong Kong in 2012, and bachelors and masters degrees in computer science from the Department of Computer Science and Technology in Nanjing University, China, in 2005 and 2008, respectively. His research focuses on query processing on multi-dimensional and spatial-textual data.
\vadjust{\vspace{20pt}}
}

\vbox{%
\begin{wrapfigure}{l}{80pt}\vspace*{-21pt}
{\includegraphics[height=1.6in]{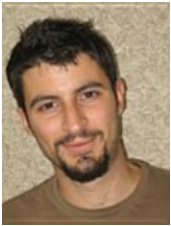}}
\vspace*{140pt}
\end{wrapfigure}
\noindent\small 
\textbf{Manolis Terrovitis} is an associate researcher at the Institute for the Management of Information Systems (IMIS) of the Research and Innovation Centre in Information, Communication and Knowledge Technologies ``Athena''. He received his PhD in 2007 from the National Technical University of Athens. His main research interests lie in the areas of data privacy, indexing and query evaluation.
}
}
\end{document}